\documentclass{article}
\usepackage[utf8]{inputenc}
\usepackage{amsmath}
\usepackage{amssymb} \usepackage{graphics}
\usepackage[all]{xy}
\usepackage{chemarrow}

\title{Spin Geometry and Some Applications \thanks{Based on the lectures given at METU Physics Department between the dates 27 October 2017 and 5 January 2018.}}
\author{\"Umit Ertem \thanks{umitertemm@gmail.com} \thanks{Department of Physics,
Ankara University, Faculty of Sciences, 06100, Tando\u gan, Ankara, Turkey}}

\begin{document}

\maketitle

\begin{abstract}

In this review, basic definitions of spin geometry are given and some of its applications to supersymmetry, supergravity and condensed matter physics are summarized. Clifford algebras and spinors are defined and the first-order differential operators on spinors which lead to the definitions of twistor and Killing spinors are discussed. Holonomy classification for manifolds admitting parallel and Killing spinors are given. Killing-Yano and conformal Killing-Yano forms resulting from the spinor bilinears of Killing and twistor spinors are introduced and the symmetry operators of special spinor equations are constructed in terms of them. Spinor bilinears and symmetry operators are used for constructing the extended superalgebras from twistor and Killing spinors. A method to obtain harmonic spinors from twistor spinors and potential forms is given and its implications on finding solutions of the Seiberg-Witten equations are discussed. Supergravity Killing spinors defined in bosonic supergravity theories are considered and possible Lie algebra structures satisfied by their spinor bilinears are examined. Spin raising and lowering operators for massless field equations with different spins are constructed and the case for Rarita-Schwinger fields is investigated. The derivation of the periodic table of topological insulators and superconductors in terms of Clifford chessboard and index of Dirac operators is summarized.

\end{abstract}

\tableofcontents

\section{Introduction}

Spin geometry deals with manifolds admitting a spin structure which is a consistent double cover of the orientation structure, namely lifting the $SO(n)$ structure to the $\text{Spin}(n)$ structure in $n$ dimensions \cite{Lawson Michelson, Benn Tucker, Friedrich, Bourguignon et al}. These manifolds are called spin manifolds and they admit a spinor bundle corresponding to the associated bundle of the principle bundle of the Spin group which can be defined via reduction of the Clifford algebra bundle on the manifold. The sections of the spinor bundle are called spinor fields. Two first-order differential operators can be defined on the spinor bundle that are Dirac operator and twistor operator \cite{Baum Friedrich Grunewald Kath}. The index structure of the Dirac operator is related to the topological invariants of the manifold via Atiyah-Singer index theorems \cite{Atiyah Singer, Atiyah Singer2, Atiyah}.

Special spinor fields that are in the kernels of Dirac and twistor operators play important roles in mathematical physics. Spinor fields that are in the kernel of the Dirac operator are called harmonic spinors and they correspond to the solutions of the massless Dirac equation. Moreover, the eigenspinors of the Dirac operator with non-zero eigenvalue correspond to the solutions of the massive Dirac equation. Spinor fields that are in the kernel of the twistor operator are called twistor spinors \cite{Lichnerowicz, Baum Leitner}. Twistor spinors which are also eigenspinors of the Dirac operator are called Killing spinors and a subset of Killing spinors which are parallel with respect to the covariant derivative are called parallel spinors \cite{Lichnerowicz2, Baum, Leitner, Alekseevsky Cortes}. Killing spinors and parallel spinors appear as supersymmetry generators of supersymmetric field theories and supergravity theories in various dimensions \cite{Duff Nilsson Pope, Van Proeyen, Freedman Van Proeyen}. Low energy limits of ten dimensional string theories and eleven dimensional M-theory correspond to the supergravity theories in relevant dimensions and preserving the supersymmetry transformations gives rise to the existence of special spinor fields on the backgrounds corresponding to the solutions of these theories \cite{Green Schwarz Witten, Becker Becker Schwarz, OFarrill Papadopoulos, O Farrill Meessen Philip}. So, the spin geometry methods are in central importance for the supergravity and string theories. On the other hand, a recently growing field of topological phases of matter also appear to have close connections with spin geometry methods \cite{Kitaev, Stone Chiu Roy, Abramovici Kalugin}. The periodic table of topological insulators and superconductors, which classifies these materials in different dimensions and symmetry classes, are originated from the Clifford algebra chessboard and the index of Dirac operators \cite{Ertem}. The Dirac operators in this context arise as the Dirac Hamiltonians representing low energy behaviour of these topological materials.

In this review, we concentrate on the basic properties of twistor, Killing and parallel spinors and their applications to different areas of mathematics and physics. We investigate the relations between special types of spinors and the hidden symmetries of a manifold called Killing-Yano (KY) and conformal Killing-Yano (CKY) forms which are antisymmetric generalizations of Killing and conformal Killing vector fields to higher degree forms \cite{Acik Ertem}. Symmetry operators of special type of spinors which are written in terms of hidden symmetries and the construction of extended superalgebras are also considered \cite{Ertem2, Ertem3, Ertem4, Ertem5, Ertem6}. Relations between harmonic spinors and twistor spinors are used to obtain the solutions of the Seiberg-Witten equations by generalizing these relations to the gauged case \cite{Ertem7, Benn Kress}. Supergravity Killing spinors which arise as the supersymmetry generators of supergravity theories are connected to the parallel and Killing spinors for different flux choices in the relevant theory and the solutions of the massless field equations for different spin particles are connected to each other via spin raising and lowering methods by using twistor spinors \cite{Ertem8, Acik Ertem2, Acik Ertem3}. Moreover, the derivation of the periodic table of topological insulators and superconductors via Clifford algebra chessboard and the index of Dirac operators is also discussed \cite{Ertem}.

The paper is organized as follows. In section 1, we define the basic concepts about Clifford algebras and spinors with their bundle structures on manifolds. Section 2 includes the construction of Dirac and twistor operators and the definitions of the twistor, Killing and parallel spinors. The integrability conditions related to the existence of these special types of spinors on relevant manifolds are also derived in terms of the curvature characteristics of the manifolds. Special holonomy manifolds and their relation to the existence of parallel spinors are considered in Section 3. Cone construction, holonomy classification for Killing spinors and the relations between the special holonomy manifolds and supergravity backgrounds are also discussed. In section 4, the construction of spinor bilinears and the definitions of Dirac currents and $p$-form Dirac currents of spinors are given. It is shown that the $p$-form Dirac currents of twistor spinors correspond to CKY forms while the $p$-form Dirac currents of Killing spinors are either KY forms or their Hodge duals. Relations between Maxwell-like and Duffin-Kemmer-Petiau (DKP) field equations and the $p$-form Dirac currents of Killing spinors are also discussed. Symmetry operators of Killing and twistor spinors as generalizations of the Lie derivative on spinor fields are considered in section 5. It is shown that these symmetry operators can be written in terms of KY and CKY forms in constant curvature manifolds. In section 6, the superalgebra structures of KY and CKY forms are discussed and the construction of extended superalgebras from twistor and Killing spinors are given. Section 7 deals with a method to construct harmonic spinors from twistor spinors by using potential forms. It is generalized to the case of gauged harmonic spinors and gauged twistor spinors which are defined in terms of Spin$^c$ structures. A discussion about finding the solutions of the Seiberg-Witten equations from gauged twistor spinors is also included. Supergravity Killing spinors arising from the fermionic variations of bosonic supergravity theories is the topic of section 8. Their relations to parallel and Killing spinors and the Lie algebra structures satisfied by their spinor bilinears are also considered. In section 9, spin raising and lowering operators constructed out of twistor spinors for massless field equations with different spins are investigated. The conditions to construct spin raising and lowering operators for massless spin-$\frac{3}{2}$ Rarita-Schwinger fields are obtained. Section 10 concerns with the topic of topological insulators. After giving basic definitions about Bloch bundles, Chern insulators and $\mathbb{Z}_2$ insulators are considered with the examples of Haldane and Kane-Mele models. The classification of topological insulators in terms of a periodic table is given and the derivation of this periodic table by using Clifford algebra chessboard, index of Dirac operators and K-theory arguments is discussed.

\section{Clifford Algebras and Spinors}

In this section, we give the basic definitions of Clifford algebras and Pin and Spin groups. Clifford and spinor bundles and basic operations on them are also defined.

\subsection{Clifford algebra and Spin group}

Let us consider an $n$-dimensional vector space $V$ defined on a field $\mathbb{K}$, with a quadratic form $Q:V\times V\rightarrow\mathbb{K}$. The dual space is defined by $V^*=\{T\,|\,T:V\rightarrow\mathbb{K}\}$. Given a basis $\{X_a\}$ for $V$ with $a=1,...,n$, the dual basis $\{e^a\}$ for $V^*$ is defined as $e^a(X_b)=\delta^a_b$ where $\delta^a_b$ is the Kronecker delta.

A tensor $N$ of type $(p,q)$ is a multilinear mapping
\[
N:\underbrace{V^*\times V^*\times...\times V^*}_p\times \underbrace{V\times
V\times...\times V}_q\longrightarrow\mathbb{K}
\]
The space of such tensors is denoted by $T^p_q(V)$ and with the tensor product $\otimes$ it is the tensor algebra on $V$. When $q$ is zero, we say that $N$ has degree $p$. The space of totally antisymmetric tensors of degree $p$ is denoted by $\Lambda_p(V)$, and its elements are called $p$-forms. The space of $p$-forms has dimension $\dim\Lambda_p(V)=\binom{n}{p}$. The space of exterior forms is $\Lambda(V)=\bigoplus_{p=0}^n\Lambda_p(V)$, where $\Lambda_0(V)=\mathbb{K}$ and $\Lambda_1(V)\cong V$.

We can define two different algebra structures on $\Lambda(V)$;

i) The exterior (wedge) product
\begin{eqnarray}
\wedge : \Lambda_p(V)\times\Lambda_q(V)&\longrightarrow&\Lambda_{p+q}(V)\nonumber\\
\omega\,,\,\phi&\longmapsto&\omega\wedge\phi=\frac{1}{2}\left(\omega\otimes\phi-\phi\otimes\omega\right)\nonumber
\end{eqnarray}
gives $\Lambda(V)$ the exterior algebra structure.

ii) The Clifford product
\begin{eqnarray}
. : \Lambda_1(V)\times\Lambda_1(V)&\longrightarrow&\Lambda(V)\nonumber\\
x\,,\,y&\longmapsto&x.y+y.x=2Q(x,y)\nonumber
\end{eqnarray}
gives $\Lambda(V)$ the Clifford algebra structure which is denoted by $Cl(V,Q)$.

We can define two automorphisms on $\Lambda(V)$;

i) The automorphism $\eta:\Lambda_p(V)\longrightarrow\Lambda_p(V)$ with $\eta^2=1$ is defined by $\eta\omega=(-1)^p\omega$ and we have $\eta(\omega\circ\phi)=\eta\omega\circ\eta\phi$ for $\omega,\phi\in\Lambda_p(V)$ where $\circ$ denotes the wedge or Clifford product.

ii) The anti-automorphism $\xi:\Lambda_p(V)\longrightarrow\Lambda_p(V)$ with $\xi^2=1$ is defined by $\xi\omega=(-1)^{\lfloor p/2\rfloor}\omega$ and we have $\xi(\omega\circ\phi)=\xi\phi\circ\xi\omega$ for $\omega,\phi\in\Lambda_p(V)$ where $\circ$ denotes the wedge or Clifford product and $\lfloor\rfloor$ is the floor function which takes the integer part of the argument.

By the automorphism $\eta$, $\Lambda(V)$ gains a $\mathbb{Z}_2$-grading
\[
\Lambda(V)=\Lambda^{\text{even}}(V)\oplus\Lambda^{\text{odd}}(V)
\]
where for $\omega\in\Lambda^{\text{even}}(V)$ we have $\eta\omega=\omega$ and for $\omega\in\Lambda^{\text{odd}}(V)$ we have $\eta\omega=-\omega$.
The automorphism $\eta$ also induces a $\mathbb{Z}_2$-grading on the Clifford algebra
\[
Cl(V,Q)=Cl^0(V,Q)\oplus Cl^1(V,Q)
\]
where for $\phi\in Cl^0(V,Q)$ we have $\eta\phi=\phi$ and for $\phi\in Cl^1(V,Q)$ we have $\eta\phi=-\phi$. So, $Cl^0(V,Q)\cong\Lambda^{\text{even}}(V)$ and $Cl^1(V,Q)\cong\Lambda^{\text{odd}}(V)$ and $Cl(V,Q)$ is a $\mathbb{Z}_2$-graded algebra or a superalgebra with the even part $Cl^0(V,Q)$ and the odd part $Cl^1(V,Q)$.

Certain elements of $Cl(V,Q)$ are invertible. For example, from the definition of the Clifford product, one has $x^2=Q(x,x)$ for $x\in V$. If $Q(x,x)\neq 0$, then we have $x^{-1}=\frac{x}{Q(x,x)}$. The group of invertible elements in $Cl(V,Q)$ is denoted by
\[
Cl^{\times}(V,Q)=\left\{\phi\in Cl(V,Q)\,|\,\phi.\phi^{-1}=\phi^{-1}.\phi=1\right\}.
\]
The Clifford group $P(V,Q)$ consists of the invertible elements in $Cl(V,Q)$ which commutes with the elements of $V\cong\Lambda_1(V)$;
\[
P(V,Q)=\left\{\phi\in Cl^{\times}(V,Q)\,|\,\phi V\phi^{-1}=V\right\}.
\]
Pin group is a subgroup of the Clifford group which defined by the anti-automorphism $\xi$ as
\[
\text{Pin}(V,Q)=\left\{\phi\in P(V,Q)\,|\,\phi^{\xi}.\phi=1\right\}.
\]
Spin group consists of the elements of the Pin group which are in the even subalgebra $Cl^0(V,Q)$ of the Clifford algebra
\[
\text{Spin}(V,Q)=Pin(V,Q)\cap Cl^0(V,Q).
\]
Pin group is the double cover of the orthogonal group whose elements correspond to rotations
\[
\text{Pin}(V,Q)\xrightarrow{2:1}O(V,Q)
\]
and Spin group is the double cover of the special orthogonal group whose elements correspond to rotations without reflections (for $n>2$)
\[
\text{Spin}(V,Q)\xrightarrow{2:1}SO(V,Q).
\]
The spaces that carry the representations of Pin and Spin groups lead to the definition of pinors and spinors. Representations of the Pin group are called pinors, $\rho:\text{Pin}(V,Q)\rightarrow \text{End}\,{\Xi}$ where $\Xi$ is the pinor space and representations of the Spin group are called spinors, $\rho:\text{Spin}(V,Q)\rightarrow \text{End}\,{\Sigma}$ where $\Sigma$ is the spinor space.

\subsection{Clifford and spinor bundles}

Let us consider an $n$-manifold $M$ with metric $g$, tangent bundle $TM$ and the cotangent bundle $T^*M$. We denote the frame basis on $TM$ as $\{X_a\}$ and co-frame basis on $T^*M$ as $e^a$ with the property $e^a(X_b)=\delta^a_b$.

We can define the $p$-form bundle $\Lambda^pM$ on $M$ with the wedge product $\wedge$ and $\Lambda M=\bigoplus_{p=0}^n\Lambda^p(M)$ corresponds to the exterior bundle on $M$. From the Levi-Civita connection $\nabla$ on $M$, we can define the exterior derivative
\[
d:\Lambda^pM\longrightarrow\Lambda^{p+1}M
\]
with $d=e^a\wedge\nabla_{X_a}$ and the co-derivative
\[
\delta:\Lambda^pM\longrightarrow\Lambda^{p-1}M
\]
with $\delta=-i_{X^a}\nabla_{X_a}$ where $i_X:\Lambda^pM\longrightarrow\Lambda^{p-1}M$ is the interior derivative (contraction) operation defined by
\begin{equation}
\omega(X_1,...,X_p)=i_{X_1}\omega(X_2,...,X_p)
\end{equation}
for $\omega\in\Lambda^pM$. We can also define the Hodge star operation $*:\Lambda^pM\longrightarrow\Lambda^{n-p}M$ as
\begin{equation}
*(e^{a_1}\wedge...\wedge e^{a_p})=\frac{1}{(n-p)!}{\epsilon^{a_1...a_p}}_{a_{p+1}...a_n}e^{a_{p+1}}\wedge...\wedge e^{a_n}
\end{equation}
where ${\epsilon^{a_1...a_p}}_{a_{p+1}...a_n}$ denotes the totally antisymmetric tensor.

If we change the product rule for $e^a$ from the wedge product to the Clifford product
\begin{equation}
e^a.e^b+e^b.e^a=2g^{ab}
\end{equation}
for the components of the metric $g^{ab}$, then we obtain the Clifford bundle $Cl(M)$ on $M$ whose sections correspond to the inhomogeneous differential forms.

There is a relation between Clifford and exterior bundles. The action of $Cl(M)$ on $\Lambda M$ is given for $x\in\Lambda^1M$ and $\omega\in\Lambda^pM$ as
\begin{eqnarray}
x.\omega&=&x\wedge\omega+i_{\widetilde{x}}\omega\\
\omega.x&=&x\wedge\eta\omega-i_{\widetilde{x}}\eta\omega
\end{eqnarray}
where $\widetilde{x}$ is the vector field metric dual to $x$ which is defined by $x(Y)=g(\widetilde{x},Y)$ for $Y\in TM$.

Although the Spin group $\text{Spin}(M)$ is a subgroup in $Cl(M)$, there are topological obstructions to define a spin bundle on $M$. In general, a basis $\{X_a\}$ of $TM$ transforms to another basis under the group $GL(n)$ (this is called as $GL(n)$-structure). If $\{X_a\}$ is an orthogonal basis, then it transforms under the rotation group $O(n)$ ($O(n)$-structure).

For a bundle $\pi:E\rightarrow M$ on $M$, one can define some characteristic classes to characterize its topological structure. A bundle $E$ is always locally trivial $\pi:E=F\times M\rightarrow M$, but globally it can be non-trivial. The non-triviality of a bundle is measured by the characteristic classes \cite{Nakahara}.

For the tangent bundle $TM\rightarrow M$, if the first Stiefel-Whitney class vanishes $w_1(M)=0$, then one can define an orientation on $M$ and the frame basis $\{X_a\}$ transforms under the group $SO(n)$ ($SO(n)$-structure). If the second Stiefel-Whitney class also vanishes $w_2(M)=0$, then one can define a lifting of $SO(n)$-structure to $Spin(n)$-structure by $Spin(n)\xrightarrow{2:1}SO(n)$. Then, it is said that the manifold $M$ has a {\bf{spin structure}} and $M$ is called as a {\bf{spin manifold}}.

So, in that case, one can define a spinor bundle $\Sigma M$ on $M$ induced from the Clifford bundle $Cl(M)$. Let $\Sigma$ denotes the representation space of $\text{Spin}(n)$, the associated bundle of the $\text{Spin}(n)$ bundle $P\rightarrow M$ which is defined by
\[
\Sigma M=P\times_{\text{Spin}(n)}\Sigma
\]
is called the spinor bundle $\Sigma M\rightarrow M$ on $M$. The sections of the spinor bundle are called spinor fields.

One can also define an action of $Cl(M)$ on $\Sigma M$ via the Clifford multiplication. For $\phi\in Cl(M)$ and $\psi\in\Sigma M$
\begin{eqnarray}
c:Cl(M)\times\Sigma M&\longrightarrow&\Sigma M\nonumber\\
\phi\,,\psi\,&\longmapsto&c(\phi,\psi)=\phi.\psi
\end{eqnarray}

\section{Twistor and Killing Spinors}

This section includes the construction of the basic first-order differential operators on the spinor bundle which are Dirac and twistor operators. After giving the definitions of harmonic, twistor, Killing and parallel spinors, the integrability conditions that constrain the curvature characteristic of manifolds admitting these special types of spinors are obtained.

\subsection{Dirac and twistor operators}

Let us consider the Levi-Civita connection $\nabla$ on $TM$. $\nabla$ can be induced onto the bundles $\Lambda M$, $Cl(M)$ and $\Sigma M$. So, we have the connection on spinor fields induced from the Levi-Civita connection;
\begin{eqnarray}
\nabla:\Sigma M&\longrightarrow&T^*M\otimes\Sigma M\nonumber\\
\psi&\longmapsto&e^a\otimes\nabla_{X_a}\psi\nonumber
\end{eqnarray}
From the Clifford action of $Cl(M)$ on $\Sigma M$, for $T^*M\cong\Lambda_1M\subset Cl(M)$ we have
\begin{eqnarray}
c:T^*M\otimes\Sigma M&\longrightarrow&\Sigma M\nonumber\\
\widetilde{X}\otimes\psi&\longmapsto&\widetilde{X}.\psi\nonumber
\end{eqnarray}
where $\widetilde{X}$ is the 1-form which is the metric dual of the vector field $X$. Note that the composition of the Clifford action and the connection corresponds to the Dirac operator;
\begin{eqnarray}
c\circ\nabla:\Sigma M&\xrightarrow{\nabla}&T^*M\otimes\Sigma M\xrightarrow{c}\Sigma M\nonumber\\
\psi&\mapsto&e^a\otimes\nabla_{X_a}\psi\mapsto e^a.\nabla_{X_a}\psi=\displaystyle{\not}D\psi.\nonumber
\end{eqnarray}
However, this is not the only first-order differential operator that we can define on spinor fields.

We can write $T^*M\otimes\Sigma M$ as a direct sum of two components
\[
T^*M\otimes\Sigma M=S_1\oplus S_2
\]
by defining two projection operators $P_1$ and $P_2$ that satisfy $P_1+P_2=\mathbb{I}$;
\begin{eqnarray}
P_1:T^*M\otimes\Sigma M&\longrightarrow&S_1\nonumber\\
\widetilde{X}\otimes\psi&\longmapsto&\frac{1}{n}e^a\otimes(e_a.\widetilde{X}.\psi)\nonumber
\end{eqnarray}
and
\begin{eqnarray}
P_2:T^*M\otimes\Sigma M&\longrightarrow&S_2\nonumber\\
\widetilde{X}\otimes\psi&\longmapsto&\widetilde{X}\otimes\psi-\frac{1}{n}e^a\otimes(e_a.\widetilde{X}.\psi).\nonumber
\end{eqnarray}
Let us take the compositions of $P_1$ and $P_2$ with $c$
\begin{eqnarray}
c\circ P_1:T^*M\otimes\Sigma M&\xrightarrow{P_1}&S_1\xrightarrow{c}\Sigma M\nonumber\\
\widetilde{X}\otimes\psi&\mapsto&\frac{1}{n}e^a\otimes(e_a.\widetilde{X}.\psi)\mapsto\frac{1}{n}e^a.e_a.\widetilde{X}.\psi=\widetilde{X}.\psi\nonumber
\end{eqnarray}
and
\begin{eqnarray}
c\circ P_2:T^*M\otimes\Sigma M&\xrightarrow{P_2}&S_2\xrightarrow{c}\Sigma M\nonumber\\
\widetilde{X}\otimes\psi&\mapsto&\widetilde{X}\otimes\psi-\frac{1}{n}e^a\otimes(e_a.\widetilde{X}.\psi)\mapsto\nonumber\\
&\mapsto&\widetilde{X}.\psi-\frac{1}{n}e^a.e_a.\widetilde{X}.\psi=0.\nonumber
\end{eqnarray}
where we have used $e^a.e_a=e^a\wedge e_a-i_{X^a}e_a=n$ since $e^a\wedge e_a=0$ and $i_{X^a}e_a=e_a(X^a)=\delta^a_a=n$. So, this implies that we have $S_1=\text{im}(c)$ and $S_2=\text{ker}(c)$. Then, we have $T^*M\otimes\Sigma M=\Sigma M\oplus \text{ker}(c)$. We will call $S_1=\Sigma M$ as the spinor bundle and $S_2=\text{ker}(c)$ as the twistor bundle.

Now, we will see that we can define two first-order differential operators on $\Sigma M$ by considering two projections of $\nabla$ on $S_1$ and $S_2$ and their composition with the Clifford action $c$ \cite{Baum Friedrich Grunewald Kath, Acik};

i) The projection to the first component
\begin{eqnarray}
c\circ P_1\circ\nabla:\Sigma M&\xrightarrow{\nabla}&T^*M\otimes\Sigma M\xrightarrow{P_1}S_1\,\xrightarrow{c}\,\Sigma M\nonumber\\
\psi&\mapsto&e^a\otimes\nabla_{X_a}\psi\mapsto\nonumber\\
&\mapsto&\frac{1}{n}e^b\otimes(e_b.e^a.\nabla_{X_a}\psi)\mapsto\frac{1}{n}e^b.e_b.e^a.\nabla_{X_a}\psi=\displaystyle{\not}D\psi\nonumber
\end{eqnarray}
is the {\bf{Dirac operator}}. The spinors which are in the kernel of the Dirac operator
\begin{equation}
\displaystyle{\not}D\psi=0
\end{equation}
are called {\bf{harmonic spinors}} and are solutions of the massless Dirac equation. The spinors which are eigenspinors of the Dirac operator
\begin{equation}
\displaystyle{\not}D\psi=m\psi
\end{equation}
are solutions of the massive Dirac equation, where $m$ is a constant corresponding to the mass.

ii) The projection to the second component is
\begin{eqnarray}
c\circ P_2\circ\nabla:\Sigma M&\xrightarrow{\nabla}&T^*M\otimes\Sigma M\xrightarrow{P_2}S_2\,\xrightarrow{c}\,ker(c)\nonumber\\
\psi&\mapsto&e^a\otimes\nabla_{X_a}\psi\mapsto\nonumber\\
&\mapsto&e^a\otimes\nabla_{X_a}\psi-\frac{1}{n}e^b\otimes(e_b.e^a.\nabla_{X_a}\psi)\mapsto\nonumber\\
&\mapsto&e^.\nabla_{X_a}\psi-\frac{1}{n}e^b.e_b.e^a.\nabla_{X_a}\psi=0.\nonumber
\end{eqnarray}
So, the projection onto $S_2$ gives $e^a\otimes\left(\nabla_{X_a}\psi-\frac{1}{n}e_a.\displaystyle{\not}D\psi\right)$. Then, we have the {\bf{twistor (Penrose) operator}}
\begin{equation}
P_X:=\nabla_X-\frac{1}{n}\widetilde{X}.\displaystyle{\not}D
\end{equation}
for $X\in TM$ and $\widetilde{X}\in T^*M$ its metric dual. The spinors $\psi$ which are in the kernel of the twistor operator and satisfy the following equation
\begin{equation}
\nabla_X\psi=\frac{1}{n}\widetilde{X}.\displaystyle{\not}D\psi
\end{equation}
are called {\bf{twistor spinors}}. If a spinor $\psi$ is a solution of both massive Dirac equation and twistor equation, then it is called as a {\bf{Killing spinor}} and satisfies the following equation
\begin{equation}
\nabla_X\psi=\lambda\widetilde{X}.\psi
\end{equation}
where $\lambda$ is the Killing number, which can be real or pure imaginary. From $\nabla_X\psi=\frac{1}{n}\widetilde{X}.\displaystyle{\not}D\psi=\frac{m}{n}\widetilde{X}.\psi$, we have $\lambda:=\frac{m}{n}$. The spinors which are in the kernel of $\nabla$
\begin{equation}
\nabla_X\psi=0
\end{equation}
are called {\bf{parallel spinors}} (or covariantly constant spinors). This is a special case $\lambda=0$ of Killing spinors.

\subsection{Integrability conditions}

We investigate the integrability conditions for the existence of special types of spinors on a spin manifold $M$. We consider the cases of a general spinor and twistor, Killing and parallel spinors.

i) Schr\"{o}dinger-Lichnerowicz-Weitzenb\"{o}ck formula;\\
From the definition of the Dirac operator, we have for any spinor $\psi$
\begin{eqnarray}
\displaystyle{\not}D^2\psi&=&e^b.\nabla_{X_b}\left(e^a.\nabla_{X_a}\psi\right)\nonumber\\
&=&e^b.\left(\nabla_{X_b}e^a.\nabla_{X_a}\psi+e^a.\nabla_{X_b}\nabla_{X_a}\psi\right)\nonumber.
\end{eqnarray}
We will use the normal coordinates for which the connection coefficients are zero and we have $\nabla_{X_a}e_b=0=[X_a,X_b]$. So, we can write
\begin{eqnarray}
\displaystyle{\not}D^2\psi&=&e^b.e^a.\nabla_{X_b}\nabla_{X_a}\psi\nonumber\\
&=&\frac{1}{2}(e^b.e^a+e^a.e^b).\nabla_{X_b}\nabla_{X_a}\psi+\frac{1}{2}(e^b.e^a-e^a.e^b).\nabla_{X_b}\nabla_{X_a}\psi\nonumber\\
&=&\nabla_{X^a}\nabla_{X_a}\psi+\frac{1}{4}(e^b.e^a-e^a.e^b).\left(\nabla_{X_b}\nabla_{X_a}-\nabla_{X_a}\nabla_{X_b}\right)\psi\nonumber\\
&=&\nabla_{X^a}\nabla_{X_a}\psi-\frac{1}{2}e^b.e^a.R(X_a,X_b)\psi\nonumber
\end{eqnarray}
where we have divided the symmetric and antisymmetric parts and used the identities $e^b.e^a+e^a.e^b=2g^{ab}$ and $e^a.e^b=-e^b.e^a$ for $a\neq b$ in the second and third lines. The curvature operator is defined by
\begin{equation}
R(X,Y)=[\nabla_X,\nabla_Y]-\nabla_{[X,Y]}
\end{equation}
for $X,Y\in TM$. In normal coordinates, we have $R(X_a,X_b)=[\nabla_{X_a},\nabla_{X_b}]$. For any spinor $\psi$, the action of the curvature operator can be written in terms of the curvature 2-forms $R_{ab}$ as \cite{Benn Tucker}
\begin{equation}
R(X_a,X_b)\psi=\frac{1}{2}R_{ab}.\psi.
\end{equation}
So, we have
\[
\displaystyle{\not}D^2\psi=\nabla_{X^a}\nabla_{X_a}\psi-\frac{1}{4}e^b.e^a.R_{ab}.\psi
\]
We can define the Laplacian $\nabla^2=\nabla_{X^a}\nabla_{X_a}$, Ricci 1-forms $P_a=i_{X^a}R_{ab}$ and the curvature scalar ${\cal{R}}=i_{X^a}P_a$. For zero torsion, we have the identities $R_{ab}\wedge e^b=0$ and $P_a\wedge e^a=0$. So, we can write the identities $e^a.R_{ab}=e^a\wedge R_{ab}+i_{X^a}R_{ab}=P_b$ and $e^b.P_b=e^b\wedge P_b+i_{X^b}P_b={\cal{R}}$. Then, we have
\begin{equation}
\displaystyle{\not}D^2\psi=\nabla^2\psi-\frac{1}{4}{\cal{R}}\psi.
\end{equation}
This is called the Schr\"{o}dinger-Lichnerowicz-Weitzenb\"{o}ck formula.

ii) Twistor spinors;\\
The existence of special types of spinors constrain the geometry of $M$. Let us consider a twistor spinor $\psi$, namely we have $\nabla_{X_a}\psi=\frac{1}{n}e_a.\displaystyle{\not}D\psi$. By taking the second covariant derivative
\[
\nabla_{X_b}\nabla_{X_a}\psi=\frac{1}{n}\nabla_{X_b}e_a.\displaystyle{\not}D\psi+\frac{1}{n}e_a.\nabla_{X_b}\displaystyle{\not}D\psi
\]
and in normal coordinates, we have
\[
\nabla_{X_b}\nabla_{X_a}\psi=\frac{1}{n}e_a.\nabla_{X_b}\displaystyle{\not}D\psi.
\]
Similarly for the reversed order of indices
\[
\nabla_{X_a}\nabla_{X_b}\psi=\frac{1}{n}e_b.\nabla_{X_a}\displaystyle{\not}D\psi.
\]
From the difference of the last two equations and the definition of the curvature operator (13), we obtain
\[
R(X_a,X_b)\psi=\frac{1}{n}\left(e_b.\nabla_{X_a}\displaystyle{\not}D\psi-e_a.\nabla_{X_b}\displaystyle{\not}D\psi\right)
\]
and the action of the curvature operator on spinors in (14) gives
\begin{equation}
R_{ab}.\psi=\frac{2}{n}\left(e_b.\nabla_{X_a}\displaystyle{\not}D\psi-e_a.\nabla_{X_b}\displaystyle{\not}D\psi\right).
\end{equation}
By Clifford multiplying with $e^a$ from the left (from $e^a.R_{ab}=P_b$), we can write
\[
P_b.\psi=\frac{2}{n}\left(e^a.e_b.\nabla_{X_a}\displaystyle{\not}D\psi-e^a.e_a.\nabla_{X_b}\displaystyle{\not}D\psi\right)
\]
and using the identities $e^a.e_b+e_b.e^a=2g^a_b$ and $e^a.e_a=n$
\[
P_b.\psi=\frac{2}{n}\left(-e_b.e^a.\nabla_{X_a}\displaystyle{\not}D\psi+2\nabla_{X_b}\displaystyle{\not}D\psi-n\nabla_{X_b}\displaystyle{\not}D\psi\right).
\]
From the definition of the Dirac operator, we have
\begin{equation}
P_b.\psi=-\frac{2}{n}e_b.\displaystyle{\not}D^2\psi-\frac{2(n-2)}{n}\nabla_{X_b}\displaystyle{\not}D\psi.
\end{equation}
By Clifford multiplying with $e^b$ from the left (from $e^b.P_b={\cal{R}}$)
\begin{eqnarray}
{\cal{R}}&=&-\frac{2}{n}e^b.e_b.\displaystyle{\not}D^2\psi-\frac{2(n-2)}{n}e^b.\nabla_{X_b}\displaystyle{\not}D\psi\nonumber\\
&=&-\frac{4(n-1)}{n}\displaystyle{\not}D^2\psi
\end{eqnarray}
where we have used $e^b.e_b=n$ and the definition of the Dirac operator. Hence, we obtain the first integrability condition for twistor spinors
\begin{equation}
\displaystyle{\not}D^2\psi=-\frac{n}{4(n-1)}{\cal{R}}\psi.
\end{equation}
By substituting this equality in (17), we have
\[
\nabla_{X_a}\displaystyle{\not}D\psi=\frac{n}{4(n-1)(n-2)}{\cal{R}}e_a.\psi-\frac{n}{2(n-2)}P_a.\psi
\]
or by defining the Schouten 1-form
\begin{equation}
K_a=\frac{1}{n-2}\left(\frac{\cal{R}}{2(n-1)}e_a-P_a\right)
\end{equation}
we obtain the second integrability condition for twistor spinors
\begin{equation}
\nabla_{X_a}\displaystyle{\not}D\psi=\frac{n}{2}K_a.\psi.
\end{equation}
Moreover, we can define the conformal 2-forms as (for $n>2$)
\[
C_{ab}=R_{ab}-\frac{1}{n-2}\left(P_a\wedge e_b-P_b\wedge e_a\right)+\frac{1}{(n-1)(n-2)}{\cal{R}}e_a\wedge e_b
\]
or in terms of Clifford products
\begin{equation}
C_{ab}=R_{ab}-\frac{1}{n-2}\left(e_a.P_b-e_b.P_a\right)+\frac{1}{(n-1)(n-2)}{\cal{R}}e_a.e_b.
\end{equation}
By using (16), (17) and (18), one can easily found the following third integrability condition for twistor spinors
\begin{equation}
C_{ab}.\psi=0
\end{equation}
This means that, if a twistor spinor $\psi$ exists on a manifold $M$, then it must be in the kernel of the conformal 2-forms $C_{ab}$. In conformally-flat manifolds, we have $C_{ab}=0$ for $n>2$. So, the third integrability condition is automatically satisfied in conformally-flat manifolds. Hence, conformally-flat manifolds admit twistor spinors. But, they can also exist on non-conformally-flat manifolds with $C_{ab}.\psi=0$.

iii) Killing spinors;\\
Now, consider a Killing spinor $\psi$ that satisfies $\nabla_{X_a}\psi=\lambda e_a.\psi$. By taking the second covariant derivative and from the definition of the curvature operator, we have
\begin{eqnarray}
R(X_a,X_b)\psi&=&\nabla_{X_a}\nabla_{X_b}\psi-\nabla_{X_b}\nabla_{X_a}\psi\nonumber\\
&=&-\lambda^2(e_a.e_b-e_b.e_a).\psi.\nonumber
\end{eqnarray}
The identity $e_a.e_b-e_b.e_a=e_a\wedge e_b+i_{X_a}e_b-e_b\wedge e_a-i_{X_b}e_a=2e_a\wedge e_b$ and the action of the curvature operator on spinors in (14) gives
\begin{equation}
R_{ab}.\psi=-4\lambda^2(e_a\wedge e_b).\psi.
\end{equation}
By Clifford multiplying with $e^a$ from the left, we obtain (from $e^a.R_{ab}=P_b$)
\begin{eqnarray}
P_b.\psi&=&-4\lambda^2e^a.(e_a\wedge e_b).\psi\nonumber\\
&=&-4\lambda^2(n-1)e_b.\psi
\end{eqnarray}
where we have used $e^a.(e_a\wedge e_b)=e^a\wedge e_a\wedge e_b+i_{X^a}(e_a\wedge e_b)=(i_{X^a}e_a)e_b-e^a\wedge i_{X_a}e_b=(n-1)e_b$ since we have $i_{X^a}e_a=n$ and $e^a\wedge i_{X_a}e_b=e_b$. Again, by Clifford multiplying with $e^b$ from the left
\[
{\cal{R}}\psi=-4\lambda^2n(n-1)\psi
\]
Since the coefficients of $\psi$ in both sides are scalars, we have the identity
\begin{equation}
{\cal{R}}=-4\lambda^2n(n-1).
\end{equation}
${\cal{R}}$ has to be positive or negative (if non-zero), then $\lambda$ must be real or pure imaginary. The integrability conditions of Killing spinors implies that Einstein manifolds ($P_a=ce_a$ with $c$ constant) admit Killing spinors. If the manifold is Riemannian, then the existence of Killing spinors requires that it is Einstein.

iv) Parallel spinors;\\
For a parallel spinor $\psi$, that is $\nabla_X\psi=0$, the integrability conditions give
\begin{equation}
P_a.\psi=0.
\end{equation}
So, Ricci-flat manifolds admit parallel spinors. In the Riemannian case, the existence of parallel spinors requires that the manifold to be Ricci-flat.

\section{Holonomy Classification}

Special holonomy manifolds which admit special types of spinors are considered in this section. By defining the holonomy group of a manifold, the classification of special holonomy manifolds admitting parallel and Killing spinors are given. The appearance of these special manifolds in string theories and M-theory are also summarized.

\subsection{Holonomy groups}

Let $\gamma$ be a loop on $M$, that is
\[
\gamma:[0,1]\longrightarrow M\quad\text{and}\quad \gamma(0)=\gamma(1).
\]
Take a vector $X\in T_pM$ and parallel transport $X$ via connection $\nabla$ along $\gamma$. After a trip along $\gamma$, we end up with a new vector $Y\in T_pM$. Thus, the loop $\gamma$ and the connection $\nabla$ induce a linear transformation
\begin{eqnarray}
g_{\gamma}:T_pM&\longrightarrow& T_pM\nonumber\\
X&\longmapsto&Y\nonumber
\end{eqnarray}
The set of these transformations constitute a group called the {\bf{holonomy group}} at $p\in M$;
\[
\text{Hol}(p)=\left\{g_{\gamma}|\gamma:[0,1]\rightarrow M,\gamma(0)=\gamma(1)=p\right\}.
\]
If $M$ is simply connected ($\pi_1(M)=0$), then $\text{Hol}(p)$ is independent of $p$ and $\gamma$ and we denote the holonomy group of $M$ as $\text{Hol}(M)$. In general, $\text{Hol}(M)$ is a subgroup of $GL(n)$. If we choose the connection $\nabla$ as the Levi-Civita connection, then we have the metric compatibility $\nabla g=0$ and $\nabla$ preserves the lengths. So, $\text{Hol}(M)\subset O(n)$. If $M$ is orientable, then $\text{Hol}(M)\subset SO(n)$.

The restriction of the holonomy group of a manifold is related to the parallel sections of a bundle on $M$. This is called the holonomy principle and it can be stated as follows. The reduction of the holonomy group $\text{Hol}(M)$ of $M$ to a subgroup of $GL(n)$ is equivalent to the existence of a covariantly constant (parallel) section of a bundle $E$ on $M$. This parallel section is invariant under the action of $\text{Hol}(M)$. The problem of finding possible holonomy groups of compact Riemannian and Lorentzian manifolds is solved \cite{Joyce, Besse}. However, for arbitrary signature manifolds, it is an unsolved problem.

For the compact Riemannian case, we have the Berger's table \cite{Berger, Bryant}

\quad\\
{\centering{
\begin{tabular}{c c c}

$n\quad$ & $\quad \text{Hol}(M)\subset SO(n)\quad$ & \quad Geometry \\ \hline
$n\quad$ & $\quad SO(n)\quad$ & \quad generic \\
$2m\quad$ & $\quad U(m)\quad$ & \quad K\"{a}hler \\
$2m\quad$ & $\quad SU(m)\quad$ & \quad Calabi-Yau \\
$4m\quad$ & $\quad Sp(m)\quad$ & \quad hyperk\"{a}hler \\
$4m\quad$ & $\quad Sp(m).Sp(1)$ & \quad quaternionic K\"{a}hler \\
$7\quad$ & $\quad G_2\quad$ & \quad exceptional \\
$8\quad$ & $\quad Spin(7)\quad$ & \quad exceptional \\
$16\quad$ & $\quad Spin(9)\quad$ & \quad - \\

\end{tabular}}
\quad\\
\quad\\
\quad\\}

where $Sp(m).Sp(1)=Sp(m)\times Sp(1)/\mathbb{Z}_2$. The special geometries arising from the special holonomy manifolds can be summarized as follows.

\underline{K\"{a}hler manifolds}: For a $n=2m$ dimensional manifold $M$, if one can define an almost complex structure $J:TM\longrightarrow TM$ with $J^2=-1$ and the property $g(JX,JY)=g(X,Y)$ for $X,Y\in TM$, then $M$ is called as an almost complex manifold. Moreover, if we have $\nabla J=0$, then $J$ is called as a complex structure and $M$ is called as a complex manifold. If we can define a symplectic 2-form $\omega(X,Y)=g(JX,Y)$ which is parallel (and hence closed) $\nabla\omega=0$, then $M$ is a K\"{a}hler manifold and $\omega$ is called as the K\"{a}hler form. Covariantly constant nature of $\omega$ implies the restriction of holonomy to $U(m)$. One can define $(p,q)$-forms on $M$ with $p$ holomorphic and $q$ anti-holomorphic components.

\underline{Calabi-Yau manifolds}: Since $SU(m)\subset U(m)$, manifolds with $SU(m)$ holonomy are K\"{a}hler. Calabi-Yau manifolds are Ricci-flat K\"{a}hler manifolds with vanishing first Chern class. Besides the parallel complex structure $J$ and the parallel K\"{a}hler form $\omega$, there is also a parallel complex volume form $\Theta\in\Lambda^{(p,0)}M$. This implies the restriction of holonomy to $SU(m)$.

\underline{hyperk\"{a}hler manifolds}: A hyperk\"{a}hler structure is a triple $\{I,J,K\}$ of complex (K\"{a}hler) structures with the property $IJ=-JI=-K$ and three closed forms $\omega_i$ for $i=I,J,K$. Hyperk\"{a}hler manifolds are Ricci-flat.

\underline{quaternionic K\"{a}hler manifolds}: The structure of quaternionic K\"{a}hler manifolds is similar to the hyperk\"{a}hler case. However, quaternionic K\"{a}hler manifolds are Einstein manifolds.

\underline{$G_2$-holonomy manifolds}: $G_2 \subset SO(7)$ is the automorphism group of octonions. There is a parallel 3-form $\phi$ which is invariant under $G_2$ and this implies the restriction of holonomy to $G_2$. The 4-form $*\phi$ is also parallel. Manifolds of $G_2$-holonomy are Ricci-flat.

\underline{$Spin(7)$-holonomy manifolds}: $Spin(7)\subset SO(8)$ and there is a parallel 4-form $\Psi$ which is invariant under $Spin(7)$. This implies the restriction of holonomy to $Spin(7)$. Manifolds of $Spin(7)$-holonomy are also Ricci-flat.

Both parallel forms in $G_2$ and $Spin(7)$-holonomy manifolds can be constructed from parallel spinors.  This is the reason of Ricci-flatness. For more details, see \cite{Besse, Joyce}.

\subsection{Holonomy classification for parallel and Killing spinors}

We saw in Section 3 that the Ricci-flatness is an integrability condition for the existence of parallel spinors. Because of the holonomy principle, the existence of parallel spinors also implies the reduction of the holonomy group. So, in the Riemannian case, parallel spinors exist on Ricci-flat special holonomy manifolds. Then, from the Berger's table, we obtain Wang's table of manifolds admitting parallel spinors \cite{Wang}\\

\quad\\
{\centering{
\begin{tabular}{c c c c}

$n\quad$ & $\quad Hol(M)\quad$ & \quad Geometry\quad & \quad parallel spinors \\ \hline
$4m+2\quad$ & $\quad SU(2m+1)\quad$ & \quad Calabi-Yau \quad & \quad $(1,1)$ \\
$4m\quad$ & $\quad SU(2m)\quad$ & \quad Calabi-Yau \quad & \quad $(2,0)$ \\
$4m\quad$ & $\quad Sp(m)\quad$ & \quad hyperk\"{a}hler \quad & \quad $(m+1,0)$ \\
$7\quad$ & $\quad G_2\quad$ & \quad exceptional \quad & \quad $1$ \\
$8\quad$ & $\quad Spin(7)\quad$ & \quad exceptional \quad & \quad $(1,0)$ \\

\end{tabular}}
\quad\\
\quad\\
\quad\\}

In even dimensions, complex spinor bundle decomposes into two chiral subbundles $\Sigma M=\Sigma^+M\oplus\Sigma^-M$. For $\psi\in\Sigma^+M$, $iz.\psi=\psi$ and for $\psi\in\Sigma^-M$, $iz.\psi=-\psi$ where $z$ is the volume form. Spinors in $\Sigma^{\pm}M$ are called Weyl spinors. $(p,q)$ in the table denotes the number of spinors in $\Sigma^{\pm}M$ respectively.

Those special holonomy manifolds, have importance in string/M-theory compactifications. In string theory, we have 4 space-time and 6 compact Riemannian dimensions; $M_{10}=M_4\times M_6$. In the absence of fluxes, $M_4$ is Minkowski space-time. On the other hand, the supersymmetry transformations require the existence of parallel spinors on $M_6$. Then, $M_6$ must be a Calabi-Yau ($SU(3)$-holonomy) 3-fold from the table ($m=1$ in the first row);
\[
M_{10}=\underbrace{M_4}_{\text{Minkowski}}\times
\underbrace{M_6}_{\text{Calabi-Yau}}
\]
In M-theory, we have 4 space-time dimensions and 7 compact Riemannian dimensions; $M_{11}=M_4\times M_7$. If $M_4$ is Minkowski, then the existence of parallel spinors on $M_7$ requires that $M_7$ must be a $G_2$-holonomy manifold (fourth row in the table);
\[
M_{11}=\underbrace{M_4}_{\text{Minkowski}}\times
\underbrace{M_7}_{\text{$G_2$-holonomy}}
\]
If we let the decomposition $M_{11}=M_3\times M_{8}$, then for $M_3$ three dimensional Minkowski space-time, $M_8$ must be a $Spin(7)$-holonomy manifold (fifth row in the table);
\[
M_{11}=\underbrace{M_3}_{\text{Minkowski}}\times
\underbrace{M_8}_{\text{$Spin(7)$-holonomy}}
\]
Similarly, in F-theory in 12 dimensions, we have
\[
M_{12}=\underbrace{M_4}_{\text{Minkowski}}\times
\underbrace{M_8}_{\text{$Spin(7)$-holonomy}}
\]

To obtain the classification of compact Riemannian manifolds admitting Killing spinors, we need to discuss the cone construction. For a manifold $(M,g)$, the warped product manifold $\widetilde{M}=\mathbb{R}^+\times_{r^2}M$ with the metric $\widetilde{g}=dr^2+r^2g$ is called the metric cone of $M$. The vector fields $X\in TM$ can be lifted to $T\widetilde{M}$. The basis vector on $T\mathbb{R}^+$ will be denoted by $E=r\frac{\partial}{\partial r}$ which is called as the Euler vector. The Levi-Civita connection $\widetilde{\nabla}$ on $T\widetilde{M}$ is related to the Levi-Civita connection $\nabla$ on $TM$ as follows, for $X,Y\in T\widetilde{M}$
\begin{eqnarray}
\widetilde{\nabla}_{E}X=\widetilde{\nabla}_XE=X\quad ,\quad \widetilde{\nabla}_EE=E\nonumber\\
\widetilde{\nabla}_XY=\nabla_XY-g(X,Y)E.
\end{eqnarray}
The relations between curvatures are
\begin{eqnarray}
\widetilde{R}(X,Y)Z&=&R(X,Y)Z-\left(g(Y,Z)X-g(X,Z)Y\right)\nonumber\\
\widetilde{Ric}(X,Y)&=&Ric(X,Y)-(n-1)g(X,Y)\\
\widetilde{\cal{R}}&=&\frac{1}{r}^2\left({\cal{R}}-n(n-1)\right)\nonumber
\end{eqnarray}
So, if $M$ is Einstein, then $\widetilde{M}$ is Ricci-flat. If $\widetilde{M}$ has parallel spinors, then $M$ has Killing spinors corresponding to them. This can also be seen from the relations between connection 1-forms $\widetilde{\omega}_{ab}$ and $\omega_{ab}$ and the action of connection on spinors $\nabla_X\psi=X(\psi)+\frac{1}{4}\omega_{ab}(X).e^a.e^b.\psi$. In that case, $\widetilde{\nabla}_X\psi$ transforms into $\nabla_X\psi\sim\widetilde{X}.\psi$.

In that way, we have turned the classification problem of manifolds admitting Killing spinors to the classification of manifolds admitting parallel spinors via cone construction. So, from the Wang's table, we obtain B\"{a}r's table \cite{Bar};

\quad\\
{\centering{
\begin{tabular}{c c c c}

$n\quad$ & Geometry of $M$ & \quad Cone $\widetilde{M}$\quad & \quad Killing spinors \\ \hline
$n\quad$ & \quad round sphere \quad & \quad flat \quad & \quad $(2^{\lfloor n/2\rfloor},2^{\lfloor n/2\rfloor})$\\
$4m-1\quad$ & \quad 3-Sasaki \quad & \quad hyperk\"{a}hler \quad & \quad $(m+1,0)$ \\
$4m-1\quad$ & \quad Sasaki-Einstein \quad & \quad Calabi-Yau \quad & \quad $(2,0)$ \\
$4m+1\quad$ & \quad Sasaki-Einstein \quad & \quad Calabi-Yau \quad & \quad $(1,1)$ \\
$6\quad$ & \quad nearly K\"{a}hler \quad & \quad $G_2$ \quad & \quad $(1,1)$ \\
$7\quad$ & \quad weak $G_2$ \quad & \quad $Spin(7)$ \quad & \quad $(1,0)$ \\

\end{tabular}}
\quad\\
\quad\\
\quad\\}

The properties of manifolds appearing in the table can be summarized as follows.

\underline{Sasaki-Einstein manifolds}: A Sasakian structure on a Riemannian manifold $M$ is a Killing vector field $K$ of unit norm with the property
\begin{equation}
\nabla_X\nabla_YK=-g(X,Y)K+\widetilde{K}(Y)X\quad\text{for}\, X,Y\in TM
\end{equation}
If a Sasakian manifold is also Einstein, then we have a Sasaki-Einstein manifold.

\underline{3-Sasaki manifolds}: A 3-Sasakian structure on a Riemannian manifold $M$ is a triple $K_i,i=1,2,3$ of Sasakian structures with the relations
\begin{equation}
\nabla_{K_i}K_j=\epsilon_{ijk}K_k.
\end{equation}

\underline{nearly K\"{a}hler manifolds}: A nearly K\"{a}hler structure on a Riemannian manifold $M$ is an almost complex structure $J:TM\rightarrow TM$ with the property
\begin{equation}
i_X\nabla_XJ=0
\end{equation}
but $\nabla J\neq 0$. So, it is not K\"{a}hler.\\

\underline{weak $G_2$ manifolds}: On a weak $G_2$ manifold, there is a 3-form $\phi$ with the property
\begin{equation}
d\phi=\lambda*\phi
\end{equation}
So, it is not closed as in the $G_2$-holonomy case. Indeed, $\phi$ is a Killing-Yano form which will be defined in the next section.

These manifolds also have importance in string/M-theory compactifications. In string theory, we have $M_{10}=M_4\times M_6$. In the presence of fluxes, $M_4$ is Anti-de Sitter (AdS) space-time and the supersymmetry transformations require the existence of Killing spinors on $M_6$. Then, $M_6$ can be $S^6$ or nearly K\"{a}hler which can be seen from the table;
\[
M_{10}=\underbrace{M_4}_{AdS_4}\times
\underbrace{M_6}_{S^6}\quad\quad\text{or}\quad\quad M_{10}=\underbrace{M_5}_{AdS_4}\times
\underbrace{M_5}_{\text{nearly K\"{a}hler}}
\]
If we let $M_{10}=M_5\times M_5$, then we have
\[
M_{10}=\underbrace{M_5}_{AdS_5}\times
\underbrace{M_5}_{S^5}\quad\quad\text{or}\quad\quad M_{10}=\underbrace{M_5}_{AdS_5}\times
\underbrace{M_5}_{\text{Sasaki-Einstein}}
\]
In M-theory, we have $M_{11}=M_4\times M_7$. In the presence of fluxes, $M_4$ is AdS and if there is no internal flux, then $M_7$ is $S^7$. However, in the presence of internal fluxes, the existence of Killing spinors requires that $M_7$ is weak $G_2$-holonomy manifold;
\[
M_{11}=\underbrace{M_4}_{AdS_4}\times
\underbrace{M_7}_{S^7}\quad\quad\text{or}\quad\quad M_{11}=\underbrace{M_4}_{AdS_4}\times
\underbrace{M_7}_{\text{weak $G_2$}}
\]
Similarly, if we let $M_{11}=M_5\times M_6$, then we have
\[
M_{11}=\underbrace{M_5}_{AdS_5}\times
\underbrace{M_6}_{S^6}\quad\quad\text{or}\quad\quad M_{11}=\underbrace{M_5}_{AdS_5}\times
\underbrace{M_6}_{\text{nearly K\"{a}hler}}
\]

For the Lorentizan case, the manifolds admitting parallel spinors are either $\text{Ricci-flat}\times\mathbb{R}^+$ or Brinkmann spaces including $pp$-wave space-times \cite{Bryant2, Baum Kath}. The classification of Lorentzian manifolds admitting Killing spinors and twistor spinors are also investigated in the literature \cite{Bohle, Baum Leitner}.

\section{Spinor Bilinears}

In this section, $p$-form Dirac currents of spinors are defined via spinor inner products. It is proved that the $p$-form Dirac currents of twistor spinors correspond to CKY forms and for Killing spinors, $p$-form Dirac currents are KY forms or Hodge duals of KY forms. Moreover, the equalities satisfied by $p$-form Dirac currents of Killing spinors are compared with Maxwell and DKP equations.

\subsection{Dirac currents}

We can define a Spin-invariant inner product on spinor fields \cite{Benn Tucker}. For $\psi,\phi\in\Sigma M$, we denote the inner product as
\begin{equation}
(\psi,\phi)=\pm(\phi,\psi)^j
\end{equation}
where $j$ is an involution in $\mathbb{D}=\mathbb{R},\mathbb{C}\text{ or }\mathbb{H}$ (depending on the Clifford algebra in relevant dimension). So, $j$ is identity in $\mathbb{R}$, identity or complex conjugation (${}^*$) in $\mathbb{C}$, quaternionic conjugation or reversion in $\mathbb{H}$. The inner product can be symmetric or anti-symmetric (symplectic) depending on the dimension. For any $\alpha\in Cl(M)$ and $c\in\mathbb{D}$, we have
\begin{eqnarray}
(\psi,\alpha.\phi)&=&(\alpha^{\cal{J}}.\psi,\phi)\\
(c\psi,\phi)&=&c^j(\psi,\phi)
\end{eqnarray}
where ${\cal{J}}$ can be $\xi$, $\xi\eta$, $\xi^*$ or $\xi\eta^*$ depending on the dimension. From this inner product, we can define the space of dual spinors $\Sigma^*M$. For $\overline{\psi}\in\Sigma^*M$, we have
\begin{equation}
\overline{\psi}(\phi)=(\psi,\phi)
\end{equation}

Let us consider the tensor product $\Sigma M\otimes\Sigma^*M$. Its action on $\Sigma M$ is given by
\begin{equation}
(\psi\otimes\overline{\phi})\kappa=(\phi,\kappa)\psi
\end{equation}
for $\psi,\kappa\in\Sigma M$ and $\overline{\phi}\in\Sigma^*M$. So, the elements of $\Sigma M\otimes\Sigma^*M$ are linear transformations on $\Sigma M$. It is isomorphic to $Cl(M)$ (to its simple part) since we have $c:Cl(M)\otimes\Sigma M\rightarrow\Sigma M$. Then, we can write $\psi\otimes\overline{\phi}\in\Sigma M\otimes\Sigma^*M$ in terms of differential forms which are elements of $Cl(M)$.

For any orthonormal co-frame basis $\{e^a\}$, we have the Fierz identity in terms of the inner product
\begin{eqnarray}
\psi\overline{\phi}&=&(\phi,\psi)+(\phi,e_a.\psi)e^a+(\phi,e_{ba}.\psi)e^{ab}+...+\nonumber\\
&&+(\phi,e_{a_p...a_2a_1}.\psi)e^{a_1a_2...a_p}+...+(-1)^{\lfloor n/2\rfloor}(\phi,z.\psi)z
\end{eqnarray}
where we denote $\psi\overline{\phi}=\psi\otimes\overline{\phi}$, $e^{a_1a_2...a_p}=e^{a_1}\wedge e^{a_2}\wedge...\wedge e^{a_p}$ and $z$ is the volume form. We can define the $p$-form projections of $\psi\overline{\phi}$ as $(\psi\overline{\phi})_p$ which are called {\bf{spinor bilinears}}. If we take $\phi=\psi$, we can define the Dirac currents of a spinor as follows. The vector field $V_{\psi}$ which is the metric dual of the 1-form $(\psi\overline{\psi})_1$ is called the Dirac current of the spinor $\psi$. In general, the map
\begin{eqnarray}
(\quad)_1:\Sigma M&\longrightarrow& T^*M\cong TM\nonumber\\
\psi&\longmapsto&(\psi\overline{\psi})_1\cong V_{\psi}\nonumber
\end{eqnarray}
is called the squaring map of a spinor. The $p$-form component of $\psi\overline{\psi}$;
\begin{equation}
(\psi\overline{\psi})_p=(\psi,e_{a_p...a_2a_1}.\psi)e^{a_1a_2...a_p}
\end{equation}
is called the $p$-form Dirac current of the spinor $\psi$.

\subsection{CKY and KY forms}

We investigate the properties of $p$-form Dirac currents for special spinors.

i) Twistor spinors;

The Levi-Civita connection $\nabla$ is compatible with the spinor inner product and the duality operation. Namely, we have
\begin{eqnarray}
\nabla_X(\psi,\phi)&=&(\nabla_X\psi,\phi)+(\psi,\nabla_X\phi)\\
\nabla_X\overline{\psi}&=&\overline{\nabla_X\psi}
\end{eqnarray} 
for $X\in TM$, $\psi,\phi\in\Sigma M$. For a twistor spinor $\psi$, we can calculate the covariant derivative of the $p$-form Dirac current $(\psi\overline{\psi})_p$ as
\begin{eqnarray}
\nabla_{X_a}(\psi\overline{\psi})_p&=&\left((\nabla_{X_a}\psi)\overline{\psi}\right)_p+\left(\psi(\overline{\nabla_{X_a}\psi})\right)_p\nonumber\\
&=&\frac{1}{n}\left((e_a.\displaystyle{\not}D\psi)\overline{\psi}\right)_p+\frac{1}{n}\left(\psi(\overline{e_a.\displaystyle{\not}D\psi})\right)_p\nonumber
\end{eqnarray}
where we have used the twistor equation (10). From the following properties of the tensor products of spinors and dual spinors, for $\alpha\in Cl(M)$, $\psi\in\Sigma M$ and $\overline{\phi}\in\Sigma^*M$
\begin{eqnarray}
\alpha.(\psi\otimes\overline{\phi})&=&\alpha.\psi\otimes\overline{\phi}\\
(\psi\otimes\overline{\phi}).\alpha &=&\psi\otimes\overline{\alpha^{\cal{J}}.\phi}
\end{eqnarray}
we can write $\psi(\overline{e_a.\displaystyle{\not}D\psi})=(\psi\overline{\displaystyle{\not}D\psi}).e_a^{\cal{J}}$. The definition of the Dirac operator on spinors $\displaystyle{\not}D=e^b.\nabla_{X_b}$ and the property ${e^b}^{\cal{J}}.{e_a}^{\cal{J}}=e^b.e_a$ for ${\cal{J}}=\xi\text{ or }\xi\eta$ gives
\[
\nabla_{X_a}(\psi\overline{\psi})_p=\frac{1}{n}\left(e_a.e^b.(\nabla_{X_b}\psi)\overline{\psi}+\psi(\overline{\nabla_{X_b}\psi}).e^b.e_a\right)_p.
\]
Since $\nabla$ is compatible with the tensor product, we have $(\nabla_{X_b}\psi)\overline{\psi}=\nabla_{X_b}(\psi\overline{\psi})-\psi(\overline{\nabla_{X_b}\psi})$. So, we can write
\[
\nabla_{X_a}(\psi\overline{\psi})_p=\frac{1}{n}\left(e_a.e^b.\nabla_{X_b}(\psi\overline{\psi})-e_a.e^b.\psi(\overline{\nabla_{X_b}\psi})+\psi(\overline{\nabla_{X_b}\psi}).e^b.e_a\right)_p.
\]
Now, we analyze each term on the right hand side of the above equation. For the first term, we can use the definition of the Hodge-de Rham operator $\displaystyle{\not}d=e^a.\nabla_{X_a}$ on differential forms which are elements of $Cl(M)$ and we can write
\[
e_a.e^b.\nabla_{X_b}(\psi\overline{\psi})=e_a.\displaystyle{\not}d(\psi\overline{\psi})
\]
For the second and third terms, we can use the explicit expansion of Clifford product in terms of wedge product and interior product as $e_a.\alpha=e_a\wedge\alpha+i_{X_a}\alpha$ and $\alpha.e_a=e_a\wedge\eta\alpha-i_{X_a}\eta\alpha$. So, we have
\begin{eqnarray}
e_a.e^b.\psi(\overline{\nabla_{X_b}\psi})&=&e_a\wedge e^b\wedge \psi(\overline{\nabla_{X_b}\psi})-i_{X_a}\left(e^b\wedge\psi(\overline{\nabla_{X_b}\psi})\right)\nonumber\\
&&+e_a\wedge i_{X^b}\left(\psi(\overline{\nabla_{X_b}\psi})\right)+i_{X_a}i_{X^b}\left(\psi(\overline{\nabla_{X_b}\psi})\right)\nonumber
\end{eqnarray}
and
\begin{eqnarray}
\psi(\overline{\nabla_{X_b}\psi}).e^b.e_a&=&e_a\wedge\eta\left(e^b\wedge\eta\left(\psi(\overline{\nabla_{X_b}\psi})\right)\right)-i_{X_a}\eta\left(e^b\wedge\eta\left(\psi(\overline{\nabla_{X_b}\psi})\right)\right)\nonumber\\
&&-e_a\wedge\eta i_{X^b}\eta\left(\psi(\overline{\nabla_{X_b}\psi})\right)+i_{X_a}\eta i_{X^b}\eta\left(\psi(\overline{\nabla_{X_b}\psi})\right).\nonumber
\end{eqnarray}
By summing all the terms and considering the degrees of differential forms, we obtain
\begin{eqnarray}
\nabla_{X_a}(\psi\overline{\psi})_p&=&\frac{1}{n}\bigg[\left(e_a.\displaystyle{\not}d(\psi\overline{\psi})\right)_p-2e_a\wedge e^b\wedge\left(\psi(\overline{\nabla_{X_b}\psi})\right)_{p-2}\nonumber\\
&&-2i_{X_a}i_{X^b}\left(\psi(\overline{\nabla_{X_b}\psi})\right)_{p+2}\bigg].\nonumber
\end{eqnarray}
By using $\left(e_a.\displaystyle{\not}d(\psi\overline{\psi})\right)_p=e_a\wedge\left(\displaystyle{\not}d(\psi\overline{\psi})\right)_{p-1}+i_{X_a}\left(\displaystyle{\not}d(\psi\overline{\psi})\right)_{p+1}$, we find
\begin{eqnarray}
\nabla_{X_a}(\psi\overline{\psi})_p&=&\frac{1}{n}\bigg[e_a\wedge\left(\displaystyle{\not}d(\psi\overline{\psi})-2e^b\wedge\left(\psi(\overline{\nabla_{X_b}\psi})\right)\right)_{p-1}\nonumber\\
&&+i_{X_a}\left(\displaystyle{\not}d(\psi\overline{\psi})-2i_{X^b}\left(\psi(\overline{\nabla_{X_b}\psi})\right)\right)_{p+1}\bigg].
\end{eqnarray}
Taking the wedge product of (45) with $e^a\wedge$ from the left and using $e^a\wedge\nabla_{X_a}=d$ and $e^a\wedge i_{X_a}\alpha=p\alpha$ for a $p$-form $\alpha$ gives
\begin{equation}
d(\psi\overline{\psi})_p=\frac{p+1}{n}\left(\displaystyle{\not}d(\psi\overline{\psi})-2i_{X^b}\left(\psi(\overline{\nabla_{X_b}\psi})\right)\right)_{p+1}.
\end{equation}
By taking the interior derivative $i_{X^a}$ of (45) and using $-i_{X^a}\nabla_{X_a}=\delta$, one finds
\begin{equation}
\delta(\psi\overline{\psi})_p=-\frac{n-p+1}{n}\left(\displaystyle{\not}d(\psi\overline{\psi})-2e^b\wedge\left(\psi(\overline{\nabla_{X_b}\psi})\right)\right)_{p-1}.
\end{equation}
So, by comparing the last three equations, we obtain
\begin{equation}
\nabla_{X_a}(\psi\overline{\psi})_p=\frac{1}{p+1}i_{X_a}d(\psi\overline{\psi})_p-\frac{1}{n-p+1}e_a\wedge\delta(\psi\overline{\psi})_p.
\end{equation}
This equation has a special meaning. If a $p$-form $\omega$ satisfies the equation
\begin{equation}
\nabla_X\omega=\frac{1}{p+1}i_Xd\omega-\frac{1}{n-p+1}\widetilde{X}\wedge\delta\omega
\end{equation}
for any $X\in TM$, then $\omega$ is called as a {\bf{conformal Killing-Yano}} (CKY) $p$-form \cite{Semmelmann}. So, $p$-form Dirac currents of twistor spinors are CKY forms \cite{Acik Ertem}. CKY forms are antisymmetric generalizations of conformal Killing vector fields to higher degree forms. For $p=1$, $\omega$ is metric dual of a conformal Killing vector.

ii) Killing spinors;

For a Killing spinor $\psi$, the covariant derivative of $(\psi\overline{\psi})_p$ gives
\begin{eqnarray}
\nabla_{X_a}(\psi\overline{\psi})_p&=&\left((\nabla_{X_a}\psi)\overline{\psi}\right)_p+\left(\psi(\overline{\nabla_{X_a}\psi})\right)_p\nonumber\\
&=&\left(\lambda e_a.\psi\overline{\psi}\right)_p+\left(\psi(\overline{\lambda e_a.\psi})\right)_p\nonumber
\end{eqnarray}
where we have used the Killing spinor equation $\nabla_{X_a}\psi=\lambda e_a.\psi$. From the properties of the inner product, we have $\psi(\overline{\lambda e_a.\psi})=\lambda^j(\psi\overline{\psi}).e_a^{\cal{J}}$. So, we can write
\begin{eqnarray}
\nabla_{X_a}(\psi\overline{\psi})_p&=&\left(\lambda e_a.\psi\overline{\psi}\right)_p+\left(\lambda^j(\psi\overline{\psi}).e_a^{\cal{J}}\right)_p\nonumber\\
&=&\lambda e_a\wedge(\psi\overline{\psi})_{p-1}+\lambda i_{X_a}(\psi\overline{\psi})_{p+1}\nonumber\\
&&+\lambda^je_a^{\cal{J}}\wedge(\psi\overline{\psi})_{p-1}^{\eta}-\lambda^ji_{\widetilde{e_a^{\cal{J}}}}(\psi\overline{\psi})_{p+1}^{\eta}\nonumber
\end{eqnarray}
where we have used the expansion of the Clifford product in terms of the wedge product and interior derivative. By taking wedge product with $e^a\wedge$ from the left, one finds
\begin{equation}
d(\psi\overline{\psi})_p=\lambda (p+1)(\psi\overline{\psi})_{p+1}-\lambda^j\text{sgn}(e_a^{\cal{J}})(p+1)(\psi\overline{\psi})_{p+1}^{\eta}
\end{equation}
and by taking interior derivative with $i_{X^a}$, we obtain
\begin{equation}
\delta(\psi\overline{\psi})_p=-\lambda(n-p+1)(\psi\overline{\psi})_{p-1}-\lambda^j\text{sgn}(e_a^{\cal{J}})(n-p+1)(\psi\overline{\psi})_{p-1}^{\eta}.
\end{equation}
Here, we have four parameters to choose for a complex spinor $\psi$; $\lambda$ real or pure imaginary, $j=Id$ or $*$, ${\cal{J}}=\xi,\xi^*,\xi\eta$ or $\xi\eta^*$ and $p$ even or odd. So, there are 16 possibilities to choose;

\quad\\
{\centering{
\begin{tabular}{c c c | c c}

$\lambda\quad$ & \quad $j$\quad & \quad ${\cal{J}}$\quad & \quad $p$\quad & \quad {} \\ \hline
$\text{Re}\quad$ & \quad Id \quad & \quad $\xi$ \quad & \quad odd \quad & \quad even\\
$\text{Re}\quad$ & \quad Id \quad & \quad $\xi^*$ \quad & \quad odd \quad & \quad even\\
$\text{Re}\quad$ & \quad * \quad & \quad $\xi$ \quad & \quad odd \quad & \quad even\\
$\text{Re}\quad$ & \quad * \quad & \quad $\xi^*$ \quad & \quad odd \quad & \quad even\\
$\text{Im}\quad$ & \quad Id \quad & \quad $\xi$ \quad & \quad odd \quad & \quad even\\
$\text{Im}\quad$ & \quad Id \quad & \quad $\xi^*$ \quad & \quad odd \quad & \quad even\\
$\text{Im}\quad$ & \quad * \quad & \quad $\xi$ \quad & \quad even \quad & \quad odd\\
$\text{Im}\quad$ & \quad * \quad & \quad $\xi^*$ \quad & \quad even \quad & \quad odd\\
$\text{Re}\quad$ & \quad Id \quad & \quad $\xi\eta$ \quad & \quad even \quad & \quad odd\\
$\text{Re}\quad$ & \quad Id \quad & \quad $\xi\eta^*$ \quad & \quad even \quad & \quad odd\\
$\text{Re}\quad$ & \quad * \quad & \quad $\xi\eta$ \quad & \quad even \quad & \quad odd\\
$\text{Re}\quad$ & \quad * \quad & \quad $\xi\eta^*$ \quad & \quad even \quad & \quad odd\\
$\text{Im}\quad$ & \quad Id \quad & \quad $\xi\eta$ \quad & \quad even \quad & \quad odd\\
$\text{Im}\quad$ & \quad Id \quad & \quad $\xi\eta^*$ \quad & \quad even \quad & \quad odd\\
$\text{Im}\quad$ & \quad * \quad & \quad $\xi\eta$ \quad & \quad odd \quad & \quad even\\
$\text{Im}\quad$ & \quad * \quad & \quad $\xi\eta$ \quad & \quad odd \quad & \quad even\\

\end{tabular}}
\quad\\
\quad\\
\quad\\}

By the detailed considerations of the possibilities, one can see that we have two different cases. The first column of $p$ gives Case 1 and the second column of $p$ gives Case 2 in which the equations above transforms into the following equalities.

\underline{Case 1}:
\begin{eqnarray}
\nabla_{X_a}(\psi\overline{\psi})_p&=&2\lambda e_a\wedge (\psi\overline{\psi})_{p-1}\nonumber\\
d(\psi\overline{\psi})_p&=&0\\
\delta(\psi\overline{\psi})_p&=&-2\lambda(n-p+1)(\psi\overline{\psi})_{p-1}\nonumber
\end{eqnarray}

\underline{Case 2}:
\begin{eqnarray}
\nabla_{X_a}(\psi\overline{\psi})_p&=&2\lambda i_{X_a}(\psi\overline{\psi})_{p+1}\nonumber\\
d(\psi\overline{\psi})_p&=&2\lambda(p+1)(\psi\overline{\psi})_{p+1}\\
\delta(\psi\overline{\psi})_p&=&0\nonumber
\end{eqnarray}

By comparing the equations in Case 1, we find
\begin{equation}
\nabla_{X_a}(\psi\overline{\psi})_p=-\frac{1}{n-p+1}e_a\wedge\delta(\psi\overline{\psi})_p
\end{equation}
and in Case 2, we have
\begin{equation}
\nabla_{X_a}(\psi\overline{\psi})_p=\frac{1}{p+1}i_{X_a}d(\psi\overline{\psi})_p.
\end{equation}
By comparing these equalities with the CKY equation (49), one can see that Case 1 and Case 2 correspond to two parts of the CKY equation;
\begin{equation}
\nabla_X\omega=\underbrace{\frac{1}{p+1}i_Xd\omega}_{\text{KY part}}\underbrace{-\frac{1}{n-p+1}\widetilde{X}\wedge\delta\omega}_{\text{CCKY part}}.
\end{equation}
Case 1 corresponds to the CCKY part and Case 2 corresponds to the KY part. {\bf{Killing-Yano}} (KY) forms are co-closed ($\delta\omega=0$) CKY forms and satisfy
\begin{equation}
\nabla_X\omega=\frac{1}{p+1}i_Xd\omega.
\end{equation}
They are antisymmetric generalizations of Killing vector fields to higher degree forms \cite{Yano}. For $p=1$, $\omega$ is the metric dual of a Killing vector. Closed CKY (CCKY) forms are CKY forms with $d\omega=0$ and satisfy
\begin{equation}
\nabla_X\omega=-\frac{1}{n-p+1}\widetilde{X}\wedge\delta\omega.
\end{equation}
CKY equation has Hodge duality invariance, namely if $\omega$ is a CKY $p$-form, then $*\omega$ is also a CKY $(n-p)$-form. To see this, let us apply the Hodge star operator * to the CKY equation
\[
*\nabla_X\omega=\frac{1}{p+1}*i_Xd\omega-\frac{1}{n-p+1}*(\widetilde{X}\wedge\delta\omega).
\]
Since $\nabla$ is metric compatible, we have $*\nabla_X=\nabla_X*$. From the definition $\delta=*^{-1}d*\eta$ and the identity $*(\alpha\wedge\widetilde{X})=i_X*\alpha$ for any $p$-form $\alpha$, we can write
\begin{eqnarray}
*i_Xd\omega&=&*i_X**^{-1}d**^{-1}\omega\nonumber\\
&=&**(\delta\eta*^{-1}\omega\widetilde{X})\nonumber\\
&=&**(-\widetilde{X}\wedge\delta*^{-1}\omega)\nonumber\\
&=&**^{-1}(-\widetilde{X}\wedge\delta*\omega)\nonumber\\
&=&-\widetilde{X}\wedge\delta*\omega
\end{eqnarray}
where we have used that $*^{-1}$ is proportional to * from the identity $**\alpha=(-1)^{p(n-p)}\frac{det g}{|det g|}\alpha$. We can also write
\begin{eqnarray}
*(\widetilde{X}\wedge\delta\omega)&=&*(-\delta\eta\omega\wedge\widetilde{X})\nonumber\\
&=&-i_X*\delta\eta\omega\nonumber\\
&=&-i_X**^{-1}d*\eta\eta\omega\nonumber\\
&=&-i_Xd*\omega.
\end{eqnarray}
So, we obtain
\begin{equation}
\nabla_X*\omega=\underbrace{\frac{1}{n-p+1}i_Xd*\omega}_{\text{*KY part}}\underbrace{-\frac{1}{p+1}\widetilde{X}\wedge\delta*\omega}_{\text{*CCKY part}}.
\end{equation}
However, as can be seen from the above analysis, the two parts of the CKY equation transform into each other under *. Namely, we have
\[
\text{KY}\longrightarrow\text{*CCKY}\quad\text{and}\quad\text{CCKY}\longrightarrow\text{*KY}
\]
This means that CCKY forms are Hodge duals of KY forms and vice versa. Then, $p$-form Dirac currents of Killing spinors correspond to KY forms (in Case 2) or the Hodge duals of $p$-form Dirac cuurents of Killing spinors correspod to KY forms (in Case 1) depending on the inner product in relevant dimension \cite{Acik Ertem};\\
Case 1:
\[
\nabla_{X_a}*(\psi\overline{\psi})_p=\frac{1}{n-p+1}i_{X_a}d*(\psi\overline{\psi})_p.
\]
Case 2:
\[
\nabla_{X_a}(\psi\overline{\psi})_p=\frac{1}{p+1}i_{X_a}d(\psi\overline{\psi})_p.
\]

The equations satisfied by the $p$-form Dirac currents of Killing spinors have an analogous structure with (generalized) Maxwell equations
\begin{equation}
dF=0 \quad\text{,}\quad d*F=j.
\end{equation}
In Case 1, we have
\begin{eqnarray}
d(\psi\overline{\psi})_p&=&0\nonumber\\
d*(\psi\overline{\psi})_p&=&-2(-1)^p\lambda(n-p+1)*(\psi\overline{\psi})_{p-1}.
\end{eqnarray}
Hence, $p$-form Dirac currents $(\psi\overline{\psi})_p$ behave like the field strength $F$ of the $p$-form Maxwell field and the source term $j$ is constructed from the Hodge duals of the one lower degree Dirac currents $*(\psi\overline{\psi})_{p-1}$.\\
In Case 2, we have
\begin{eqnarray}
d(\psi\overline{\psi})_p&=&2\lambda(p+1)(\psi\overline{\psi})_{p+1}\nonumber\\
d*(\psi\overline{\psi})_p&=&0.
\end{eqnarray}
So, $*(\psi\overline{\psi})_p$ behave like the $(n-p)$-form Maxwell field strength $F$ and source term $j$ is the Dirac current of one higher degree $(\psi\overline{\psi})_{p+1}$.

Two sets of equations in Case 1 and Case 2 have also interesting relations with Duffin-Kemmer-Petiau (DKP) equations \cite{Lichnerowicz3}. DKP equations are first-order equations describing integer spin particles and are written as
\begin{equation}
d\phi_+-\delta\phi_-=\mu\phi
\end{equation}
where $\phi\in Cl(M)$ is the integer spin field which can be written as a sum of even and odd parts $\phi=\phi_++\phi_-$ with $\phi_{\pm}=\frac{1}{2}(1\pm\eta)\phi$ and $\mu$ is the mass. Between Case 1 and Case 2, the parity (oddness or evenness) of $p$ changes. So, if $(\psi\overline{\psi})_p$ satisfies Case 1, then $(\psi\overline{\psi})_{p-1}$ satisfies Case 2. Hence, we have
\begin{eqnarray}
d(\psi\overline{\psi})_{p-1}=2\lambda p(\psi\overline{\psi})_p\quad&,&\quad\delta(\psi\overline{\psi})_{p-1}=0\nonumber\\
\delta(\psi\overline{\psi})_p=-2\lambda(n-p+1)(\psi\overline{\psi})_{p-1}\quad&,&\quad d(\psi\overline{\psi})_p=0.\nonumber
\end{eqnarray}
If we choose $p=\frac{n+1}{2}$ and define $\phi_{\pm}=(\psi\overline{\psi})_{p-1}$, $\phi_{\mp}=(\psi\overline{\psi})_p$ for $p$ odd or even with $\mu=2\lambda p$, we obtain
\begin{eqnarray}
d\phi_{\pm}=\mu\phi_{\mp}\quad&,&\quad\delta\phi_{\pm}=0\nonumber\\
\delta\phi_{\mp}=-\mu\phi_{\pm}\quad&,&\quad d\phi_{\mp}=0
\end{eqnarray}
These are equivalent to the DKP equations.

Moreover, $p$-form Dirac currents of Killing spinors are eigenforms of the Laplace-Beltrami operator $\Delta=-d\delta-\delta d$. If $(\psi\overline{\psi})_p$ satisfies Case 1, then $(\psi\overline{\psi})_{p-1}$ satisfies Case 2 and we have
\begin{eqnarray}
\Delta(\psi\overline{\psi})_p=-d\delta(\psi\overline{\psi})_p&=&2\lambda(n-p+1)d(\psi\overline{\psi})_{p-1}\nonumber\\
&=&4\lambda^2p(n-p+1)(\psi\overline{\psi})_p.
\end{eqnarray}
If $(\psi\overline{\psi})_p$ satisfies Case 2, then $(\psi\overline{\psi})_{p+1}$ satisfies Case 1 and we have
\begin{eqnarray}
\Delta(\psi\overline{\psi})_p=-\delta d(\psi\overline{\psi})_p&=&-2\lambda(p+1)\delta(\psi\overline{\psi})_{p+1}\nonumber\\
&=&4\lambda^2(p+1)(n-p+2)(\psi\overline{\psi})_p.
\end{eqnarray}

iii) Parallel spinors;

If $\psi$ is a parallel spinor $\nabla_X\psi=0$, then $p$-form Dirac currents are parallel forms
\begin{equation}
\nabla_X(\psi\overline{\psi})_p=0.
\end{equation}
Indeed, they are harmonic forms which are in the kernel of $\Delta$
\begin{equation}
\Delta(\psi\overline{\psi})_p=0.
\end{equation}
More details can be found in \cite{Acik Ertem}.

\section{Symmetry Operators}

Symmetry operators of Dirac, twistor and Killing spinors are considered in this section. They transform one solution of a differential equation to another solution and correspond to the generalizations of the Lie derivatives on spinor fields. CKY and KY forms are used in the construction of symmetry operators.

\subsection{Lie derivatives}

For two vector fields $X,Y\in TM$, the Lie derivative of $Y$ with respect to $X$ which is denoted by ${\cal{L}}_XY$ is the change of $Y$ with respect to the flow of $X$. It can be written as
\begin{equation}
{\cal{L}}_XY=[X,Y]=X(Y)-Y(X).
\end{equation}
Lie derivative has the property
\begin{equation}
[{\cal{L}}_X,{\cal{L}}_Y]={\cal{L}}_{[X,Y]}.
\end{equation}
For a differential form $\alpha\in\Lambda M$, the Lie derivative ${\cal{L}}_X$ with respect to $X$ can be written in terms of $d$ and $i_X$ as
\begin{equation}
{\cal{L}}_X\alpha=i_Xd\alpha+di_X\alpha.
\end{equation}
We consider Lie derivatives with respect to Killing and conformal Killing vector fields. They are generators of isometries and conformal isometries
\begin{eqnarray}
{\cal{L}}_Kg=0\quad\quad&;&\quad K\text{  Killing vector}\nonumber\\
{\cal{L}}_Cg=2\mu g\quad\quad&;&\quad C\text{  conformal Killing vector}\quad(\mu\text{ function})\nonumber
\end{eqnarray}
If $\alpha\in Cl(M)$, namely an inhomogeneous differential form, then the Lie derivative with respect to a Killing vector $K$ is written as
\begin{equation}
{\cal{L}}_K\alpha=\nabla_K\alpha+\frac{1}{4}[d\widetilde{K},\alpha]_{Cl}
\end{equation}
where $\widetilde{K}$ is the metric dual of $K$ and $[\,]_{Cl}$ is the Clifford bracket which is defined for $\alpha,\beta\in Cl(M)$ as $[\alpha,\beta]_{Cl}=\alpha.\beta-\beta.\alpha$. For a spinor field $\psi\in\Sigma M$, the Lie derivative with respect to $K$ is
\begin{equation}
\pounds_K\psi=\nabla_K\psi+\frac{1}{4}d\widetilde{K}.\psi.
\end{equation}
It has a derivative property on spinors $\phi\in\Sigma M$, $\overline{\psi}\in\Sigma^*M$
\begin{equation}
\pounds_K(\phi\overline{\psi})=(\pounds_K\phi)\overline{\psi}+\phi(\overline{\pounds_K\psi})
\end{equation}
and compatible with the spinor duality operation. This is also true for conformal Killing vectors $C$, but not true for an arbitrary vector field $X$. So, the Lie derivative on spinors is only defined with respect to isometries \cite{Benn Tucker, Kosmann}.

\subsection{Symmetry operators}

A symmetry operator is an operator that maps a solution of an equation to another solution. We consider first-order symmetry operators for spinor field equations.

i) Dirac equation;\\
For the massive Dirac equation $\displaystyle{\not}D\psi=m\psi$, the Lie derivative with respect to a Killing vector $K$ is a symmetry operator. So, if $\psi$ is a solution, then $\pounds_K\psi$ is also a solution
\begin{equation}
\displaystyle{\not}D\pounds_K\psi=m\pounds_K\psi.
\end{equation}
For the massless Dirac equation $\displaystyle{\not}D\psi=0$, the operator $\pounds_C+\frac{1}{2}(n-1)\mu$ defined with respect to a conformal Killing vector $C$ with ${\cal{L}}_Cg=2\mu g$ is a symmetry operator in $n$ dimensions, namely we have
\begin{equation}
\displaystyle{\not}D\left(\pounds_C-\frac{1}{2}(n-1)\mu\right)\psi=0.
\end{equation}
These can be generalized to higher-degree KY and CKY forms. For the massive Dirac equation, the following operator written in terms of a KY $p$-form $\omega$ (for $p$ odd)
\begin{equation}
L_{\omega}\psi=(i_{X^a}\omega).\nabla_{X_a}\psi+\frac{p}{2(p+1)}d\omega.\psi
\end{equation}
is a symmetry operator which reduces to $\pounds_K$ for $p=1$ \cite{Benn Kress2, Acik Ertem Onder Vercin, Cariglia Krtous Kubiznak}. If $p$ is even, then $L_{\omega} z$ is a symmetry operator with $z$ is the volume form. For the massless Dirac equation, the operator written in terms of a CKY $p$-form $\omega$
\begin{equation}
L_{\omega}\psi=(i_{X^a}\omega).\nabla_{x_a}\psi+\frac{p}{2(p+1)}d\omega.\psi-\frac{n-p}{2(n-p+1)}\delta\omega.\psi
\end{equation}
is a symmetry operator which reduces to $\pounds_C-\frac{1}{2}(n-1)\mu$ for $p=1$ since $\delta\widetilde{C}=-n\mu$ \cite{Benn Charlton}.

ii) Killing spinors;\\
Killing spinors are solutions of the massive Dirac equation wihch are also twistor spinors. However, this does not mean that the symmetry operators of the massive Dirac equation must preserve the subset of Killing spinors. For a Killing vector $K$, the Lie derivative $\pounds_K$ is a symmetry operator for the Killing spinor equation;
\begin{equation}
\nabla_X\pounds_K\psi=\lambda\widetilde{X}.\pounds_K\psi.
\end{equation}
For higher-degree KY forms $\omega$, we investigate that in which circumstances the operator $L_{\omega}$ defined for the massive Dirac equation is also a symmetry operator for Killing spinors;
\begin{equation}
\nabla_{X}L_{\omega}\psi=\lambda\widetilde{X}.L_{\omega}\psi.
\end{equation}
For a Killing spinor $\psi$ and a KY $p$-form $\omega$, we can write
\begin{eqnarray}
L_{\omega}\psi&=&(i_{X^a}\omega).\nabla_{X_a}\psi+\frac{p}{2(p+1)}d\omega.\psi\nonumber\\
&=&\lambda(i_{X^a}\omega).e_a.\psi+\frac{p}{2(p+1)}d\omega.\psi\nonumber\\
&=&(-1)^{p-1}\lambda p\omega.\psi+\frac{p}{2(p+1)}d\omega.\psi
\end{eqnarray}
where we have used the Killing spinor equation and the identities $(i_{X^a}\omega).e_a=e_a\wedge\eta(i_{X^a}\omega)-i_{X_a}\eta i_{X^a}\omega$ (second term is zero since $i_Xi_Y$ is antisymmetric) and $e_a\wedge i_{X^a}\omega=p\omega$. If we take $p$ odd, then
\begin{equation}
L_{\omega}\psi=\lambda p\omega.\psi+\frac{p}{2(p+1)}d\omega.\psi.
\end{equation}
By calculating the covariant derivative, we obtain
\begin{eqnarray}
\nabla_{X_a}L_{\omega}\psi&=&\nabla_{X_a}\left(\lambda p\omega.\psi+\frac{p}{2(p+1)}d\omega.\psi\right)\\
&=&\lambda p\nabla_{X_a}\omega.\psi+\lambda p\omega.\nabla_{X_a}\psi+\frac{p}{2(p+1)}\nabla_{X_a}d\omega.\psi\nonumber\\
&&+\frac{p}{2(p+1)}d\omega.\nabla_{X_a}\psi.\nonumber
\end{eqnarray}
We can use the Killing spinor equation $\nabla_{X_a}\psi=\lambda e_a.\psi$, KY equation $\nabla_{X_a}\omega=\frac{1}{p+1}i_{X_a}d\omega$ and the integrability condition for the KY equation which can be calculated as
\begin{eqnarray}
\nabla_{X_b}\nabla_{X_a}\omega&=&\frac{1}{p+1}\nabla_{X_b}i_{X_a}d\omega\nonumber\\
&=&\frac{1}{p+1}i_{X_a}\nabla_{X_b}d\omega\nonumber
\end{eqnarray}
from $[i_{X_a},\nabla_{X_b}]=0$ in normal coordinates (in general, we have $[\nabla_X,i_Y]=i_{\nabla_XY}$). By using the curvature operator $R(X_a,X_b)=\nabla_{X_a}\nabla_{X_b}-\nabla_{X_b}\nabla_{X_a}$, we can write
\[
R(X_a,X_b)\omega=\frac{1}{p+1}\left(i_{X_b}\nabla_{X_a}-i_{X_a}\nabla_{X_b}\right)d\omega.
\]
By multiplying $e^a\wedge$ from the left, we have
\begin{eqnarray}
e^a\wedge R(X_a,X_b)\omega&=&\frac{1}{p+1}\left(e^a\wedge i_{X_b}\nabla_{X_a}d\omega-e^a\wedge i_{X_a}\nabla_{X_b}d\omega\right)\nonumber\\
&=&\frac{1}{p+1}\left(e^a\wedge\nabla_{X_a}i_{X_b}d\omega-(p+1)\nabla_{X_b}d\omega\right)\nonumber
\end{eqnarray}
where we have used $[i_{X_b},\nabla_{X_a}]=0$ and $e^a\wedge i_{X_a}\nabla_{X_b}d\omega=(p+1)\nabla_{X_b}d\omega$. From the identities $e^a\wedge\nabla_{X_a}=d$, $d^2=0$ and $\nabla_{X_b}=i_{X_b}d+di_{X_b}$ in normal coordinates, we can write
\begin{equation}
e^a\wedge R(X_a,X_b)\omega=-\frac{p}{p+1}\nabla_{X_b}d\omega.
\end{equation}
The action of the curvature operator on a $p$-form $\omega$ is \cite{Acik Ertem Onder Vercin2}
\begin{equation}
R(X_a,X_b)\omega=-i_{X^c}R_{ab}\wedge i_{X_c}\omega.
\end{equation}
So, we have
\begin{eqnarray}
e^a\wedge R(X_a,X_b)\omega&=&-e^a\wedge i_{X^c}R_{ab}\wedge i_{X_c}\omega\nonumber\\
&=&\left(i_{X^c}(e^a\wedge R_{ab}-R^c_b)\right)\wedge i_{X_c}\omega\nonumber\\
&=&-R_{cb}\wedge i_{X^c}\omega.\nonumber
\end{eqnarray}
Then, we obtain the integrability condition for KY forms $\omega$ as follows
\begin{equation}
\nabla_{X_a}d\omega=\frac{p+1}{p}R_{ba}\wedge i_{X^b}\omega.
\end{equation}
Now, we can write the covariant derivative of $L_{\omega}\psi$ from (85) as
\begin{eqnarray}
\nabla_{X_a}L_{\omega}\psi&=&\lambda\frac{p}{p+1}i_{X_a}d\omega.\psi+\lambda^2p\omega.e_a.\psi\nonumber\\
&&+\frac{1}{2}(R_{ba}\wedge i_{X^b}\omega).\psi+\lambda\frac{p}{2(p+1)}d\omega.e_a.\psi.
\end{eqnarray}
On the other hand, for the right hand side of the Killing spinor equation (82), we have
\begin{equation}
\lambda e_a.L_{\omega}\psi=\lambda^2pe_a.\omega.\psi+\lambda\frac{p}{2(p+1)}e_a.d\omega.\psi.
\end{equation}
Obviously, the right hand sides of (89) and (90) are not equal to each other. Let us consider special KY forms which have the following special integrability condition
\begin{equation}
\nabla_Xd\omega=-c(p+1)\widetilde{X}\wedge\omega
\end{equation}
where $c$ is a constant. For example, in constant curvature manifolds we have $R_{ab}=ce_a\wedge e_b$ and the special integrability condition is satified by all KY forms since the right hand side of the ordinary integrability condition (88) gives
\begin{eqnarray}
\frac{p}{p+1}R_{ba}\wedge i_{X^b}\omega&=&c\frac{p+1}{p}e_b\wedge e_a\wedge i_{X^b}\omega\nonumber\\
&=&-c(p+1)e_a\wedge\omega\nonumber
\end{eqnarray}
where we have used $e_b\wedge e_a=-e_a\wedge e_b$ and $e_b\wedge i_{X^b}\omega=p\omega$. So, all KY forms are special KY forms in constant curvature manifolds. If $R_{ab}=ce_a\wedge e_b$, then $P_b=i_{X^a}R_{ab}=c(n-1)e_b$ and ${\cal{R}}=i_{X^b}P_b=cn(n-1)$, so we have
\[
c=\frac{\cal{R}}{n(n-1)}\quad\quad \text{and}\quad\quad R_{ab}=\frac{\cal{R}}{n(n-1)}e_a\wedge e_b.
\]
Moreover, we know from the integrability condition of Killing spinors that ${\cal{R}}=-4\lambda^2n(n-1)$ which implies that
\[
c=-4\lambda^2\quad\quad\text{and}\quad\quad R_{ab}=-4\lambda^2e_a\wedge e_b.
\]
Then, the covariant derivative of $L_{\omega}\psi$ turns into
\begin{eqnarray}
\nabla_{X_a}L_{\omega}\psi&=&\lambda\frac{p}{p+1}i_{X_a}d\omega.\psi+\lambda^2p(-e_a\wedge\omega+i_{X_a}\omega).\psi\nonumber\\
&&+2\lambda^2p(e_a\wedge\omega).\psi+\lambda\frac{p}{2(p+1)}(e_a\wedge d\omega-i_{X_a}d\omega).\psi\nonumber
\end{eqnarray}
where we have expanded the Clifford product in terms of wedge and interior products and used the above equality for $R_{ab}$. Finally, we obtain
\begin{eqnarray}
\nabla_{X_a}L_{\omega}\psi&=&\lambda^2p(e_a\wedge\omega+i_{X_a}\omega).\psi+\lambda\frac{p}{2(p+1)}(e_a\wedge d\omega+i_{X_a}d\omega).\psi\nonumber\\
&=&\lambda^2pe_a.\omega.\psi+\lambda\frac{p}{2(p+1)}e_a.d\omega.\psi\nonumber\\
&=&\lambda e_a.L_{\omega}\psi.\nonumber
\end{eqnarray}
So, we prove that for odd KY forms $\omega$ in constant curvature manifolds (or odd special KY forms with $c=\frac{\cal{R}}{n(n-1)}$ in general manifolds), the operator
\[
L_{\omega}=(i_{X^a}\omega).\nabla_{X_a}\psi+\frac{p}{2(p+1)}d\omega.\psi
\]
is a symmetry operator for Killing spinors \cite{Ertem2}.

iii) Twistor spinors;\\
For a conformal Killing vector $C$ with ${\cal{L}}_Cg=2\mu g$, the operator $\pounds_C-\frac{1}{2}\mu$ is a symmetry operator for the twistor equation
\begin{equation}
\nabla_X\left(\pounds_C-\frac{1}{2}\mu\right)\psi=\frac{1}{n}\widetilde{X}.\displaystyle{\not}D\left(\pounds_C-\frac{1}{2}\mu\right)\psi.
\end{equation}
Note that it is different from the case of massless Dirac equation which was $\pounds_C+\frac{1}{2}(n-1)\mu$. For higher-degree CKY forms $\omega$, we consider the operator
\begin{equation}
L_{\omega}\psi=(i_{X^a}\omega).\nabla_{X_a}\psi+\frac{p}{2(p+1)}d\omega.\psi+\frac{p}{2(n-p+1)}\delta\omega.\psi
\end{equation}
which reduces to $\pounds_C-\frac{1}{2}\mu$ for $p=1$ since $\delta\widetilde{C}=-n\mu$. We investigate that in which condition the operator $L_{\omega}$ is a symmetry operator for a twistor spinor $\psi$
\begin{equation}
\nabla_XL_{\omega}\psi=\frac{1}{n}\widetilde{X}.\displaystyle{\not}DL_{\omega}\psi.
\end{equation}
By using the twistor equation, $L_{\omega}$ can be written as
\begin{equation}
L_{\omega}\psi=-(-1)^p\frac{p}{n}\omega.\displaystyle{\not}D\psi+\frac{p}{2(p+1)}d\omega.\psi+\frac{p}{2(n-p+1)}\delta\omega.\psi.
\end{equation}
By taking the covariant derivative
\begin{eqnarray}
\nabla_{X_a}L_{\omega}\psi&=&-(-1)^p\frac{p}{n}\nabla_{X_a}\omega.\displaystyle{\not}D\psi-(-1)^p\frac{p}{n}\omega.\nabla_{X_a}\displaystyle{\not}D\psi\nonumber\\
&&+\frac{p}{2(p+1)}\nabla_{X_a}d\omega.\psi+\frac{p}{2(p+1)}d\omega.\nabla_{X_a}\psi\\
&&+\frac{p}{2(n-p+1)}\nabla_{X_a}\delta\omega.\psi+\frac{p}{2(n-p+1)}\delta\omega.\nabla_{X_a}\psi.\nonumber
\end{eqnarray}
Here, we can use the twistor equation (10) and the integrability condition (21). By also expanding Clifford products in terms of wedge products and interior derivatives, we obtain
\begin{eqnarray}
\nabla_{X_a}L_{\omega}\psi&=&\bigg[-(-1)^p\frac{p}{2n(p+1)}i_{X_a}d\omega-(-1)^p\frac{p}{2n(p+1)}e_a\wedge d\omega\nonumber\\
&&+(-1)^p\frac{p}{2n(n-p+1)}e_a\wedge\delta\omega+(-1)^p\frac{p}{2n(n-p+1)}i_{X_a}\delta\omega\bigg].\displaystyle{\not}D\psi\nonumber\\
&&+\bigg[-(-1)^p\frac{p}{2}\omega.K_a+\frac{p}{2(p+1)}\nabla_{X_a}d\omega+\frac{p}{2(n-p+1)}\nabla_{X_a}\delta\omega\bigg].\psi.\nonumber
\end{eqnarray}
A similar calculation for the right hand side of the twistor equation gives \cite{Ertem4}
\begin{eqnarray}
\frac{1}{n}e_a.\displaystyle{\not}DL_{\omega}\psi&=&\bigg[-(-1)^p\frac{p}{2n(p+1)}i_{X_a}d\omega-(-1)^p\frac{p}{2n(p+1)}e_a\wedge d\omega\nonumber\\
&&+(-1)^p\frac{p}{2n(n-p+1)}e_a\wedge\delta\omega+(-1)^p\frac{p}{2n(n-p+1)}i_{X_a}\delta\omega\bigg].\displaystyle{\not}D\psi\nonumber\\
&&+\bigg[-(-1)^p\frac{p}{2n}e_a.e^b.\omega.K_b+\frac{p}{2n(p+1)}e_a.e^b.\nabla_{X_b}d\omega\nonumber\\
&&+\frac{p}{2n(n-p+1)}e_a.e^b.\nabla_{X_b}\delta\omega\bigg].\psi
\end{eqnarray}
By comparing the last two equalities, one can see that the coefficients of $\displaystyle{\not}D\psi$ are equal to each other. However, to check the equivalence of coefficients of $\psi$, we need to use the integrability conditions of CKY forms. They can be calculated as \cite{Ertem3}
\begin{equation}
\nabla_{X_a}d\omega=\frac{p+1}{p(n-p+1)}e_a\wedge d\delta\omega+\frac{p+1}{p}R_{ba}\wedge i_{X^b}\omega
\end{equation}
\begin{eqnarray}
\nabla_{X_a}\delta\omega&=&-\frac{n-p+1}{(p+1)(n-p)}i_{X_a}\delta d\omega\nonumber\\
&&+\frac{n-p+1}{n-p}\left(i_{X_b}P_a\wedge i_{X^b}\omega+i_{X_b}R_{ca}\wedge i_{X^c}i_{X^b}\omega\right)
\end{eqnarray}
\begin{equation}
\frac{p}{p+1}\delta d\omega+\frac{n-p}{n-p+1}=P_a\wedge i_{X^a}\omega+R_{ab}\wedge i_{X^a}i_{X^b}\omega.
\end{equation}
In constant curvature manfiolds, these equalities satisfy the equivalence of the coefficients of $\psi$ in (96) and (97) \cite{Ertem4}. So, $L_{\omega}$ is a symmetry operator of twistor spinors in constant curvature manifolds. Moreover, we can also consider normal CKY forms in Einstein manifolds which have the following integrability conditions \cite{Leitner2}
\begin{equation}
\nabla_{X_a}d\omega=\frac{p+1}{p(n-p+1)}e_a\wedge d\delta\omega+2(p+1)K_a\wedge\omega
\end{equation}
\begin{equation}
\nabla_{X_a}\delta\omega=-\frac{n-p+1}{(p+1)(n-p)}i_{X_a}\delta d\omega-2(n-p+1)i_{X_b}K_a\wedge i_{X_b}\omega
\end{equation}
\begin{equation}
\frac{p}{p+1}\delta d\omega+\frac{n-p}{n-p+1}=-2(n-p)K_a\wedge i_{X^a}\omega.
\end{equation}
In that case also we have the equality $\nabla_{X_a}L_{\omega}\psi=\frac{1}{n}e_a.\displaystyle{\not}DL_{\omega}\psi$. So, $L_{\omega}$ is a symmetry operator in Einstein manifolds with $\omega$ are normal CKY forms \cite{Ertem4}. In constant curvature manifolds, all CKY forms are normal CKY forms \cite{Ertem3}.

In summary, we have the following diagrams of symmetry operators
\begin{displaymath}
\xymatrix{\displaystyle{\not}D\psi=m\psi \ar[d] & \nabla_X\psi=\lambda\widetilde{X}.\psi \ar[d] \\
\pounds_K \ar[d]^{\text{odd} \,p\, (\text{even}\,p\,,\,L_{\omega}z)} & \pounds_K \ar[d]^{\text{odd} \,p,\,\text{constant curvature}} \\
L_{\omega}=i_{X^a}\omega.\nabla_{X_a}+\frac{p}{2(p+1)}d\omega \ar@/^/[u]^{p=1} & L_{\omega}=i_{X^a}\omega.\nabla_{X_a}+\frac{p}{2(p+1)}d\omega \ar@/^/[u]^{p=1}}
\end{displaymath}
and
\begin{displaymath}
\xymatrix{\displaystyle{\not}D\psi=0 \ar[d] & \nabla_X\psi=\frac{1}{n}\widetilde{X}.\displaystyle{\not}D\psi \ar[d] \\
\pounds_C+\frac{1}{2}(n-1)\mu \ar[d] & \pounds_C-\frac{1}{2}\mu \ar[d]^{\text{const. curv.}\,,\,\text{Einstein}} \\
L_{\omega}=i_{X^a}\omega.\nabla_{X_a}+\frac{p}{2(p+1)}d\omega-\frac{n-p}{2(n-p+1)}\delta\omega \ar@/^/[u]^{p=1} & L_{\omega}=i_{X^a}\omega.\nabla_{X_a}+\frac{p}{2(p+1)}d\omega+\frac{p}{2(n-p+1)}\delta\omega \ar@/^/[u]^{p=1}}
\end{displaymath}

\section{Extended Superalgebras}

In this section, we define KY and CKY superalgebras in constant curvature manifolds. By using spinor bilinears of Killing and twistor spinors and the symmetry operators defined in the previous section, extended Killing and conformal superalgebras are also constructed.

\subsection{Symmetry superalgebras}

A superalgebra $\mathfrak{g}$ is a $\mathbb{Z}_2$-graded algebra which can be written as a direct sum of two components;
\[
\mathfrak{g}=\mathfrak{g}_0\oplus\mathfrak{g}_1
\]
The even part $\mathfrak{g}_0$ is a subalgebra of $\mathfrak{g}$ and the odd part $\mathfrak{g}_1$ is a module of $\mathfrak{g}_0$, i.e., $\mathfrak{g}_0$ acts on $\mathfrak{g}_1$. The product rule is given by a bilinear operation $[\,,\,]$;
\begin{equation}
[\,,\,]:\mathfrak{g}_i\times\mathfrak{g}_j\longrightarrow\mathfrak{g}_{i+j}\quad,\quad i,j=0,1.
\end{equation}
So, we have
\[
[\,,\,]:\mathfrak{g}_0\times\mathfrak{g}_0\longrightarrow\mathfrak{g}_0
\]
\[
[\,,\,]:\mathfrak{g}_0\times\mathfrak{g}_1\longrightarrow\mathfrak{g}_1
\]
\[
[\,,\,]:\mathfrak{g}_1\times\mathfrak{g}_1\longrightarrow\mathfrak{g}_0
\]
If $[\,,\,]$ is antisymmetric on $\mathfrak{g}_0$, namely $[a,b]=-[b,a]$ for $a,b\in\mathfrak{g}_0$ and satisfies the Jacobi identity $[a,[b,c]]+[b,[c,a]]+[c,[a,b]]=0$ for $a,b,c\in\mathfrak{g}_0$, then $\mathfrak{g}_0$ is a Lie algebra. Moreover, if $[\,,\,]$ is symmetric on $\mathfrak{g}_1$, $[a,b]=[b,a]$ for $a,b\in\mathfrak{g}_1$ and $[\,,\,]$ on $\mathfrak{g}=\mathfrak{g}_0\oplus\mathfrak{g}_1$ satisfies the graded Jacobi identity
\begin{equation}
[a,[b,c]]-[[a,b],c]-(-1)^{|a||b|}[b,[a,c]]=0
\end{equation}
for $a,b,c\in\mathfrak{g}$ and $|a|$ denotes the degree of $a$ which corresponds to 0 or 1 depending on that $a$ is in $\mathfrak{g}_0$ or $\mathfrak{g}_1$, respectively, then $\mathfrak{g}$ is called as a Lie superalgebra. The dimension of a superalgebra is denoted by $(\alpha|\beta)$ where $\alpha$ is the dimension of $\mathfrak{g}_0$ and $\beta$ is the dimension of $\mathfrak{g}_1$.

We consider the superalgebra structures constructed from KY and CKY forms.

i) Killing-Yano superalgebra;\\
Killing vector fields on a manifold $M$ have a Lie algebra structure with respect to the Lie bracket of vector fields;
\begin{equation}
[K_i,K_j]=c_{ijk}K_k
\end{equation}
which is called the isometry algebra where $c_{ijk}$ are structure constants. For KY forms $\omega$ which satisfy (57), we can define the Schouten-Nijenhuis (SN) bracket $[\,,\,]_{SN}$ \cite{Schouten, Nijenhuis}. For a $p$-form $\alpha$ and a $q$-form $\beta$, $[\,,\,]_{SN}$ is defined as follows
\begin{equation}
[\alpha,\beta]_{SN}=i_{X^a}\alpha\wedge\nabla_{X_a}\beta+(-1)^{pq}i_{X^a}\beta\wedge\nabla_{X_a}\alpha
\end{equation}
and gives a $(p+q-1)$-form. For $p=q=1$, $[\,,\,]_{SN}$ reduces to the Lie bracket of vector fields
\begin{equation}
[X,Y]=\nabla_XY-\nabla_YX\quad,\quad X,Y\in TM.
\end{equation}
SN bracket satisfies the following graded Lie bracket properties
\begin{eqnarray}
[\alpha,\beta]_{SN}&=&(-1)^{pq}[\beta,\alpha]\\
(-1)^{p(r+1)}[\alpha,[\beta,\gamma]_{SN}]_{SN}&+&(-1)^{q(p+1)}[\beta,[\gamma,\alpha]_{SN}]_{SN}\nonumber\\
&+&(-1)^{r(q+1)}[\gamma,[\alpha,\beta]_{SN}]_{SN}=0
\end{eqnarray}
where $\gamma$ is a $r$-form.\\

Since Killing vector fields satisfy a Lie algebra structure, we can ask if KY forms satisfy a Lie algebra under $[\,,\,]_{SN}$. For a KY $p$-form $\omega_1$ and KY $q$-form $\omega_2$; we investigate that in which conditions we have the following equality;
\begin{eqnarray}
\nabla_{X_a}[\omega_1,\omega_2]_{SN}=\frac{1}{p+q}i_{X_a}d[\omega_1,\omega_2]_{SN}.
\end{eqnarray}
For the left hand side of (111), we have from the definition (107)
\begin{eqnarray}
\nabla_{X_a}[\omega_1,\omega_2]_{SN}&=&\nabla_{X_a}\left(i_{X^b}\omega_1\wedge\nabla_{X_b}\omega_2+(-1)^{pq}i_{X^b}\omega_2\wedge\nabla_{X_b}\omega_1\right)\nonumber\\
&=&\nabla_{X_a}i_{X^b}\omega_1\wedge\nabla_{X_b}\omega_2+i_{X^b}\omega_1\wedge\nabla_{X_a}\nabla_{X_b}\omega_2\nonumber\\
&&+(-1)^{pq}\nabla_{X_a}i_{X^b}\omega_2\wedge\nabla_{X_b}\omega_1+(-1)^{pq}i_{X^b}\omega_2\wedge\nabla_{X_a}\nabla_{X_b}\omega_1.\nonumber
\end{eqnarray}
Here, we can use $[i_{X^b},\nabla_{X_a}]=0$ and the KY equation (57) to obtain
\begin{eqnarray}
\nabla_{X_a}[\omega_1,\omega_2]_{SN}&=&\frac{1}{(p+1)(q+1)}\big(i_{X^b}i_{X_a}d\omega_1\wedge i_{X_b}d\omega_2\nonumber\\
&&+(-1)^{pq}i_{X^b}i_{X_a}d\omega_2\wedge i_{X_b}d\omega_1\big)\nonumber\\
&&+\frac{1}{q+1}i_{X^b}\omega_1\wedge\nabla_{X_a}i_{X_b}d\omega_2\nonumber\\
&&+\frac{(-1)^{pq}}{p+1}i_{X^b}\omega_2\wedge\nabla_{X_a}i_{X_b}d\omega_1.\nonumber
\end{eqnarray}
By using $\nabla_{X_a}i_{X_b}=i_{X_b}\nabla_{X_a}$ again and from the integrability condition of KY forms in (88), we find
\begin{eqnarray}
\nabla_{X_a}[\omega_1,\omega_2]_{SN}&=&\frac{1}{(p+1)(q+1)}\big(i_{X^b}i_{X_a}d\omega_1\wedge i_{X_b}d\omega_2\nonumber\\
&&+(-1)^{pq}i_{X^b}i_{X_a}d\omega_2\wedge i_{X_b}d\omega_1\big)\nonumber\\
&&+\frac{1}{q}i_{X^b}\omega_1\wedge i_{X_b}\left(R_{ca}\wedge i_{X^c}\omega_2\right)\nonumber\\
&&+\frac{(-1)^{pq}}{p}i_{X^b}\omega_2\wedge i_{X_b}\left(R_{ca}\wedge i_{X^c}\omega_1\right).
\end{eqnarray}
For the right hand side of (111), we have
\[
\frac{1}{p+q}i_{X_a}d[\omega_1,\omega_2]_{SN}=\frac{1}{p+q}i_{X_a}d\left(i_{X^a}\omega_1\wedge\nabla_{X_a}\omega_2+(-1)^{pq}i_{X^a}\omega_2\wedge\nabla_{X_a}\omega_1\right)
\]
and from the properties $d(\alpha\wedge\beta)=d\alpha\wedge\beta+(-1)^p\alpha\wedge d\beta$, $i_X(\alpha\wedge\beta)=i_X\alpha\wedge\beta+(-1)^p\alpha\wedge i_X\beta$ and $[\nabla_{X_a},d]=0$ in normal coordinates, we obtain
\begin{eqnarray}
\frac{1}{p+q}i_{X_a}d[\omega_1,\omega_2]_{SN}&=&\frac{1}{(p+1)(q+1)}\big(i_{X^b}i_{X_a}d\omega_1\wedge i_{X_b}d\omega_2\nonumber\\
&&+(-1)^{pq}i_{X^b}i_{X_a}d\omega_2\wedge i_{X_b}d\omega_1\big)\nonumber\\
&&-\frac{(-1)^p}{pq}i_{X_a}\left(i_{X^b}\omega_1\wedge R_{cb}\wedge i_{X^c}\omega_2\right).
\end{eqnarray}
In general, we do not have the equality (111) as can be seen by comparing (112) and (113). However, for constant curvature manifolds $R_{ab}=c e_a\wedge e_b$, the curvature terms in (112) and (113) are equal to each other and we have the equality (111). So, KY forms satisfy a Lie superalgebra $\mathfrak{k}$ in constant cuvature manifolds \cite{Kastor Ray Traschen};
\[
\mathfrak{k}=\mathfrak{k}_0\oplus\mathfrak{k}_1
\]
The even part $\mathfrak{k}_0$ corresponds to odd degree KY forms and the odd part $\mathfrak{k}_1$ corresponds to even degree KY forms.
\[
[\,,\,]_{SN}:\mathfrak{k}_0\times\mathfrak{k}_0\longrightarrow\mathfrak{k}_0 \quad \text{(if $p$ and $q$ odd, then $p+q-1$ odd)}
\]
\[
[\,,\,]_{SN}:\mathfrak{k}_0\times\mathfrak{k}_1\longrightarrow\mathfrak{k}_1 \quad \text{(if $p$ odd and $q$ even, then $p+q-1$ even)}
\]
\[
[\,,\,]_{SN}:\mathfrak{k}_1\times\mathfrak{k}_1\longrightarrow\mathfrak{k}_0 \quad \text{(if $p$ and $q$ even, then $p+q-1$ odd)}
\]
and $[\,,\,]_{SN}$ satisfies the graded Jacobi identity.

From the integrabilty condition of KY forms in (88), the second and higher order derivatives of KY forms can be written in terms of themselves. So, the maximal number of KY $p$-forms in $n$-dimensions can be found by counting the independent degrees of freedom of $\omega$ and $d\omega$ as
\begin{eqnarray}
K_p=\left(
             \begin{array}{c}
               n \\
               p \\
             \end{array}
           \right)+\left(
             \begin{array}{c}
               n \\
               p+1 \\
             \end{array}
           \right)=\left(
             \begin{array}{c}
               n+1 \\
               p+1 \\
             \end{array}
           \right)=\frac{(n+1)!}{(p+1)!(n-p)!}
\end{eqnarray}
and this number is achived in constant curvature manifolds. Then, the dimension of the KY superalgebra is $(K_{odd}|K_{even})$ where
\[
K_{odd}=\sum_{k=1}^{\lfloor\frac{n}{2}\rfloor}\left(
             \begin{array}{c}
               n+1 \\
               2k \\
             \end{array}
           \right) \quad \text{and} \quad K_{even}=\sum_{k=1}^{\lfloor\frac{n-1}{2}\rfloor}\left(
             \begin{array}{c}
               n+1 \\
               2k+1 \\
             \end{array}
           \right)
\]

ii) Conformal Killing-Yano superalgebra;

Conformal Killing vector fields also satisfy a Lie algebra with respect to the Lie bracket of vector fields
\begin{equation}
[C_i,C_j]=f_{ijk}C_k
\end{equation}
which is called the conformal algebra where $f_{ijk}$ are structure constants. For CKY forms $\omega$ satisfying (49), we can define the CKY bracket $[\,,\,]_{CKY}$ \cite{Ertem3}. For a CKY $p$-form $\omega_1$ and a CKY $q$-form $\omega_2$, $[\,,\,]_{CKY}$ is defined as
\begin{eqnarray}
[\omega_1,\omega_2]_{CKY}&=&\frac{1}{q+1}i_{X^a}\omega_1\wedge i_{X_a}d\omega_2+\frac{(-1)^p}{p+1}i_{X^a}d\omega_1\wedge i_{X_a}\omega_2\nonumber\\
&&+\frac{(-1)^p}{n-q+1}\omega_1\wedge\delta\omega_2+\frac{1}{n-p+1}\delta\omega_1\wedge\omega_2.
\end{eqnarray}
Note that it is different from the SN bracket, but a slight modification of it. $[\,,\,]_{CKY}$ also satisfies the graded Lie bracket properties. For a CKY $p$-form $\omega_1$ and a CKY $q$-form $\omega_2$, $[\omega_1,\omega_2]_{CKY}$ is a CKY $(p+q-1)$-form, i.e.
\[
\nabla_{X_a}[\omega_1,\omega_2]_{CKY}=\frac{1}{p+q}i_{X_a}d[\omega_1,\omega_2]_{CKY}-\frac{1}{n-p-q+2}e_a\wedge\delta[\omega_1,\omega_2]_{CKY}
\]
in constant curvature manifolds and in Einstein manifolds for normal CKY forms \cite{Ertem3}. So, CKY forms satisfy a Lie superalgebra $\mathfrak{c}$ in these cases;
\[
\mathfrak{c}=\mathfrak{c}_0\oplus\mathfrak{c}_1
\]
The even part $\mathfrak{c}_0$ corresponds to odd degree CKY forms and the odd part $\mathfrak{c}_1$ corresponds to even degree CKY forms;
\[
[\,,\,]_{CKY}:\mathfrak{c}_0\times\mathfrak{c}_0\longrightarrow\mathfrak{c}_0
\]
\[
[\,,\,]_{CKY}:\mathfrak{c}_0\times\mathfrak{c}_1\longrightarrow\mathfrak{c}_1
\]
\[
[\,,\,]_{CKY}:\mathfrak{c}_1\times\mathfrak{c}_1\longrightarrow\mathfrak{c}_0
\]
From the integrability condition of CKY forms in (100), the counting of the independent degrees of freedom of $\omega$, $d\omega$, $\delta\omega$ and $d\delta\omega$ gives the maximal number of CKY $p$-forms in $n$-dimensions which is achieved in constant curvature manifolds
\begin{eqnarray}
C_p&=&2\left(
             \begin{array}{c}
               n \\
               p \\
             \end{array}
           \right)+\left(
             \begin{array}{c}
               n \\
               p-1 \\
             \end{array}
           \right)+\left(
             \begin{array}{c}
               n \\
               p+1 \\
             \end{array}
           \right)=\left(
             \begin{array}{c}
               n+2 \\
               p+1 \\
             \end{array}
           \right)\nonumber\\
&=&\frac{(n+2)!}{(p+1)!(n-p+1)!}.
\end{eqnarray}
So, the dimension of the CKY superalgebra is $(C_{odd}|C_{even})$ where
\[
C_{odd}=\sum_{k=1}^{\lfloor\frac{n}{2}\rfloor}\left(
             \begin{array}{c}
               n+2 \\
               2k \\
             \end{array}
           \right)\quad\text{and}\quad C_{even}=\sum_{k=1}^{\lfloor\frac{n-1}{2}\rfloor}\left(
             \begin{array}{c}
               n+2 \\
               2k+1 \\
             \end{array}
           \right).
\]
All KY forms are CKY forms at the same time and the number of CKY $p$-forms that do not correspond to KY $p$-forms is given by
\begin{equation}
C_p-K_p=\left(
             \begin{array}{c}
               n+2 \\
               p+1 \\
             \end{array}
           \right)-\left(
             \begin{array}{c}
               n+1 \\
               p+1 \\
             \end{array}
           \right)=\frac{(n+1)!}{p!(n-p+1)!}
\end{equation}

\subsection{Extended superalgebras}

Now, we construct extended superalgebras by using Killing and twistor spinors \cite{Ertem5}.

i) Killing superalgebras;

We can define a superalgebra structure by using Killing vectors and Killing spinors as
\[
\mathfrak{k}=\mathfrak{k}_0\oplus\mathfrak{k}_1
\]
The even part $\mathfrak{k}_0$ corresponds to Lie algebra of Killing vector fields and the odd part $\mathfrak{k}_1$ corresponds to the set of Killing spinors. The brackets of the superalgebra is defined as follows;

The even-even bracket is the Lie bracket of Killing vector fields
\begin{eqnarray}
[\,,\,]:\mathfrak{k}_0\times\mathfrak{k}_0&\longrightarrow&\mathfrak{k}_0\nonumber\\
(K_1,K_2)&\longmapsto&[K_1,K_2],
\end{eqnarray}
the even-odd bracket is the Lie derivative on spinor fields
\begin{eqnarray}
\pounds:\mathfrak{k}_0\times\mathfrak{k}_1&\longrightarrow&\mathfrak{k}_1\nonumber\\
(K,\psi)&\longmapsto&\pounds_K\psi,
\end{eqnarray}
and the odd-odd bracket is the Dirac currents of Killling spinors
\begin{eqnarray}
(\,)_1:\mathfrak{k}_1\times\mathfrak{k}_1&\longrightarrow&\mathfrak{k}_0\nonumber\\
(\psi,\phi)&\longmapsto&(\psi\overline{\phi})_1\cong V_{\psi,\phi}.
\end{eqnarray}
The Jacobi identities correspond to
\begin{eqnarray}
&&[K_1,[K_2,K_3]]+[K_2,[K_3,K_1]]+[K_3,[K_1,K_2]]=0\nonumber\\
&&[\pounds_{K_1},\pounds_{K_2}]\psi=\pounds_{[K_1,K_2]}\psi\nonumber\\
&&{\cal{L}}_K(\psi\overline{\phi})=(\pounds_K\psi)\overline{\phi}+\psi(\overline{\pounds_K\phi})\\
&&\pounds_{V_{\psi}}\psi=0.\nonumber
\end{eqnarray}
The first three identities are satisfied automatically from the properties of the Lie derivative, but the last one is not satisfied automatically. For manifolds on which the last identity is satisfied, Killing superalgebra is a Lie superalgebra. We can extend this superalgebra structure to include higher-degree KY forms in constant curvature manifolds,
\[
\bar{\mathfrak{k}}=\bar{\mathfrak{k}}_0\oplus\bar{\mathfrak{k}}_1
\]
The even part $\bar{\mathfrak{k}}_0$ corresponds to the Lie algebra of odd KY forms and the odd part $\bar{\mathfrak{k}}_1$ corresponds to the set of Killing spinors. The brackets of the extended superalgebra are defined as follows;

The even-even bracket is the SN bracket of KY forms
\begin{eqnarray}
[\,,\,]_{SN}:\bar{\mathfrak{k}}_0\times\bar{\mathfrak{k}}_0&\longrightarrow&\bar{\mathfrak{k}}_0\nonumber\\
(\omega_1,\omega_2)&\longmapsto&[\omega_1,\omega_2]_{SN},
\end{eqnarray}
the even-odd bracket is the symmetry operators of Killing spinors
\begin{eqnarray}
L:\bar{\mathfrak{k}}_0\times\bar{\mathfrak{k}}_1&\longrightarrow&\bar{\mathfrak{k}}_1\nonumber\\
(\omega,\psi)&\longmapsto&L_{\omega}\psi=(i_{X^a}\omega).\nabla_{X_a}\psi+\frac{p}{2(p+1)}d\omega.\psi,
\end{eqnarray}
and the odd-odd bracket is the $p$-form Dirac currents of Killling spinors
\begin{eqnarray}
(\,)_p:\bar{\mathfrak{k}}_1\times\bar{\mathfrak{k}}_1&\longrightarrow&\bar{\mathfrak{k}}_0\nonumber\\
(\psi,\phi)&\longmapsto&(\psi\overline{\phi})_p.
\end{eqnarray}
The Jacobi identities are
\begin{eqnarray}
&&[\omega_1,[\omega_2,\omega_3]_{SN}]_{SN}+[\omega_2,[\omega_3,\omega_1]_{SN}]_{SN}+[\omega_3,[\omega_1,\omega_2]_{SN}]_{SN}=0\nonumber\\
&&[L_{\omega_1},L_{\omega_2}]\psi=L_{[\omega_1,\omega_2]_{SN}}\psi\nonumber\\
&&[\omega,(\psi\overline{\phi})]_{SN}=(L_{\omega}\psi)\overline{\phi}+\psi(\overline{L_{\omega}\phi})\\
&&L_{(\psi\overline{\psi})_p}\psi=0.\nonumber
\end{eqnarray}
The first identity is satisfied from the properties of the SN bracket, but the last three identities are not satisfied. So, $\bar{\mathfrak{k}}$ is a superalgebra in constant curvature manifolds, but not a Lie superalgebra. For $n\leq 5$, we can also define a new bracket for KY forms (for a $p$-form $\omega_1$ and $q$-form $\omega_2$),
\begin{equation}
[\omega_1,\omega_2]_{KY}=\frac{pq}{p+q-1}[\omega_1,\omega_2]_{SN}-\frac{pq}{p+q}[i_{X^a}i_{x^b}\omega_1,i_{X_a}i_{X_b}\omega_2]_{SN}
\end{equation}
which is a Lie bracket and satisfies the second Jacobi identity \cite{Ertem2}
\begin{equation}
[L_{\omega_1},L_{\omega_2}]\psi=L_{[\omega_1,\omega_2]_{KY}}\psi.
\end{equation}

ii) Conformal superalgebras;

We can also define a superalgebra structure by using conformal Killing vectors and twistor spinors,
\[
\mathfrak{c}=\mathfrak{c}_0\oplus\mathfrak{c}_1
\]
The even part $\mathfrak{c}_0$ corresponds to the Lie algebra of conformal Killing vectors and the odd part $\mathfrak{c}_1$ corresponds to the set of twistor spinors. The brackets of the superalgebra are defined as follows;

The even-even bracket is the Lie bracket of conformal Killing vector fields
\begin{eqnarray}
[\,,\,]:\mathfrak{c}_0\times\mathfrak{c}_0&\longrightarrow&\mathfrak{c}_0\nonumber\\
(C_1,C_2)&\longmapsto&[C_1,C_2],
\end{eqnarray}
the even-odd bracket is the Lie derivative on twistor spinors
\begin{eqnarray}
\pounds-\frac{1}{2}\mu:\mathfrak{c}_0\times\mathfrak{c}_1&\longrightarrow&\mathfrak{c}_1\nonumber\\
(C,\psi)&\longmapsto&\pounds_C\psi-\frac{1}{2}\mu\psi,
\end{eqnarray}
and the odd-odd bracket is the Dirac currents of twistor spinors
\begin{eqnarray}
(\,)_1:\mathfrak{c}_1\times\mathfrak{c}_1&\longrightarrow&\mathfrak{c}_0\nonumber\\
(\psi,\phi)&\longmapsto&(\psi\overline{\phi})_1\cong V_{\psi,\phi}.
\end{eqnarray}
The Jacobi identities are not satisfied automatically. We can extend this superalgebra structure to include higher-degree CKY forms in constant curvature manifolds or Einstein manifolds with normal CKY forms
\[
\bar{\mathfrak{c}}=\bar{\mathfrak{c}}_0\oplus\bar{\mathfrak{c}}_1
\]
The even part $\bar{\mathfrak{c}}_0$ corresponds to the Lie algebra of CKY forms (or normal CKY forms) and the even part $\bar{\mathfrak{c}}_1$ corresponds to the set of twistor spinors. The brackets of the extended conformal superalgebra are defined in the following form.

The even-even bracket is the CKY bracket of CKY forms
\begin{eqnarray}
[\,,\,]_{CKY}:\bar{\mathfrak{c}}_0\times\bar{\mathfrak{c}}_0&\longrightarrow&\bar{\mathfrak{c}}_0\nonumber\\
(\omega_1,\omega_2)&\longmapsto&[\omega_1,\omega_2]_{CKY},
\end{eqnarray}
the even-odd bracket is the symmetry operators of twistor spinors
\begin{eqnarray}
L:\bar{\mathfrak{c}}_0\times\bar{\mathfrak{c}}_1&\longrightarrow&\bar{\mathfrak{c}}_1\\
(\omega,\psi)&\longmapsto&L_{\omega}\psi=(i_{X^a}\omega).\nabla_{X_a}\psi+\frac{p}{2(p+1)}d\omega.\psi+\frac{p}{2(n-p+1)}\delta\omega.\psi,\nonumber
\end{eqnarray}
and the odd-odd bracket is the $p$-form Dirac currents of twistor spinors
\begin{eqnarray}
(\,)_p:\bar{\mathfrak{c}}_1\times\bar{\mathfrak{c}}_1&\longrightarrow&\bar{\mathfrak{c}}_0\nonumber\\
(\psi,\phi)&\longmapsto&(\psi\overline{\phi})_p.
\end{eqnarray}
The Jacobi identities are not satisfied automatically. So, $\bar{\mathfrak{c}}$ is a superalgebra in constant curvature manifolds (or in Einstein manifolds with normal CKY forms), but not a Lie superalgebra.

\section{Harmonic Spinors from Twistors}

We construct transformation operators that give harmonic spinors from twistor spinors via solutions of the conformal Laplace equation and potential forms in this section. By defining $Spin^c$ structures, we generalize these transformation operators to gauged harmonic spinors and gauged twistor spinors. This gives way to obtain the solutions of the Seiberg-Witten equations from gauged twistor spinors.

\subsection{Twistors to harmonic spinors}

Let us start with a twistor spinor $\psi$ which satisfies (10) and consider a function $f$ which is a solution of the conformally generalized Laplace equation in $n$-dimensions;
\begin{equation}
\Delta f-\frac{n-2}{4(n-1)}{\cal{R}}f=0
\end{equation}
where $\Delta=-d\delta-\delta d$ is the Laplace-Beltrami operator and ${\cal{R}}$ is the curvature scalar. We can define an operator
\begin{equation}
L_f=\frac{n-2}{n}f\displaystyle{\not}D+df
\end{equation}
and the action of this operator on a twistor spinor gives a harmonic spinor, i.e., we have $\displaystyle{\not}DL_f\psi=0$. This can be seen as follows
\begin{eqnarray}
\displaystyle{\not}DL_f\psi&=&e^a.\nabla_{X_a}\left(\frac{n-2}{n}f\displaystyle{\not}D\psi+df.\psi\right)\nonumber\\
&=&e^a.\left(\frac{n-2}{n}(\nabla_{X_A}f)\displaystyle{\not}D\psi+\frac{n-2}{n}f\nabla_{X_a}\displaystyle{\not}D\psi+\nabla_{X_a}df.\psi+df.\nabla_{X_a}\psi\right)\nonumber\\
&=&\frac{n-2}{n}\displaystyle{\not}df.\displaystyle{\not}D\psi+\frac{n-2}{n}f\displaystyle{\not}D^2\psi+\displaystyle{\not}ddf.\psi+e^a.df.\nabla_{X_a}\psi\nonumber
\end{eqnarray}
where we have used $e^a.\nabla_{X_a}f=(d-\delta)f=\displaystyle{\not}df$, $e^a.\nabla_{X_a}\displaystyle{\not}D\psi=\displaystyle{\not}D^2\psi$ and $e^a.\nabla_{X_a}df=\displaystyle{\not}ddf$. Since $f$ is a function; $\delta f=0$, $\displaystyle{\not}df=df$ and $\displaystyle{\not}ddf=-\delta df=\Delta f$ (since $d^2=0$). From the twistor equation (10) and the integrability condition of twistor spinors in (19), we can write
\[
\displaystyle{\not}DL_f\psi=\frac{n-2}{n}\left(df.\displaystyle{\not}D\psi-\frac{n}{4(n-1)}{\cal{R}}f\psi\right)+\Delta f.\psi+\frac{1}{n}e^a.df.e_a.\displaystyle{\not}D\psi.
\]
We know that $f$ satisfies the generalized Laplace equation and we have the property $e^a.\alpha.e_a=(-1)^p(n-2p)\alpha$ for a $p$-form $\alpha$. So, we have
\begin{equation}
\displaystyle{\not}DL_f\psi=\left(\Delta f-\frac{n-2}{4(n-1)}{\cal{R}}f\right).\psi=0.
\end{equation}

The generalization of the operator $L_f$ to higher-degree differential forms can be investigated. We propose the following operator for a $p$-form $\alpha$ and an inhomogeneous Clifford form $\Omega$;
\begin{equation}
L_{\alpha}=\alpha.\displaystyle{\not}D+\Omega.
\end{equation}
We investigate the conditions that we can have a harmonic spinor by applying $L_{\alpha}$ to a twistor spinor $\psi$, i.e., $\displaystyle{\not}DL_{\alpha}\psi=0$. By calculating explicitly as
\begin{eqnarray}
\displaystyle{\not}DL_{\alpha}\psi&=&e^a.\nabla_{X_a}L_{\alpha}\psi\nonumber\\
&=&e^a.\nabla_{X_a}\alpha.\displaystyle{\not}D\psi+e^a.\alpha.\nabla_{X_a}\displaystyle{\not}D\psi+e^a.\nabla_{X_a}\Omega.\psi+e^a.\Omega.\nabla_{X_a}\psi.\nonumber
\end{eqnarray}
From the definition $\displaystyle{\not}d=e^a.\nabla_{X_a}$, twistor equation (10) and the integrability condition (21), we can write
\begin{eqnarray}
\displaystyle{\not}DL_{\alpha}\psi&=&\displaystyle{\not}d\alpha.\displaystyle{\not}D\psi+\frac{n}{2}e^a.\alpha.K_a.\psi+\displaystyle{\not}d\Omega.\psi+\frac{1}{n}e^a.\Omega.e_a.\displaystyle{\not}D\psi\nonumber\\
&=&\left(\displaystyle{\not}d\alpha+\frac{n-2\Pi}{n}\eta\Omega\right).\displaystyle{\not}D\psi+\left(\displaystyle{\not}d\Omega+\frac{n}{2}e^a.\alpha.K_a\right).\psi.
\end{eqnarray}
Here we have used $e^a.\Omega.e_a=(n-2\Pi)\eta\Omega$ where $\Pi\Omega=e^a\wedge i_{X_a}\Omega$. Hence, if $L_{\alpha}\psi$ would be a harmonic spinor, then we have two conditions
\begin{eqnarray}
\displaystyle{\not}d\alpha+\frac{n-2\Pi}{n}\eta\Omega&=&0\\
\displaystyle{\not}d\Omega+\frac{n}{2}e^a.\alpha.K_a&=&0.
\end{eqnarray}
As a special case, in even dimensions $n=2k$, we can choose $\Omega$ as a $k$-form (middle form) and (140) gives $d\alpha=-\delta\alpha$ and this can be possible for $\alpha=0$ since $\alpha$ is a homogeneous $p$-form. In that case, (141) gives that $\Omega$ is a harmonic form; $d\Omega=\delta\Omega=0$. So, in even dimensions, for a harmonic middle form $\Omega$ and a twistor spinor $\psi$, $\Omega.\psi$ is a harmonic spinor \cite{Benn Kress}.

In general, if we choose $\alpha$ a non-zero $p$-form, then (140) gives that $\Omega$ is a sum of $(p+1)$ and $(p-1)$-forms. So, we have from (140)
\begin{equation}
\Omega=\frac{(-1)^pn}{n-2(p+1)}d\alpha-\frac{(-1)^pn}{n-2(p-1)}\delta\alpha.
\end{equation}
By applying the Hodge-de Rham operator $\displaystyle{\not}d=d-\delta$
\begin{eqnarray}
\displaystyle{\not}d\Omega&=&\frac{(-1)^pn}{n-2(p+1)}\displaystyle{\not}dd\alpha-\frac{(-1)^pn}{n-2(p-1)}\displaystyle{\not}d\delta\alpha\nonumber\\
&=&-\frac{(-1)^pn}{n-2(p+1)}\delta d\alpha-\frac{(-1)^pn}{n-2(p-1)}d\delta\alpha\nonumber
\end{eqnarray}
and by substituting in (141), we have
\begin{equation}
\frac{(-1)^p}{n-2(p+1)}\delta d\alpha+\frac{(-1)^p}{n-2(p-1)}d\delta\alpha=\frac{1}{2}e^a.\alpha.K_a.
\end{equation}
From the definiton of $K_a$ in (20), the right hand side of (143) can be written explicitly as
\begin{equation}
e^a.\alpha.K_a=(-1)^p\frac{2}{n-2}P_a\wedge i_{X^a}\alpha-(-1)^p\frac{n+2(p-1)}{2(n-1)(n-2)}{\cal{R}}\alpha.
\end{equation}
So, we obtain
\begin{eqnarray}
\frac{1}{n-2(p+1)}\delta d\alpha+\frac{1}{n-2(p-1)}d\delta\alpha=\frac{1}{n-2}P_a\wedge i_{X^a}\alpha-\frac{n+2(p-1)}{4(n-1)(n-2)}{\cal{R}}\alpha.\nonumber\\
\end{eqnarray}
A $p$-form $\alpha$ that satisfies this equation is called as a potential form. The reason of the nomenclature is that, in even dimensions, for $p=\frac{n}{2}-1$, $\alpha$ is a potential form for middle-form Maxwell equations and for $p=\frac{n}{2}+1$ it is a co-potential form for the same equations. For $p=0$, potential form equation reduces to the conformally generalized Laplace equation. So, it is generalization of the conformal Laplace equation. Then, we obtain that for a potential $p$-form $\alpha$ and a twistor spinor $\psi$
\begin{equation}
L_{\alpha}\psi=\alpha.\displaystyle{\not}D\psi+\frac{(-1)^pn}{n-2(p+1)}d\alpha.\psi-\frac{(-1)^pn}{n-2(p-1)}\delta\alpha.\psi
\end{equation}
is a harmonic spinor \cite{Benn Kress, Ertem7}.

Remember the symmetry operators of twistor spinors in (95) which are defined in terms of CKY forms $\omega$ in constant curvature manifolds and normal CKY forms in Einstein manifolds;
\[
L_{\omega}=-(-1)^p\frac{p}{n}\omega.\displaystyle{\not}D+\frac{p}{2(p+1)}d\omega+\frac{p}{2(n-p+1)}\delta\omega
\]
and the symmetry operators of harmonic spinors (massless Dirac equation) in (80) which are written in terms of CKY forms $\omega$
\[
{\cal{L}}_{\omega}=(i_{x^a}\omega).\nabla_{X_a}+\frac{p}{2(p+1)}d\omega-\frac{n-p}{2(n-p+1)}\delta\omega.
\]
So, by considering the last three operators, we have the following relations
\[
\textrm{twistor spinors}\quad\autorightarrow{$L_{\omega}$}{CKY forms}\quad\textrm{twistor spinors}
\]

\[
\textrm{twistor spinors}\quad\autorightarrow{$L_{\alpha}$}{potential forms}\quad\textrm{harmonic spinors}
\]

\[
\textrm{harmonic spinors}\quad\autorightarrow{${\cal{L}}_{\omega}$}{CKY forms}\quad\textrm{harmonic spinors}
\]

\subsection{Gauged twistors to gauged harmonic spinors}

We generalize the constructions in the previous subsection to $Spin^c$ spinors. We consider a manifold $M$ with a $Spin^c$ structure which means that we can lift the complexified bundle $SO(n)\times_{\mathbb{Z}_2}U(1)$ to $Spin^c(n)=Spin(n)\times_{\mathbb{Z}_2}U(1)=Spin(n)\times U(1)/\mathbb{Z}_2$. Note that $Spin^c(n)\subset Cl(n)\otimes\mathbb{C}=Cl^c(n)$ is a subgroup of the complexified Clifford algebra. In that case, we have the $Spin^c$ spinor bundle $\Sigma^cM=\Sigma M\times_{\mathbb{Z}_2}U(1)$ and the connection on $\Sigma^cM$ consists of two parts; the Levi-Civita part on $\Sigma M$ and the $U(1)$-connection part. The gauged connection $\widehat{\nabla}$ on $\Sigma^cM$ is written as follows
\begin{equation}
\widehat{\nabla}_X=\nabla_X+i_XA\quad,\quad X\in TM.
\end{equation}
Gauged exterior and co-derivatives are defined as
\begin{equation}
\widehat{d}=e^a\wedge\widehat{\nabla}_{X_a}=d+A\wedge
\end{equation}
\begin{equation}
\widehat{\delta}=-i_{X^a}\widehat{\nabla}_{X_a}=\delta-i_{\widetilde{A}}
\end{equation}
where $\widetilde{A}$ is the vector field corresponding to the metric dual of $A$. However, unlike the case of $d^2=0=\delta^2$, we have
\begin{equation}
\widehat{d}^2=F\wedge
\end{equation}
\begin{equation}
\widehat{\delta}^2=-(i_{X a}i_{X^b}F)i_{X_a}i_{X_b}
\end{equation}
where $F=dA$ is the gauge curvature. The gauged Hodge-de Rham operator is
\begin{equation}
\widehat{\displaystyle{\not}d}=\widehat{d}-\widehat{\delta}=\displaystyle{\not}d+A.
\end{equation}
Gauged curvature operator is defined as
\begin{eqnarray}
\widehat{R}(X_a,X_b)&=&[\widehat{\nabla}_{X_a},\widehat{\nabla}_{X_b}]-\widehat{\nabla}_{[X_a,X_b]}\nonumber\\
&=&R(X_a,X_b)-i_{X_a}i_{X_b}F
\end{eqnarray}
and the gauged Dirac operator acting on $\Sigma^cM$ is written as
\begin{equation}
\widehat{\displaystyle{\not}D}=e^a.\widehat{\nabla}_{X_a}=\displaystyle{\not}D+A
\end{equation}
and its square is equal to
\begin{equation}
\widehat{\displaystyle{\not}D}^2=\widehat{\nabla}^2-\frac{1}{4}{\cal{R}}+F.
\end{equation}
We can define the gauged twistor equation as
\begin{equation}
\widehat{\nabla}_{X_a}\psi=\frac{1}{n}e_a.\widehat{\displaystyle{\not}D}\psi
\end{equation}
and the gauged harmonic spinor equation as
\begin{equation}
\widehat{\displaystyle{\not}D}\psi=0.
\end{equation}
The integrability conditions of gauged twistor spinors can be obtained as \cite{Ertem7}
\begin{eqnarray}
\widehat{\displaystyle{\not}D}^2\psi&=&-\frac{n}{4(n-1)}{\cal{R}}\psi+\frac{n}{n-1}F.\psi\\
\widehat{\nabla}_{X_a}\widehat{\displaystyle{\not}D}\psi&=&\frac{n}{2}K_a.\psi-\frac{n}{(n-1)(n-2)}e_a.F.\psi+\frac{n}{n-2}i_{X_a}F.\psi\\
C_{ab}.\psi&=&2(i_{X_a}i_{X_b}F)\psi+\frac{n}{n-2}\left(e_b.i_{X_a}F-e_a.i_{X_b}F\right).\psi\nonumber\\
&&+\frac{4}{(n-1)(n-2)}e_a.e_b.F.\psi.
\end{eqnarray}
In constant curvature manifolds, we have $C_{ab}=0$ and the existence of gauged twistor spinors implies $F=0$. So, we can have gauged twistor spinors with respect to flat connections ($A\neq 0$ and $F=0$). Symmetry operators of gauged twistor spinors can be constructed from CKY forms in constant curvature manifolds. For a CKY $p$-form $\omega$
\begin{equation}
\widehat{L}_{\omega}=-(-1)^p\frac{p}{n}\omega.\widehat{\displaystyle{\not}D}+\frac{p}{2(p+1)}d\omega+\frac{p}{2(n-p+1)}\delta\omega
\end{equation}
is a symmetry operator, i.e., we have $\widehat{\nabla}_{X_a}\widehat{L}_{\omega}\psi=\frac{1}{n}e_a.\widehat{\displaystyle{\not}D}\widehat{L}_{\omega}\psi$ for a gauged twistor spinor $\psi$ \cite{Ertem6}.

Symmetry operators of gauged harmonic spinors can be constructed from gauged CKY forms which satisfy
\begin{equation}
\widehat{\nabla}_X\omega=\frac{1}{p+1}i_X\widehat{d}\omega-\frac{1}{n-p+1}\widetilde{X}\wedge\widehat{\delta}\omega
\end{equation}
for any $X\in TM$. Symmetry operators are written in terms of a gauged CKY $p$-form as
\begin{equation}
\widehat{\cal{L}}_{\omega}=(i_{X^a}\omega).\widehat{\nabla}_{X_a}+\frac{p}{2(p+1)}\widehat{d}\omega-\frac{n-p}{2(n-p+1)}\widehat{\delta}\omega
\end{equation}
i.e, for a harmonic spinor $\psi$, we have $\widehat{\displaystyle{\not}D}\widehat{\cal{L}}_{\omega}\psi=0$.

We can also construct transformation operators between gauged twistor spinors and gauged harmonic spinors. For a function $f$ that satisfies the generalized gauged Laplace equation
\begin{equation}
\widehat{\Delta}f+\left[\left(1+\frac{n-2}{n-1}\right)\gamma-\frac{n-2}{4(n-1)}{\cal{R}}\right]f=0
\end{equation}
where $\gamma$ is a real number, we can construct the operator
\begin{equation}
\widehat{L}_f=\frac{n-2}{n}f\widehat{\displaystyle{\not}D}+\widehat{d}f
\end{equation}
which transforms a gauged twistor spinor $\psi$ with the property $F.\psi=\gamma\psi$ to a gauged harmonic spinor, i.e., $\widehat{\displaystyle{\not}D}\widehat{L}_f\psi=0$. For a gauged potential form $\alpha$ which satisfies
\begin{eqnarray}
\frac{1}{n-2(p+1)}\widehat{\delta}\widehat{d}\alpha+\frac{1}{n-2(p-1)}\widehat{d}\widehat{\delta}\alpha=\frac{1}{n-2}P_a\wedge i_{X^a}\alpha-\frac{n+2(p-1)}{4(n-1)(n-2)}{\cal{R}}\alpha,\nonumber\\
\end{eqnarray}
the operator
\begin{equation}
\widehat{L}_{\alpha}=\alpha.\widehat{\displaystyle{\not}D}+\frac{(-1)^pn}{n-2(p+1)}\widehat{d}\alpha-\frac{(-1)^pn}{n-2(p-1)}\widehat{\delta}\alpha
\end{equation}
transforms a gauged twistor spinor $\psi$ to a gauged harmonic spinor in constant curvature manifolds, i.e., $\widehat{\displaystyle{\not}D}\widehat{L}_{\alpha}\psi=0$ \cite{Ertem7}. So, we have the following diagram in constant curvature manifolds

\[
\textrm{gauged twistor spinors}\quad\autorightarrow{$\widehat{L}_{\omega}$}{CKY forms}\quad\textrm{gauged twistor spinors}
\]

\[
\textrm{gauged twistor spinors}\quad\autorightarrow{$\widehat{L}_{\alpha}$}{gauged potential forms}\quad\textrm{gauged harmonic spinors}
\]

\[
\textrm{gauged harmonic spinors}\quad\autorightarrow{$\widehat{\cal{L}}_{\omega}$}{gauged CKY forms}\quad\textrm{gauged harmonic spinors}
\]

\subsection{Seiberg-Witten solutions}

The transformation operators constructed in the previous subsecton can give way to obtain the solutions of the Seiberg-Witten equations. Let us consider a 4-dimensional manifold $M$. Every compact orientable 4-manifold has a $Spin^c$ structure. In four dimensions, we have $\Sigma^cM=\Sigma^+M\oplus\Sigma^-M$. For a spinor $\psi\in\Sigma^+M$, Seiberg-Witten (SW) equations are written as \cite{Witten, Seiberg Witten, Friedrich}
\begin{eqnarray}
\widehat{\displaystyle{\not}D}\psi&=&0\\
(\psi\overline{\psi})_2&=&F^+
\end{eqnarray}
where $F^+=\frac{1}{2}(F+*F)$ is the self-dual curvature 2-form. So, the solutions of the SW equations are gauged harmonic spinors with a condition on 2-form Dirac currents. We obtain gauged harmonic spinors from gauged twistor spinors in constant curvature manifolds by using (167). In that case, we have $F=0$ (with $A\neq 0$), and the second SW equation (169) turns into
\begin{equation}
(\psi\overline{\psi})_2=0.
\end{equation}
This is not a very restrictive condition since we can automatically satisfy this condition by choosing symmetric spinor inner product, i.e., $(\psi,\phi)=(\phi,\psi)$ for $\psi, \phi\in\Sigma^cM$;
\begin{eqnarray}
(\psi\overline{\psi})_2&=&(\psi,e_b.e_a.\psi)e^a\wedge e^b\nonumber\\
&=&((e_b.e_a.)^{\cal{J}}.\psi,\psi)e^a\wedge e^b\nonumber\\
&=&-(e_b.e_a.\psi,\psi)e^a\wedge e^b\nonumber\\
&=&-(\psi,e_b.e_a.\psi)e^a\wedge e^b\nonumber\\
&=&0\nonumber
\end{eqnarray}
where we have used that the choice of involution ${\cal{J}}=\xi\text{ or }\xi\eta$ gives the same sign $(e_b.e_a.)^{\cal{J}}=-e_b.e_a$. By starting with a gauged twistor spinor $\psi$, we can construct gauged harmonic spinors from gauged potential, gauged CKY and CKY forms ($\alpha$, $\widehat{\omega}$ and $\omega$) via the operators
\[
\widehat{L}_{\alpha}\psi\quad,\quad \widehat{L}_{\omega}\widehat{L}_{\alpha}\psi\quad,\quad\widehat{L}_{\alpha}\widehat{\cal{L}}_{\widehat{\omega}}\psi\quad,\quad\widehat{L}_{\omega}\widehat{L}_{\alpha}\widehat{\cal{L}}_{\widehat{\omega}}\psi.
\]
If they have vanishing 2-form Dirac currents (which is automatically true for symmetric spinor inner product), then we obtain SW solutions with respect to flat connections in constant curvature manifolds. So, we have the relations

\[
\textrm{gauged twistor spinors}\quad\autorightarrow{$\widehat{L}_{\alpha}$}{gauged potential forms}\quad\textrm{gauged harmonic spinors}
\]

\[
\textrm{gauged harmonic spinors}\quad\autorightarrow{flat connection}{vanishing 2-form Dirac current}\quad\textrm{SW solutions}
\]

\section{Supergravity Killing Spinors}

The variations of the fermionic fields in a bosonic supergravity action give rise to some conditions on the supersymmetry parameters of the theory. The solutions of these conditions are called supergravity Killing spinors and they reduce to the parallel and Killing spinors for some special choices of the fluxes in the theory \cite{Duff Nilsson Pope, Freedman Van Proeyen}. In this section, we consider the spinor bilinears of supergravity Killing spinors which correspond to supergravity Killing forms and investigate the Lie algebra structures they satisfy.

\subsection{Bosonic supergravity}

Supergravity theories are supersymmetric generalizations of General Relativity in various dimensions. A supergravity action $S$ consists of bosonic ($\phi_i$) and fermionic ($\psi_i$) fields;
\[
S=\text{bosonic }(\phi_i,\nabla\phi_i,...)+\text{fermionic }(\psi_i,\nabla\psi_i,...).
\]
Bosonic part corresponds to gravitational and gauge field degrees of freedom (graviton (metric), $p$-form fields, etc.) and fermionic part consists of matter degrees of freedom (gravitino (spin-$\frac{3}{2}$ Rarita-Schwinger field), gaugino, etc). Supersymmetry transformations relate the bosonic and fermionic fields to each other. For a spinor parameter $\epsilon$, supersymmetry transformations are in the following form
\begin{eqnarray}
\delta_{\epsilon}\phi&=&\overline{\epsilon}. \psi\\
\delta_{\epsilon}\psi&=&\left(\nabla+f(\phi)\right)\epsilon
\end{eqnarray}
where $\delta_{\epsilon}$ denotes the variation of the field. When we take the fermionic fields to be zero, we obtain the bosonic supergravity whose solutions give the consistent backgrounds of the theory. When $\psi_i\rightarrow 0$ we have $\delta_{\epsilon}\phi_i=0$ automatically, and $\delta_{\epsilon}\psi=0$ gives a constraint on the spinor parameter $\epsilon$ that is $\left(\nabla+f(\phi)\right)\epsilon=0$ which is called the supergravity Killing spinor equation. So, to obtain a consistent supergravity background, besides the field equations of the theory, we must also solve the supergravity Killing spinor equation.

There are various supergravity theories between the dimensions 4 and 11;

\quad\\
{\centering{
\begin{tabular}{c | c}

$D\text{ (dimension)}\quad$ & \quad $N\text{ (number of susy generators)}$ \\ \hline
$4\quad$ & \quad $1, 2, 3, 4, 5, 6, 8$ \\
$5\quad$ & \quad $2, 4, 6, 8$ \\
$6\quad$ & \quad $(1,0), (2,0), (1,1), (3,0), (2,1), (4,0), (3,1), (2,2)$ \\
$7\quad$ & \quad $2, 4$ \\
$8\quad$ & \quad $1, 2$ \\
$9\quad$ & \quad $1, 2$ \\
$10\quad$ & \quad $1 (\text{type I + heterotic}), 2 (\text{type IIA + IIB})$ \\
$11\quad$ & \quad $1 (\text{M})$ \\

\end{tabular}}
\quad\\
\quad\\
\quad\\}

10 and 11 dimensional supergravity theories correspond to low energy limits of string and M-theories, respectively.

For example, $D=6$, $N=(1,0)$ bosonic supergravity action is
\begin{equation}
S=\kappa\int\left(R_{ab}\wedge *(e^a\wedge e^b)-\frac{1}{2}H\wedge *H\right)
\end{equation}
where $H$ is a 3-form field. The field equations give the Einstein field equations and the conditions $dH=0$ and $H$ is anti-self-dual. Variation of the gravitino field (Rarita-Schwinger) in the fermionic action gives the following supergravity Killing spinor equation
\begin{equation}
\nabla_X\epsilon+\frac{1}{4}i_XH.\epsilon=0.
\end{equation}
$D=10$, type I and heterotic theories also give the same spinor equation. However, in those cases one also have the extra algebraic constraints due to the existence of extra gaugino fields;
\begin{eqnarray}
\left(d\phi+\frac{1}{2}H\right).\epsilon&=&0\\
F_2.\epsilon&=&0
\end{eqnarray}
where $\phi$ is the scalar dilaton and $F_2$ is a 2-form gauge field.

On the other hand, $D=11$ bosonic supergravity action is \cite{Cremmer Julia Scherk}
\begin{equation}
S=\kappa\int\left(R_{ab}\wedge *(e^a\wedge e^b)-\frac{1}{2}F\wedge *F-\frac{1}{6}A\wedge F\wedge F\right)
\end{equation}
where $F=dA$ is a 4-form gauge field. Note that the first term is the gravity action, the second term is the Maxwell-like term and the third term is the Chern-Simons term. The field equations of the theory are given by
\begin{eqnarray}
Ric(X,Y)*1&=&\frac{1}{2}i_XF\wedge *i_YF-\frac{1}{6}g(X,Y)F\wedge *F\\
d*F&=&\frac{1}{2}F\wedge F
\end{eqnarray}
where $X,Y\in TM$ and $Ric$ is the Ricci tensor. Variation of the gravitino field gives the following supergravity Killing spinor equation
\begin{equation}
\nabla_X\epsilon+\frac{1}{6}i_XF.\epsilon-\frac{1}{12}(\widetilde{X}\wedge F).\epsilon=0
\end{equation}
which can also be written in the form
\begin{equation}
\nabla_X\epsilon+\frac{1}{24}\left(\widetilde{X}.F-3F.\widetilde{X}\right).\epsilon=0.
\end{equation}
$D=5$ supergravity has also a similar structure with the coefficient $\frac{1}{24}$ is replaced by the coefficient $\frac{1}{4\sqrt{3}}$ and $F$ is a 2-form.

We can also define a supergravity connection
\begin{equation}
\widehat{\nabla}_X=\nabla_X+\frac{1}{24}\left(\widetilde{X}.F-3F.\widetilde{X}\right)
\end{equation}
to write the supergravity Killing spinor equation as a parallel spinor equation
\begin{equation}
\widehat{\nabla}_X\epsilon=0.
\end{equation}

In $D=11$, we can search for the solutions of type $M_{11}=M_4\times M_7$. So, the metric is in the form of $ds_{11}^2=ds_4^2+ds_7^2$. Co-frame basis, 4-form field $F$ and the spinor parameter $\epsilon$ are decomposed as
\[
e^a=e^{\mu}\otimes e^i\quad\quad,\quad\quad F=F_4\otimes F_7\quad\quad,\quad\quad \epsilon=\epsilon_4\otimes\epsilon_7
\]
where $\mu=0,1,2,3$ and $i=4,5,...,10$. To have $M_4$ as Minkowski space-time, we can choose the ansatz $F=0$. Field equatios give $M_7$ as a Ricci-flat manifold and the supergravity Killing spinor equation reduces to parallel spinor equations $\nabla_X\epsilon_4=0=\nabla_X\epsilon_7$. So, $M_7$ is a Ricci-flat 7-manifold with a parallel spinor which is a $G_2$-holonomy manifold \cite{Atiyah Witten, Duff, House Micu}. Namely, we have
\[
M_{11}=\underbrace{M_4}_{\text{Minkowski}}\times
\underbrace{M_7}_{G_2\text{-holonomy}}
\]

If we choose $F=\lambda z_4 \otimes F_7$ where $\lambda$ is a constant, $z_4$ is the volume form of $M_4$ and $F_7$ is a 4-form on $M_7$ which satisfies (resulting from the field equations)
\begin{eqnarray}
dF_7&=&0\nonumber\\
\delta F_7&=&-c*F_7
\end{eqnarray}
with $c$ is a constant, then these conditions say that $F_7$ is a special CCKY form. This choice gives that $M_4$ and $M_7$ are Einstein manifolds with $M_4$ corresponds to $AdS_4$ space-time, and supergravity Killing spinor equation reduces to Killing spinor equations on $M_4$ and $M_7$. This means that $M_7$ is an Einstein 7-manifold with a Killing spinor which is a weak $G_2$-holonomy manifold \cite{Behrndt Jeschek, Bilal Derendinger Sfetsos, Friedrich Kath Moroianu Semmelmann}. So, we have
\[
M_{11}=\underbrace{M_4}_{AdS_4}\times
\underbrace{M_7}_{\text{weak }G_2\text{-holonomy}}
\]
For $AdS_4\times S^7$ solution, we need to choose $F_7=0$ \cite{Duff Nilsson Pope}.

Similar procedures can be applied to other dimensions and supergravity theories.

\subsection{Supergravity Killing forms}

We can analyze the equations satisfied by the spinor bilinears of supergravity Killing spinors. For the six dimensional case, we have the supergravity Killing spinor equation
\[
\nabla_X\epsilon=-\frac{1}{4}i_XH.\epsilon.
\]
The covariant derivative of the spinor bilinears $(\epsilon\overline{\epsilon})_p$ can be calculated as
\begin{eqnarray}
\nabla_X(\epsilon\overline{\epsilon})_p&=&\left((\nabla_X\epsilon)\overline{\epsilon}\right)_p+\left(\epsilon(\overline{\nabla_X\epsilon})\right)_p\nonumber\\
&=&-\frac{1}{4}\left(i_XH.\epsilon\overline{\epsilon}\right)_p-\frac{1}{4}\left(\epsilon(\overline{i_XH.\epsilon})\right)_p\nonumber\\
&=&-\frac{1}{4}\left(i_XH.\epsilon\overline{\epsilon}\right)_p+\frac{1}{4}\left(\epsilon\overline{\epsilon}.i_XH\right)_p\nonumber
\end{eqnarray}
where we have used $\overline{i_XH.\epsilon}=\overline{\epsilon}(i_XH)^{\cal{J}}=-\overline{\epsilon}.i_XH$ for ${\cal{J}}=\xi$ or $\xi\eta$ since $i_XH$ is a 2-form. So, we have
\begin{eqnarray}
\nabla_X(\epsilon\overline{\epsilon})_p&=&-\frac{1}{4}\left([i_XH,\epsilon\overline{\epsilon}]_{Cl}\right)_p\nonumber\\
&=&-\frac{1}{4}[i_XH,(\epsilon\overline{\epsilon})_p]_{Cl}
\end{eqnarray}
since the Clifford bracket of any form $\alpha$ with a 2-form $i_XH$ can be written as $[i_XH,\alpha]_{Cl}=-2i_{X_a}i_XH\wedge i_{X^a}\alpha$. The equation satisfied by spinor bilinears reduces to the Killing equation for $p=1$;
\begin{eqnarray}
\nabla_{X_a}(\epsilon\overline{\epsilon})_1&=&-\frac{1}{4}[i_{X_a}H,(\epsilon\overline{\epsilon})_1]_{Cl}\nonumber\\
&=&\frac{1}{2}i_{X_b}i_{X_a}H\wedge i_{X^b}(\epsilon\overline{\epsilon})_1.
\end{eqnarray}
By taking wedge product with $e^a$, we have
\begin{eqnarray}
d(\epsilon\overline{\epsilon})_1&=&e^a\wedge\nabla_{X_a}(\epsilon\overline{\epsilon})_1\nonumber\\
&=&\frac{1}{2}e^a\wedge i_{X_b}i_{X_a}H\wedge i_{X^b}(\epsilon\overline{\epsilon})_1\nonumber\\
&=&-i_{X_b}H\wedge i_{X^b}(\epsilon\overline{\epsilon})_1.
\end{eqnarray}
Comparing the last two equalities, we obtain
\begin{equation}
\nabla_{X_a}(\epsilon\overline{\epsilon})_1=\frac{1}{2}i_{X_a}d(\epsilon\overline{\epsilon})_1
\end{equation}
which is the Killing equation. This means that the Dirac currents of supergravity Killing spinors are Killing vector fields. However, for higher-degree Dirac currents, we have the equation
\begin{equation}
\nabla_X\omega=-\frac{1}{4}[i_XH,\omega]_{Cl}
\end{equation}
which is different from the KY equation. For example,
\begin{equation}
\nabla_X(\epsilon\overline{\epsilon})_3=\frac{1}{2}i_Xd(\epsilon\overline{\epsilon})_3+\frac{1}{4}[H,i_X(\epsilon\overline{\epsilon})_3]_{Cl}.
\end{equation}
So, we obtain another type of generalization of Killing vector fields to higher-degree forms different from KY forms. These forms will be called as supergravity Killing forms. Supergravity Killing forms have a Lie algebra structure under the Clifford bracket. Namely, if $\omega_1$ and $\omega_2$ are supergravity Killing forms, then $[\omega_1,\omega_2]_{Cl}$ is also satisfies the supergravity Killing form equation \cite{Acik Ertem2}. This can be seen as follows
\begin{eqnarray}
\nabla_X[\omega_1,\omega_2]_{Cl}&=&[\nabla_X\omega_1,\omega_2]_{Cl}+[\omega_1,\nabla_X\omega_2]_{Cl}\nonumber\\
&=&-\frac{1}{4}[[i_XH,\omega_1]_{Cl},\omega_2]_{Cl}-\frac{1}{4}[\omega_1,[i_XH,\omega_2]_{Cl}]_{Cl}\nonumber\\
&=&-\frac{1}{4}[i_XH.\omega_1,\omega_2]_{Cl}+\frac{1}{4}[\omega_1.i_XH,\omega_2]_{Cl}\nonumber\\
&&-\frac{1}{4}[\omega_1,i_XH.\omega_2]_{Cl}+\frac{1}{4}[\omega_1,\omega_2.i_XH]_{Cl}\nonumber\\
&=&-\frac{1}{4}\left(i_XH.\omega_1.\omega_2+\omega_2.\omega_1.i_XH-i_Xh.\omega_2.\omega_1-\omega_1.\omega_2.i_XH\right)\nonumber\\
&=&-\frac{1}{4}[i_XH,[\omega_1,\omega_2]_{Cl}]_{Cl}.
\end{eqnarray}
Since the Clifford bracket is antisymmetric and satisfies the Jacobi identity, supergravity Killing forms have a Lie algebra structure. The supergravity Killing forms of ten dimensional type I and heterotic theories are defined from the same supergravity Killing spinor equation and they also have the similar Lie algebra structure.

In ten-dimensional type IIA supergravity, the supergravity Killing spinor equation is
\begin{equation}
\nabla_X\epsilon=-\frac{1}{4}i_XH.\epsilon-\frac{1}{8}e^{\phi}i_XF_2.\epsilon+\frac{1}{8}e^{\phi}(\widetilde{X}\wedge F_4).\epsilon
\end{equation}
where $\phi$ is the dilaton scalar field and $F_2$ and $F_4$ are 2-form and 4-form gauge fields. The supergravity Killing forms satisfy the equation
\begin{equation}
\nabla_X\omega=-\frac{1}{8}\left([2i_XH+e^{\phi}i_XF_2-e^{\phi}\widetilde{X}\wedge F_4,\omega]_{Cl}\right)
\end{equation}
where $\omega$ is an inhomogeneous form. Solutions of this equation also have a Lie algebra structure under the Clifford bracket.

In eleven-dimensional supergravity, the supergravity Killing spinor equation is
\[
\nabla_X\epsilon=-\frac{1}{24}(\widetilde{X}.F-3F.\widetilde{X}).\epsilon.
\]
The equation satisfied by the spinor bilinears can be found as
\begin{eqnarray}
\nabla_X(\epsilon\overline{\epsilon})_p&=&\left((\nabla_X\epsilon)\overline{\epsilon}\right)_p+\left(\epsilon(\overline{\nabla_X\epsilon})\right)_p\nonumber\\
&=&-\frac{1}{24}\left((\widetilde{X}.F-3F.\widetilde{X}).\epsilon\overline{\epsilon}\right)_p-\frac{1}{24}\left(\epsilon\overline{(\widetilde{X}.F-3F.\widetilde{X}).\epsilon}\right)_p\nonumber\\
&=&-\frac{1}{24}\left((\widetilde{X}.F-3F.\widetilde{X}).\epsilon\overline{\epsilon}\right)_p+\frac{1}{24}\left(\epsilon\overline{\epsilon}.(F.\widetilde{X}-3\widetilde{X}.F)\right)_p\nonumber
\end{eqnarray}
where we have used $\overline{(\widetilde{X}.F-3F.\widetilde{X}).\epsilon}=\overline{\epsilon}.(\widetilde{X}.F-3F.\widetilde{X})^{\cal{J}}$ and by choosing ${\cal{J}}=\xi\eta$ and using $(\widetilde{X}.F)^{\xi\eta}=F^{\xi\eta}.\widetilde{X}^{\xi\eta}=-F.\widetilde{X}$ and $(F.\widetilde{X})^{\xi\eta}=\widetilde{X}^{\xi\eta}.F^{\xi\eta}=-\widetilde{X}.F$. So, we have
\[
\nabla_X(\epsilon\overline{\epsilon})_p=-\frac{1}{24}\left\{\left((\widetilde{X}.F-3F.\widetilde{X}).\epsilon\overline{\epsilon}\right)_p-\left(\epsilon\overline{\epsilon}.(F.\widetilde{X}-3\widetilde{X}.F)\right)_p\right\}
\]
by adding and subtracting $\frac{1}{24}\epsilon\overline{\epsilon}.(4F.\widetilde{X}-4\widetilde{X}.F)-\frac{1}{24}\epsilon\overline{\epsilon}.(4F.\widetilde{X}-4\widetilde{X}.F)$, we obtain
\begin{equation}
\nabla_X(\epsilon\overline{\epsilon})_p=-\frac{1}{24}\left([(\widetilde{X}.F-3F.\widetilde{X}),\epsilon\overline{\epsilon}]_{Cl}\right)_p-\frac{1}{6}\left(\epsilon\overline{\epsilon}.[F,\widetilde{X}]_{Cl}\right)_p.
\end{equation}
For $p=1$, this reduces to the Killing equation and the Dirac currents of supergravity Killing spinors correspond to Killing vector fields. However, for higher-degree forms it is again different from the KY equation and we find another generalization of Killing vector fields to higher-degree forms. So, the supergravity Killing form equation in eleven dimensions is
\begin{equation}
\nabla_X\omega=-\frac{1}{24}[(\widetilde{X}.F-3F.\widetilde{X}),\omega]_{Cl}-\frac{1}{6}\omega.[F,\widetilde{X}]_{Cl}.
\end{equation}
But, in that case, the solutions does not have a Lie algebra structure automatically. We have
\begin{eqnarray}
\nabla_X[\omega_1,\omega_2]_{Cl}&=&-\frac{1}{24}[\Phi_X,[\omega_1,\omega_2]_{Cl}]_{Cl}-\frac{1}{6}[\omega_1,\omega_2]_{Cl}.\Psi_X\nonumber\\
&&-\frac{1}{6}(\omega_1.\Psi_X.\omega_2-\omega_2.\Psi_X.\omega_1)
\end{eqnarray}
where $\Phi_X=\widetilde{X}.F-3F.\widetilde{X}$ and $\Psi_X=[F,\widetilde{X}]_{Cl}=-2i_XF$. If the condition
\begin{equation}
\omega_1.\Psi_X.\omega_2=\omega_2.\Psi_X.\omega_1
\end{equation}
is satisfied, then the Clifford bracket of the solutions is again a solution \cite{Acik Ertem2}.

\section{Spin Raising and Lowering Operators}

In this section, we construct spin raising and lowering operators for massles field equations of particles with different spins by using twistor spinors. A spin raising operator takes a solution of a massless spin-$s$ field equation and gives a solution of a massless spin-$\left(s+\frac{1}{2}\right)$ field equation. Spin lowering operators transform the solutions of massless spin-$\left(s+\frac{1}{2}\right)$ fields to massless spin-$s$ fields. Besides the conformal Laplace, massless Dirac and source-free Maxwell equations, we also construct spin changing operators for massless Rarita-Schwinger fields by using twistor spinors with some constraints. 

\subsection{Spin raising and lowering}

We consider massless and source-free field equations for different spins
\begin{eqnarray}
\text{spin}-0\quad\quad&\longrightarrow&\quad\quad\Delta f-\frac{n-2}{4(n-1)}{\cal{R}}f=0\\
\text{spin}-\frac{1}{2}\quad\quad&\longrightarrow&\quad\quad\displaystyle{\not}D\psi=0\\
\text{spin}-1\quad\quad&\longrightarrow&\quad\quad\displaystyle{\not}dF=0
\end{eqnarray}
where spin-0 field equation (198) is the conformal Laplace equation, spin-$\frac{1}{2}$ field equation (199) is the massless Dirac equation and the spin-1 field equation (200) is the source-free Maxwell equation. By using twistor spinors satsifying (10), we can construct spin raising and spin lowering operators that transform solutions of spin-$s$ field equations to solutions of spin-$\left(s+\frac{1}{2}\right)$ field equations and vice versa \cite{Penrose Rindler}.

i) Spin raising from spin-0 to spin-$\frac{1}{2}$;\\
For a function $f$ satisfying conformal Laplace equation (198) and a twistor spinor $u$ satisfying (10), we can construct a spinor
\[
\psi=df.u+\frac{n-2}{n}f\displaystyle{\not}Du.
\]
This is equivalent to the operator written in (137) and we have shown in Section 8.1 that $\psi$ is a harmonic spinor, namely it satisfies the massless Dirac equation; $\displaystyle{\not}D\psi=0$. So, we transform spin-0 solution $f$ to spin-$\frac{1}{2}$ solution $\psi$ via a twistor spinor $u$.

ii) Spin lowering from spin-$\frac{1}{2}$ to spin-0;\\
In a reverse procedure, we can construct a function from a massless Dirac spinor $\psi$ and a twistor spinor $u$ by using the spinor inner product $(\,,\,)$ as
\begin{equation}
f=(u,\psi).
\end{equation}
We will show that $f$ is a solution of the conformal Laplace equation. By applying the Laplace-Beltrami operator $\Delta=\nabla^2=\nabla_{X_a}\nabla_{X^a}-\nabla_{\nabla_{X_a}X^a}$ to (201), we obtain
\begin{eqnarray}
\Delta f&=&\nabla_{X_a}\nabla_{X^a}(u,\psi)-\nabla_{\nabla_{X_a}X^a}(u,\psi)\nonumber\\
&=&\nabla_{X_a}\big((\nabla_{X^a}u,\psi)+(u,\nabla_{X^a}\psi)\big)-(\nabla_{\nabla_{X_a}X^a}u,\psi)-(u,\nabla_{\nabla_{X_a}X^a}\psi)\nonumber\\
&=&(\nabla_{X_a}\nabla_{X^a}u,\psi)+2(\nabla_{X^a}u,\nabla_{X_a}\psi)+(u,\nabla_{X_a}\nabla_{X^a}\psi)\nonumber\\
&&-(\nabla_{\nabla_{X_a}X^a}u,\psi)-(u,\nabla_{\nabla_{X_a}X^a}\psi)\nonumber\\
&=&(\nabla_{X_a}\nabla_{X^a}u-\nabla_{\nabla_{X_a}X^a}u,\psi)+2(\nabla_{X^a}u,\nabla_{X_a}\psi)\nonumber\\
&&+(u,\nabla_{X_a}\nabla_{X^a}\psi-\nabla_{\nabla_{X_a}X^a}\psi)\nonumber\\
&=&(\nabla^2u,\psi)+2(\nabla_{X^a}u,\nabla_{X_a}\psi)+(u,\nabla^2\psi).\nonumber
\end{eqnarray}
The square of the Dirac operator is written from (15) as
\[
\displaystyle{\not}D^2\psi=\nabla^2\psi-\frac{1}{4}{\cal{R}}\psi
\]
for any spinor $\psi$. So, we have
\begin{eqnarray}
\Delta (u,\psi)&=&(\displaystyle{\not}D^2u+\frac{1}{4}{\cal{R}}u,\psi)+2(\nabla_{X_a}u,\nabla_{X^a}\psi)+(u,\displaystyle{\not}D^2\psi+\frac{1}{4}{\cal{R}}\psi)\nonumber\\
&=&(\displaystyle{\not}D^2u,\psi)+2(\nabla_{X_a}u,\nabla_{X^a}\psi)+(u,\displaystyle{\not}D^2\psi)+\frac{1}{2}{\cal{R}}(u,\psi).\nonumber
\end{eqnarray}
From the twistor equation (10) and the integrability condition of twistor spinors in (19), we can write
\begin{equation}
\Delta (u,\psi)=\frac{n-2}{4(n-1)}{\cal{R}}(u,\psi)+\frac{2}{n}(e_a.\displaystyle{\not}Du,\nabla_{X^a}\psi)+(u,\displaystyle{\not}D^2\psi).
\end{equation}
From the identity $(e_a.\displaystyle{\not}Du,\nabla_{X^a}\psi)=(\displaystyle{\not}Du,(e_a)^{\cal{J}}.\nabla_{X^a}\psi)=\pm(\displaystyle{\not}Du,\displaystyle{\not}D\psi)$ and since $\psi$ is a massless Dirac spinor $\displaystyle{\not}D\psi=0$, we have
\begin{equation}
\Delta(u,\psi)-\frac{n-2}{4(n-1)}{\cal{R}}(u,\psi)=0
\end{equation}
which means that $f=(u,\psi)$ is a solution of the conformal Laplace equation. So, we transform the spin-$\frac{1}{2}$ soluıtion $\psi$ to spin-0 solution $f$ via a twistor spinor $u$. We can also construct symmetry operators of massless Dirac spinors by first applying spin lowering procedure from $\frac{1}{2}\rightarrow 0$ and then applying spin raising procedure from $0\rightarrow\frac{1}{2}$ \cite{Charlton, Benn Charlton Kress}. These operators are equivalent to the symmetry operators of massless Dirac spinors constructed in Section 6.

iii) Spin raising from spin-$\frac{1}{2}$ to spin-1;\\
In even dimensions $n=2p$, we can construct a spin-1 quantity from a twistor spinor $u$ and a massless Dirac spinor $\psi$ as
\begin{equation}
F=e^b.u\otimes\overline{\nabla_{X_b}\psi}+\frac{n-2}{n}\displaystyle{\not}Du\otimes\overline{\psi}+\psi\otimes\overline{\displaystyle{\not}Du}.
\end{equation}
Only the $p$-form component of $F$ will be important for us. By applying the Hodge-de Rham operator $\displaystyle{\not}d$ to (204)
\begin{eqnarray}
\displaystyle{\not}dF&=&e^a.\nabla_{X_a}F\nonumber\\
&=&e^a.\bigg(e^b.\nabla_{X_a}u\otimes\overline{\nabla_{X_b}\psi}+e^b.u\otimes\overline{\nabla_{X_a}\nabla_{X_b}\psi}+\frac{n-2}{n}\nabla_{X_a}\displaystyle{\not}Du\otimes\overline{\psi}\nonumber\\
&&+\frac{n-2}{n}\displaystyle{\not}Du\otimes\overline{\nabla_{X_a}\psi}+\nabla_{X_a}\psi\otimes\overline{\displaystyle{\not}Du}+\psi\otimes\overline{\nabla_{X_a}\displaystyle{\not}Du}\bigg)\nonumber\\
&=&e^a.e^b.\nabla_{X_a}u\otimes\overline{\nabla_{X_b}\psi}+e^a.e^b.u\otimes\overline{\nabla_{X_a}\nabla_{X_b}\psi}+\frac{n-2}{n}\displaystyle{\not}D^2u\otimes\overline{\psi}\nonumber\\
&&+\frac{n-2}{n}e^a.\displaystyle{\not}Du\otimes\overline{\nabla_{X_a}\psi}+\displaystyle{\not}D\psi\otimes\overline{\displaystyle{\not}Du}+e^a.\psi\otimes\overline{\nabla_{X_a}\displaystyle{\not}Du}
\end{eqnarray}
where we have used the normal coordinates and the definiton of the Dirac operator $\displaystyle{\not}D=e^a.\nabla_{X_a}$. From the Clifford algebra identity $e^a.e^b+e^b.e^a=2g^{ab}$, we can write
\begin{eqnarray}
e^a.e^b.\nabla_{X_a}u\otimes\overline{\nabla_{X_b}\psi}&=&(2g^{ab}-e^b.e^a).\nabla_{X_a}u\otimes\overline{\nabla_{X_b}\psi}\nonumber\\
&=&2\nabla_{X_a}u\otimes\overline{\nabla_{X^a}\psi}-e^a.\displaystyle{\not}Du\otimes\overline{\nabla_{X_a}\psi}\nonumber\\
&=&-\frac{n-2}{n}e^a.\displaystyle{\not}Du\otimes\overline{\nabla_{X_a}\psi}.
\end{eqnarray}
Because of $u$ is a twistor spinor satisfying (10) and $\psi$ is a massless Dirac spinor satisfying (199), we have
\begin{eqnarray}
\displaystyle{\not}dF&=&e^a.e^b.u\otimes\overline{\nabla_{X_a}\nabla_{X_b}\psi}+\frac{n-2}{n}\displaystyle{\not}D^2\otimes\overline{\psi}+e^a.\psi\otimes\overline{\nabla_{X_a}\displaystyle{\not}Du}\nonumber\\
&=&(e^a\wedge e^b).u\otimes\overline{\nabla_{X_a}\nabla_{X_b}\psi}+u\otimes\overline{\nabla_{X_a}\nabla_{X^a}\psi}\nonumber\\
&&+\frac{n-2}{n}\displaystyle{\not}D^2u\otimes\overline{\psi}+e^a.\psi\otimes\overline{\nabla_{X_a}\displaystyle{\not}Du}
\end{eqnarray}
where we have used $e^a.e^b=e^a\wedge e^b+\delta^a_b$. By antisymmetrizing the first term and from the definition of the curvature operator and the Laplacian, we can write
\begin{eqnarray}
\displaystyle{\not}dF&=&\frac{1}{2}(e^a\wedge e^b).u\otimes\overline{R(X_a,X_b)\psi}+u\otimes\overline{\nabla^2\psi}\nonumber\\
&&+\frac{n-2}{n}\displaystyle{\not}D^2u\otimes\overline{\psi}+e^a.\psi\otimes\overline{\nabla_{X_a}\displaystyle{\not}Du}.
\end{eqnarray}
From the identities $R(X_a,X_b)\psi=\frac{1}{2}R_{ab}.\psi$ and $\displaystyle{\not}D^2\psi=\nabla^2\psi-\frac{1}{4}{\cal{R}}\psi$, we obtain
\begin{eqnarray}
\displaystyle{\not}dF&=&\frac{1}{4}(e^a\wedge e^b).u\otimes\overline{R_{ab}.\psi}+\frac{1}{4}{\cal{R}}u\otimes\overline{\psi}\nonumber\\
&&+\frac{n-2}{n}\displaystyle{\not}D^2u\otimes\overline{\psi}+e^a.\psi\otimes\overline{\nabla_{X_a}\displaystyle{\not}Du}
\end{eqnarray}
By using the pairwise symmetry of the curvature tensor $R_{abcd}=R_{cdab}$, we can write
\begin{equation}
(e^a\wedge e^b).u\otimes\overline{R_{ab}.\psi}=R_{ab}.u\otimes\overline{(e^a\wedge e^b).\psi}.
\end{equation}
The integrability conditions of twistor spinors which are given in (16) and (19) results that 
\begin{eqnarray}
\displaystyle{\not}dF&=&\frac{1}{2n}\left(e_b.\nabla_{X_a}\displaystyle{\not}Du-e_a.\nabla_{X_b}\displaystyle{\not}Du\right)\otimes\overline{(e^a\wedge e^b).\psi}+\frac{1}{4(n-1)}{\cal{R}}u\otimes\overline{\psi}\nonumber\\
&&+e^a.\psi\otimes\overline{\nabla_{X_a}\displaystyle{\not}Du}\nonumber\\
&=&\frac{1}{2n}e_b.\nabla_{X_a}\displaystyle{\not}Du\otimes\overline{(e^a.e^b-e^b.e^a).\psi}+\frac{1}{4(n-1)}{\cal{R}}u\otimes\overline{\psi}+e^a.\psi\otimes\overline{\nabla_{X_a}\displaystyle{\not}Du}\nonumber\\
&=&-\frac{1}{n}e_b.\nabla_{X_a}\displaystyle{\not}Du\otimes\overline{e^b.e^a.\psi}+\frac{1}{n}\displaystyle{\not}D^2u\otimes\overline{\psi}+\frac{1}{4(n-1)}{\cal{R}}u\otimes\overline{\psi}\nonumber\\
&&+e^a.\psi\otimes\overline{\nabla_{X_a}\displaystyle{\not}Du}\nonumber\\
&=&-\frac{1}{n}e_b.\nabla_{X_a}\displaystyle{\not}Du\otimes\overline{e^b.e^a.\psi}+e^a.\psi\otimes\overline{\nabla_{X_a}\displaystyle{\not}Du}
\end{eqnarray}
where we have used $e^a\wedge e^b=e^a.e^b-\delta^{ab}$ and $e^ a.e^b=-e^b.e^a+2g^{ab}$. In $n=2p$ dimensions, the identity $(\psi\otimes\overline{\phi})^{\xi}=(-1)^{\lfloor r/2\rfloor}\phi\otimes\overline{\psi}$ corresponds to $\psi\otimes\overline{\phi}=\phi\otimes\overline{\psi}$ for the $p$-form component of the bilinear. Then, we can write
\begin{eqnarray}
e_b.\nabla_{X_a}\displaystyle{\not}Du\otimes\overline{e^b.e^a.\psi}&=&e_b.\left(\nabla_{X_a}\displaystyle{\not}Du\otimes\overline{e^b.e^a.\psi}\right)\nonumber\\
&=&e_b.\left(e^b.e^a.\psi\otimes\overline{\nabla_{X_a}\displaystyle{\not}Du}\right)\nonumber\\
&=&e_b.e^b.e^a.\psi\otimes\overline{\nabla_{X_a}\displaystyle{\not}Du}\nonumber\\
&=&ne^a.\psi\otimes\overline{\nabla_{X_a}\displaystyle{\not}Du}
\end{eqnarray}
and this gives the result $\displaystyle{\not}dF=0$. So, we obtain a middle-form source-free Maxwell solution $F$ from a massless Dirac spinor $\psi$ via a twistor spinor $u$.

iv) Spin lowering from spin-1 to spin-$\frac{1}{2}$;\\
In a reverse procedure, we can find a massless Dirac spinor from a middle-form source-free Maxwell solution $F$ by using a twistor spinor $u$. Let us define
\begin{equation}
\psi=F.u
\end{equation}
By applying the Dirac operator to (213), we find
\begin{eqnarray}
\displaystyle{\not}D\psi&=&e^a.\nabla_{X_a}(F.u)\nonumber\\
&=&e^a.\nabla_{X_a}F.u+e^a.F.\nabla_{X_a}u\nonumber\\
&=&\displaystyle{\not}dF.u+\frac{1}{n}e^a.F.e_a.\displaystyle{\not}Du
\end{eqnarray}
where we have used the definition of Hodge-de Rham operator $\displaystyle{\not}d=e^a.\nabla_{X_a}$ and the twistor equation (10). Since $F$ is a source-free Maxwell field, we have $\displaystyle{\not}dF=0$ and for a $p$-form in $2p$-dimensions we have $e^a.F.e_a=(-1)^p(n-2p)F=0$. Then, we obtain
\[
\displaystyle{\not}D\psi=0.
\]
We can also construct symmetry operators for source-free Maxwell fields by applying spin lowering from $1\rightarrow\frac{1}{2}$ and spin raising from $\frac{1}{2}\rightarrow 1$ \cite{Charlton, Benn Charlton Kress}.

\subsection{Rarita-Schwinger fields}

We consider spinor-valued 1-forms representing spin-$\frac{3}{2}$ particles. For a spinor field $\psi_a$ and the co-frame basis $e^a$, we define the spinor-valued 1-form as
\begin{equation}
\Psi=\psi_a\otimes e^a.
\end{equation}
Action of a Clifford form $\alpha$ on $\Psi$ is defined by
\begin{equation}
\alpha.\Psi=\alpha.\psi_a\otimes e^a.
\end{equation}
Inner product of a spinor-valued 1-form $\Psi$ and a spinor $u$ is defined as
\begin{equation}
(u,\Psi)=(u,\psi_a)e^a.
\end{equation}
We can also define the Rarita-Schwinger operator acting on spinor-valued 1-forms
\begin{equation}
\mathbb{\displaystyle{\not}D}=e^a.\nabla_{X_a}.
\end{equation}
From these definitions, the massless Rarita-Schwinger equations of spin-$\frac{3}{2}$ fields in supergravity can be written for $\Psi=\psi_a\otimes e^a$ as follows
\begin{eqnarray}
\mathbb{\displaystyle{\not}D}\Psi&=&0\\
e^a.\psi_a&=&0.
\end{eqnarray}
The first one can be seen as the generalization of the Dirac equation to spin-$\frac{3}{2}$ fields, and second one is the tracelessness condition. These equations imply a Lorentz-type condition
\begin{equation}
\nabla_{X^a}\psi_a=0.
\end{equation}

i) Spin raising from spin-1 to spin-$\frac{3}{2}$;\\
We construct a massless Rarita-Schwinger field from a source-free Maxwell field and a twistor spinor. For even dimensions $n=2p$, we propose the following spinor-valued 1-form constructed out of a $p$-form Maxwell field $F$ and a twistor spinor $u$
\begin{eqnarray}
\Psi=\left(\nabla_{X_a}F.u-\frac{1}{n}F.e_a.\displaystyle{\not}Du\right)\otimes e^a
\end{eqnarray}
where we have $\psi_a=\nabla_{X_a}F.u-\frac{1}{n}F.e_a.\displaystyle{\not}Du$. To check the tracelessness condition, we calculate $e^a.\psi_a$
\begin{eqnarray}
e^a.\psi_a&=&e^a.\nabla_{X_a}F.u-\frac{1}{n}e^a.F.e_a.\displaystyle{\not}Du\nonumber\\
&=&\displaystyle{\not}dF.u-(-1)^p\frac{n-2p}{n}F.\displaystyle{\not}Du\nonumber\\
&=&0\nonumber
\end{eqnarray}
where we have used $\displaystyle{\not}dF=0$ and $e^a.F.e_a=(-1)^p(n-2p)F=0$ since $F$ is a $p$-form. Hence, (216) is satisfied. By applying the Rarita-Schwinger operator to $\Psi$, we obtain
\begin{eqnarray}
\mathbb{\displaystyle{\not}D}\Psi&=&e^b.\nabla_{X_b}\left[\left(\nabla_{X_a}F.u-\frac{1}{n}F.e_a.\displaystyle{\not}Du\right)\otimes e^a\right]\nonumber\\
&=&\left(\displaystyle{\not}d\nabla_{X_a}F.u+e^b.\nabla_{X_a}F.\nabla_{X_b}u-\frac{1}{n}\displaystyle{\not}dF.e_a.\displaystyle{\not}Du-\frac{1}{n}e^b.F.e_a.\nabla_{X_b}\displaystyle{\not}Du\right)\otimes e^a\nonumber\\
&=&\left(\displaystyle{\not}d\nabla_{X_a}F.u-\frac{1}{2}e^b.F.e_a.K_b.u\right)\otimes e^a
\end{eqnarray}
where we have used $\displaystyle{\not}dF=0$, the twistor equation (10) and the integrability condition in (21). The identity
\begin{equation}
[\displaystyle{\not}d,\nabla_{X_a}]F=-e^b.R(X_a,X_b)F=-\frac{1}{2}e^b.[R_{ab},F]_{Cl}
\end{equation}
gives
\begin{equation}
\mathbb{\displaystyle{\not}D}\Psi=\left(-\frac{1}{2}e^b.[R_{ab},F]_{Cl}.u-\frac{1}{2}e^b.F.e_a.K_b.u\right)\otimes e^a.
\end{equation}
We can write $R_{ab}=C_{ab}+e_b.K_a-e_a.K_b$ in terms of conformal 2-forms and Schouten 1-forms and by using the integrability condition of twistor spinors $C_{ab}.u=0$ with the identity $e^b.R_{ba}=P_a$, we obtain
\begin{equation}
\mathbb{\displaystyle{\not}D}\Psi=\left(\frac{1}{2}P_a.F.u-e^b.F.e_a.K_b.u\right)\otimes e^a.
\end{equation}
Then, to have a massless Rarita-Schwinger field, the condition
\begin{equation}
P_a.F.u=2e^b.F.e_a.K_b.u
\end{equation}
must be satisfied. From the definition $K_a=\frac{1}{n-2}\left(\frac{\cal{R}}{2(n-1)}e_a-P_a\right)$ and Clifford multiplying with $e^a$ from the left, this turns into
\begin{equation}
\left(\frac{n(n-1)-2}{(n-1)(n-2)}\right){\cal{R}}F.u=0
\end{equation}
where we have used $e^a.P_a={\cal{R}}$, $e^a.e_a=n$, $e^a.e^b=-e^b.e^a+2g^{ab}$ and $e^a.F.e_a=0$.\\
So, from a $p$-form source-free Maxwell field $F$ and a twistor spinor $u$ which satisfies the condition
\begin{equation}
F.u=0
\end{equation}
we can construct a massless Rarita-Schwinger field $\Psi$ as in (222) \cite{Acik Ertem3}.

ii) Spin lowering from spin-$\frac{3}{2}$ to spin-1;\\
In four dimensions, we can also construct a source-free Maxwell field from a massless Rarita-Schwinger field via a twistor spinor satisfying a constraint. Let us define a 1-form $A$ from a massless Rarita-Schwinger field $\Psi=\psi_a\otimes e^a$ and a twistor spinor $u$ as
\begin{equation}
A=(u,\Psi)=(u,\psi_a)e^a.
\end{equation}
We consider the 2-form $F=dA$ which can be written as
\begin{eqnarray}
F&=&d(u,\Psi)\nonumber\\
&=&\left[\frac{1}{n}(e_b.\displaystyle{\not}Du,\psi_a)+(u,\nabla_{X_b}\psi_a)\right]e^b\wedge e^a
\end{eqnarray}
from $d=e^a\wedge\nabla_{X_a}$ and the twistor equation (10). Since $F$ is an exact form, we have $dF=0$. So, the action of Hodge-de Rham operator gives
\begin{eqnarray}
\displaystyle{\not}dF&=&dF-\delta F\nonumber\\
&=&i_{X^a}\nabla_{X_a}F.
\end{eqnarray}
From a direct calculation by using the properties of the inner product and the Rarita-Schwinger equations, we can find
\begin{eqnarray}
\displaystyle{\not}dF&=&\bigg[\frac{1}{n}(\nabla_{X_b}\displaystyle{\not}Du,e^c.\psi_a)-\frac{1}{n}(\nabla_{X_b}\displaystyle{\not}Du,e_a.\psi^b)\nonumber\\
&&+(u,\nabla_{X_b}\nabla_{X^b}\psi_a)-(u,\nabla_{X_b}\nabla_{X_a}\psi^b)\bigg]e^a.
\end{eqnarray}
The integrability condition of twistor spinors in (21), the square of the Dirac operator in (15) and the identities $e^a.e_a=n$, $e^a.P_a={\cal{R}}$, $e^a.K_a=-\frac{\cal{R}}{2(n-1)}$ gives
\begin{equation}
\displaystyle{\not}dF=\left[(u,\frac{n-3}{4(n-1)}{\cal{R}}\psi_a-\frac{1}{2}K_b.e_a.\psi^b-\nabla_{X_b}\nabla_{X_a}\psi^b)\right]e^a.
\end{equation}
This means that, to obtain a source-free Maxwell field, $\psi_a$ of the Rarita-Schwinger field has to satisfy
\begin{equation}
\nabla_{X_b}\nabla_{X_a}\psi^b=-\frac{1}{2}K_b.e_a.\psi^b+\frac{n-3}{4(n-1)}{\cal{R}}\psi_a
\end{equation}
which can be simplified by using $\nabla_{X_b}\nabla_{X_a}\psi^b=\nabla_{X_a}\nabla_{X_b}\psi^b+R(X_b,X_a)\psi^b$ and $R(X_b,X_a)\psi^b=\frac{1}{2}R_{ba}.\psi^b$ with $\nabla_{X_a}\psi^a=0$ as \cite{Acik Ertem3}
\begin{equation}
(R_{ba}+K_b.e_a).\psi^b=\frac{n-3}{2(n-1)}{\cal{R}}\psi_a.
\end{equation}
which is automatically satisfied in a flat background. The symmetry operators can also be constructed by applying spin lowering and spin raising procedures.

\section{Topological Insulators}

Spin geometry methods are also used in the classification problem of topological phases in condensed matter physics. In this section, we give basic definitions of Bloch bundles and topological phases and review the Chern and $Z_2$ insulators with the examples of Haldane and Kane-Mele models. We summarize the derivation of the periodic table of topological insulators and superconductors via Clifford chessboard and index of Dirac operators.

Let us consider a $d$-dimensional periodic crystal lattice. Periodicity of the lattice results that the unit cell determines all the properties of the whole lattice. Fourier transform of the unit cell that is the momentum space unit cell corresponds to the Brillouin zone (BZ) which is a $d$-dimensional torus $T^d$ (from the periodicity of the lattice).

In tight-binding approximation, energy levels of last orbital electrons determine the energy band structure of the crystal. If there is a gap in energy levels (between valence and conduction bands), it corresponds to an insulator. If there is no gap in energy levels, it corresponds to a metal.

The system is determined by the Bloch Hamiltonian $H(k)$ with eigenstates $|u_n(k)\rangle$ and eigenvalues $E_n(k)$ ($n$ corresponds to the number of bands);
\begin{equation}
H(k)|u_n(k)\rangle=E_n(k)|u_n(k)\rangle
\end{equation}
$H(k)$ is a Hermitian operator on the Hilbert space ${\cal{H}}=\mathbb{C}^{2n}$ and $|u_n(k)\rangle\in\mathbb{C}^{2n}$. For each $k\in T^d$, we have $|u_n(k)\rangle\in\mathbb{C}^{2n}$. So, we have the Bloch bundle locally $T^d\times\mathbb{C}^{2n}$ with fiber $\mathbb{C}^{2n}$ and base space $T^d$. In an insulator, we consider valence Bloch bundle which is locally $T^d\times\mathbb{C}^n$. All vector bundles are locally trivial; namely can be written as a cross product of base and fiber. However, they can be non-trivial globally. Triviality or non-triviality of the valence bundle of an insulator determine the topological character of it. For example; a cylinder $S=S^1\times\mathbb{R}$ is a trivial line bundle over $S^1$. However, a M\"{o}bius strip $M$ is a non-trivial line bundle over $S^1$. It has a twist on the bundle structure. One can define a section without zero in trivial bundles. But, this is not possible in non-trivial bundles. Non-triviality of a bundle is characterized by the characteristic classes (Chern classes, Pontryagin classes, etc.) which are topological invariants.

A topological insulator is an insulator with a non-trivial valence bundle. A topological insulator Hamiltonian in one topological class cannot be deformed continuously to a Hamiltonian in another topological class (deformation means changing Hamiltonian parameters without closing the gap). To convert one topological class Hamiltonian to another one,  there must be a gapless state between two classes. Hence, the insulating phase must disappear. For example, a topological insulator which has an interface with vacuum (which is a trivial insulator) has a gapless boundary although insulating in the bulk. These gapless boundary degrees of freedom are robust to perturbations.

There are two types of topological insulators; Chern insulators which are determined by $\mathbb{Z}$ topological invariants and $\mathbb{Z}_2$ insulators which are determined by $\mathbb{Z}_2$ topological invariants.

\subsection{Chern insulators}

For the Bloch Hamiltonian $H(k)$ with eigenstates $|u_n(k)\rangle$ and eigenvalues $E_n(k)$;
\[
H(k)|u_n(k)\rangle=E_n(k)|u_n(k)\rangle),
\]
the Berry connection is defined by
\begin{equation}
A=-i\langle u(k)|d|u(k)\rangle
\end{equation}
where $d$ is the exterior derivative, so $A$ is a 1-form. The Berry curvature is
\begin{equation}
F=dA.
\end{equation}
The topological invariant that characterizes the 2-dimensional Chern insulators is the first Chern number
\begin{equation}
C_1=\frac{1}{2\pi}\int_{BZ}F.
\end{equation}
For higher even dimensions, the higher Chern numbers which take integer values determine the topological phases. For example, a Bloch Hamiltonian in the form
\begin{equation}
H(k)={\bf{d}}(k).\bf{\sigma}
\end{equation}
with $\sigma_i$ are Pauli matrices, the first Chern number corresponds to
\begin{equation}
C_1=\frac{1}{4\pi}\int_{BZ}d^2k\left(\partial_{k_x}\widehat{\bf{d}}(k)\times\partial_{k_y}\widehat{\bf{d}}(k)\right).\widehat{\bf{d}}(k)
\end{equation}
where $\widehat{\bf{d}}(k)=\frac{\bf{d}(k)}{|{\bf{d}}(k)|}$ with $|{\bf{d}}(k)|=\sqrt{d_1^2(k)+d_2^2(k)+d_3^2(k)}$.

\underline{Example}: Haldane model;\\
Let us consider the honeycomb lattice of graphene with a magnetic flux which is non-zero locally but zero in a unit cell. Namely, we can divide the unit cell into three parts which are called $a$, $b$ and $c$ regions and define themagnetic flux as follows. The flux $\phi_a$ in the region $a$ and the flux $\phi_b$ in the region $b$ are related as $\phi_a=-\phi_b$ and the flux $\phi_c=0$ in the region $c$. Hence, the total flux is zero \cite{Haldane}. Bloch Hamiltonian of this system is given by
\[
H(k)={\bf{d}}(k).{\bf{\sigma}}
\]
with $\sigma_i$ are Pauli matrices and
\begin{eqnarray}
d_0(k)&=&2t_2\cos{\phi}\sum_i\cos{(\bf{k}.\bf{b_i})}\nonumber\\
d_1(k)&=&t_1\sum_i\cos{(\bf{k}.\bf{a_i})}\nonumber\\
d_2(k)&=&t_1\sum_i\sin{(\bf{k}.\bf{a_i})}\\
d_3(k)&=&M-2t_2\sin{\phi}\sum_i\sin{(\bf{k}.\bf{b_i})}\nonumber
\end{eqnarray}
where ${\bf{a_i}}$ are Bravais lattice basis vectors and ${\bf{b}}_1={\bf{a}}_2-{\bf{a}}_3$, ${\bf{b}}_2={\bf{a}}_3-{\bf{a}}_1$, ${\bf{b}}_3={\bf{a}}_1-{\bf{a}}_2$. $t_1$ and $t_2$ are nearest neighbour and next nearest neighbour hopping parameters, respectively and $M$ is the on-site energy. $\phi=\frac{2\pi}{\phi_0}(2\phi_a+\phi_b)$ with $\phi_0=\frac{h}{c}$. Valence and conduction bands of graphene touch each other at two different points which are called $K$ and $K'$ points. By considering low energy limit of $H(k)$ (small $k$ limit) at $K$ and $K'$ points, we obtain
\begin{eqnarray}
H(K+k)&=&-3t_2\cos{\phi}+\frac{3}{2}at_1(-k_x\sigma_1-k_y\sigma_2)+(M+3\sqrt{3}t_2\sin{\phi})\sigma_3\nonumber\\
H(K'+k)&=&-3t_2\cos{\phi}+\frac{3}{2}at_1(k_x\sigma_1-k_y\sigma_2)+(M-3\sqrt{3}t_2\sin{\phi})\sigma_3\nonumber\\
\end{eqnarray}
This is a Dirac Hamiltonian and $\sigma_3$ is the mass term which determine the gap property of the system. Chern number of the system corresponds to
\begin{eqnarray}
C_1&=&\frac{1}{2}\big[\text{sgn}(d_3(k)\text{ at }K)-\text{sgn}(d_3(k)\text{ at }K')\big]\nonumber\\
&=&\frac{1}{2}\big[\text{sgn}(M+3\sqrt{3}t_2\sin{\phi})-\text{sgn}(M-3\sqrt{3}t_2\sin{\phi})\big].
\end{eqnarray}
We consider three cases;

i) $M=3\sqrt{3}t_2\sin{\phi}$ or $M=-3\sqrt{3}t_2\sin{\phi}$;\\
In this case $H(k)$ is gapless at $K'$ and gapped at $K$ or gapless at $K$ and gapped at $K'$. So, this is a gapless phase.

ii) $M>3\sqrt{3}t_2\sin{\phi}$;\\
In this case $C_1=0$, hence we have a trivial insulator.

iii) $M<3\sqrt{3}t_2\sin{\phi}$;\\
In this case $C_1=+1$ for $\phi>0$ and $C_1=-1$ for $\phi<0$, hence we have topological insulator phases.

So, by deforming Hamiltonian parameters, we obtain different topological phases and there is a gapless transition between them which is the case (i) \cite{Haldane, Fruchart Carpentier}. For $\phi=0$, hence no magnetic fluxes, Haldane model reduces to an ordinary insulator, $C_1=0$ in all cases.

\subsection{$\mathbb{Z}_2$ insulators}

Anti-unitary symmetries of the Hamiltonian can give different topological properties to the system in various dimensions. An example is the time-reversal symmetry and the time-reversal operator $T$ maps momentum $k$ to $-k$. If the Bloch Hamiltonian $H(k)$ of a system is invariant under $T$, namely
\begin{equation}
H=THT^{-1}
\end{equation}
then, Chern number cannot characterize the topological property of the system. Some points on BZ are invariant under $T$ and these are called Time Reversal Invariamt Momentum (TRIM) points.\\

For the eigenstates $|u_n(k)\rangle$, let us define the sewing matrix
\begin{equation}
w_{mn}(k)=\langle u_m(-k)|T|u_n(k)\rangle.
\end{equation}
By using the determinant $\det(w(k))$ and the Pfaffian $\text{Pf}(w(k))$ which is defined as $\text{Pf}^2=\det$, we can construct the $\mathbb{Z}_2$ invariant of the system as
\begin{equation}
\nu=\prod_{i=\text{TRIM}}\frac{\text{Pf}(w(k_i))}{\sqrt{\det(w(k_i))}}
\end{equation}
which takes $\pm 1$ values. If $\nu=1$, we have trivial insulator and if $\nu=-1$, we have topological insulator \cite{Kane Mele}.

\underline{Example}: Kane-Mele model;\\
By adding spin-orbit coupling term to the Haldane model, we obtain Kane-Mele model which is a $\mathbb{Z}_2$ insulator \cite{Kane Mele2}. Spin degrees of freedom splits the Hamiltonian to spin up and spin down parts, and the mass terms of the Bloch Hamiltonian at $K$ and $K'$ points are given by
\begin{eqnarray}
h_{\uparrow}(K)&=&(M+3\sqrt{3}\lambda_{SO})\sigma_3+...\nonumber\\
h_{\uparrow}(K')&=&(M-3\sqrt{3}\lambda_{SO})\sigma_3+...\nonumber\\
h_{\downarrow}(K)&=&(M-3\sqrt{3}\lambda_{SO})\sigma_3+...\nonumber\\
h_{\downarrow}(K')&=&(M+3\sqrt{3}\lambda_{SO})\sigma_3+...\nonumber
\end{eqnarray}
where $\lambda_{SO}$ is the spin-orbit coupling parameter. Although the total Chern number does not characterize the topology of the system, the difference of spin up and spin down Chern numbers corresponds to the $\mathbb{Z}_2$ invariant.

We have three cases;

i) $M>3\sqrt{3}\lambda_{SO}$;\\
Chern numbers are
\begin{eqnarray}
C_{1\uparrow}&=&\frac{1}{2}\big[\text{sgn}(M+3\sqrt{3}\lambda_{SO})-\text{sgn}(M-3\sqrt{3}\lambda_{SO})\big]=0\nonumber\\
C_{1\downarrow}&=&\frac{1}{2}\big[\text{sgn}(M-3\sqrt{3}\lambda_{SO})-\text{sgn}(M+3\sqrt{3}\lambda_{SO})\big]=0.
\end{eqnarray}
So, we have $C_{1\uparrow}+C_{1\downarrow}=0$ and $C_{1\uparrow}-C_{1\downarrow}=0$, and this corresponds to the trivial insulator.

ii) $M<3\sqrt{3}\lambda_{SO}$;\\
Chern numbers are $C_{1\uparrow}=1$ and $C_{1\downarrow}=-1$. So, we have $C_{1\uparrow}+C_{1\downarrow}=0$, but $C_{1\uparrow}-C_{1\downarrow}\neq 0$. This corresponds to a topological insulator.

iii) $M=3\sqrt{3}\lambda_{SO}$;\\
The Hamiltonian is gapless in this case. Hence, there is gapless phase between two different topological phases.

Spin Chern number (difference of $C_{1\uparrow}$ and $C_{1\downarrow}$) does not work in all systems, so $\mathbb{Z}_2$ invariant is more general than this.

\subsection{Classification}

Gapped free-fermion Hamiltonians can be classified with respect to anti-unitary symmetries they can admit. We consider two anti-unitary symmetries;\\
i) Time-reversal $TH(k)T^{-1}=H(-k)$\\
ii) Charge conjugation (particle-hole) $CH(k)C^{-1}=-H(-k)$\\
and their combination\\
iii) Chiral symmetry $SH(k)S^{-1}=-H(k)$.

The square of $T$ and $C$ can be $\pm1$ (for example for fermionic systems we have $T^2=-1$). So, we have three possibilities for $T$; $T=0$, $T^2=\pm1$ and three possibilities for $C$; $C=0$, $C^2=\pm1$ where 0 means the non-existence of the symmetry. Moreover, $S$ can be 0 or 1 when both $T=C=0$. Then, we have ten possibilities given in the table which corresponds to the Altland-Zirnbauer (AZ) classification of gapped free-fermion Hamiltonians \cite{Altland Zirnbauer}

\quad\\
{\centering{
\begin{tabular}{c c c}

$T\quad$ & $C$\quad & \quad $S$\\ \hline
$0\quad$ &  0 \quad & \quad 0 \\
$0\quad$ & 0 \quad & \quad 1 \\
$+1\quad$ &  0 \quad & \quad 0 \\
$+1\quad$ &  +1 \quad & \quad 1 \\
$0\quad$ &  +1 \quad & \quad 0 \\
$-1\quad$ &  +1 \quad & \quad 1 \\
$-1\quad$ &  0 \quad & \quad 0 \\
$-1\quad$ &  -1 \quad & \quad 1 \\
$0\quad$ &  -1 \quad & \quad 0 \\
$+1\quad$ &  -1 \quad & \quad 1 \\

\end{tabular}}
\quad\\
\quad\\
\quad\\}

Indeed, these Hamiltonians are elements of Cartan symmetric spaces and have topological properties in relevant dimensions. The general picture is given as a peridoic table in the following form \cite{Kitaev, Ryu Schnyder Furusaki Ludwig, Budich Trauzettel, Kaufmann et al}

\quad\\
{\centering{
\begin{tabular}{c| c c c|c c c c c c c c}

$\text{label}$ & $T$ & $C$ & $S$ & $0$ & $1$ & $2$ & $3$ & $4$ & $5$ & $6$ & $7$ \\ \hline
$\text{A}$ & $0$ & $0$ & $0$ & $\mathbb{Z}$ & $0$ & $\mathbb{Z}$ & $0$ & $\mathbb{Z}$ & $0$ & $\mathbb{Z}$ & $0$ \\
$\text{AIII}$ & $0$ & $0$ & $1$ & $0$ & $\mathbb{Z}$ & $0$ & $\mathbb{Z}$ & $0$ & $\mathbb{Z}$ & $0$ & $\mathbb{Z}$ \\ \hline
$\text{AI}$ & $+1$ & $0$ & $0$ & $\mathbb{Z}$ & $0$ & $0$ & $0$ & $\mathbb{Z}$ & $0$ & $\mathbb{Z}_2$ & $\mathbb{Z}_2$ \\
$\text{BDI}$ & $+1$ & $+1$ & $1$ & $\mathbb{Z}_2$ & $\mathbb{Z}$ & $0$ & $0$ & $0$ & $\mathbb{Z}$ & $0$ & $\mathbb{Z}_2$ \\
$\text{D}$ & $0$ & $+1$ & $0$ & $\mathbb{Z}_2$ & $\mathbb{Z}_2$ & $\mathbb{Z}$ & $0$ & $0$ & $0$ & $\mathbb{Z}$ & $0$ \\
$\text{DIII}$ & $-1$ & $+1$ & $1$ & $0$ & $\mathbb{Z}_2$ & $\mathbb{Z}_2$ & $\mathbb{Z}$ & $0$ & $0$ & $0$ & $\mathbb{Z}$ \\
$\text{AII}$ & $-1$ & $0$ & $0$ & $\mathbb{Z}$ & $0$ & $\mathbb{Z}_2$ & $\mathbb{Z}_2$ & $\mathbb{Z}$ & $0$ & $0$ & $0$ \\
$\text{CII}$ & $-1$ & $-1$ & $1$ & $0$ & $\mathbb{Z}$ & $0$ & $\mathbb{Z}_2$ & $\mathbb{Z}_2$ & $\mathbb{Z}$ & $0$ & $0$ \\
$\text{C}$ & $0$ & $-1$ & $0$ & $0$ & $0$ & $\mathbb{Z}$ & $0$ & $\mathbb{Z}_2$ & $\mathbb{Z}_2$ & $\mathbb{Z}$ & $0$ \\
$\text{CI}$ & $+1$ & $-1$ & $1$ & $0$ & $0$ & $0$ & $\mathbb{Z}$ & $0$ & $\mathbb{Z}_2$ & $\mathbb{Z}_2$ & $\mathbb{Z}$ \\ 

\end{tabular}}
\quad\\
\quad\\
\quad}

The table repeats itself after dimension seven. The first two rows correspond to complex classes and the last eight rows correspond to real classes. The class A of dimension 2 is the Haldane model, and the class AII of dimension 2 is the Kane-Mele model. This table originates from the Clifford algebra chessboard \cite{Ertem}.

If we consider the vector spaces $V=\mathbb{R}^{n,s}$ with $n$ negative and $s$ positive generators, the Clifford algebras defined on them correspond to the matrix algebras constructed out of the division algebras $\mathbb{R}$, $\mathbb{C}$, $\mathbb{H}$ \cite{Lawson Michelson}.

\quad\\
{\centering
\resizebox{\columnwidth}{!}{
\begin{tabular}{c c c c c c c c c}

$Cl_{n,s}$ & $n=0$ & $1$ & $2$ & $3$ & $4$ & $5$ & $6$ & $7$ \\ \hline
$s=0$ & $\mathbb{R}$ & $\mathbb{C}$ & $\mathbb{H}$ & $\mathbb{H}\oplus\mathbb{H}$ & $\mathbb{H}(2)$ & $\mathbb{C}(4)$ & $\mathbb{R}(8)$ & $\mathbb{R}(8)\oplus\mathbb{R}(8)$ \\
$1$ & $\mathbb{R}\oplus\mathbb{R}$ & $\mathbb{R}(2)$ & $\mathbb{C}(2)$ & $\mathbb{H}(2)$ & $\mathbb{H}(2)\oplus\mathbb{H}(2)$ & $\mathbb{H}(4)$ & $\mathbb{C}(8)$ & $\mathbb{R}(16)$ \\
$2$ & $\mathbb{R}(2)$ & $\mathbb{R}(2)\oplus\mathbb{R}(2)$ & $\mathbb{R}(4)$ & $\mathbb{C}(4)$ & $\mathbb{H}(4)$ & $\mathbb{H}(4)\oplus\mathbb{H}(4)$ & $\mathbb{H}(8)$ & $\mathbb{C}(16)$ \\
$3$ & $\mathbb{C}(2)$ & $\mathbb{R}(4)$ & $\mathbb{R}(4)\oplus\mathbb{R}(4)$ & $\mathbb{R}(8)$ & $\mathbb{C}(8)$ & $\mathbb{H}(8)$ & $\mathbb{H}(8)\oplus\mathbb{H}(8)$ & $\mathbb{H}(16)$ \\
$4$ & $\mathbb{H}(2)$ & $\mathbb{C}(4)$ & $\mathbb{R}(8)$ & $\mathbb{R}(8)\oplus\mathbb{R}(8)$ & $\mathbb{R}(16)$ & $\mathbb{C}(16)$ & $\mathbb{H}(16)$ & $\mathbb{H}(16)\oplus\mathbb{H}(16)$ \\
$5$ & $\mathbb{H}(2)\oplus\mathbb{H}(2)$ & $\mathbb{H}(4)$ & $\mathbb{C}(8)$ & $\mathbb{R}(16)$ & $\mathbb{R}(16)\oplus\mathbb{R}(16)$ & $\mathbb{R}(32)$ & $\mathbb{C}(32)$ & $\mathbb{H}(32)$ \\
$6$ & $\mathbb{H}(4)$ & $\mathbb{H}(4)\oplus\mathbb{H}(4)$ & $\mathbb{H}(8)$ & $\mathbb{C}(16)$ & $\mathbb{R}(32)$ & $\mathbb{R}(32)\oplus\mathbb{R}(32)$ & $\mathbb{R}(64)$ & $\mathbb{C}(64)$ \\
$7$ & $\mathbb{C}(8)$ & $\mathbb{H}(8)$ & $\mathbb{H}(8)\oplus\mathbb{H}(8)$ & $\mathbb{H}(16)$ & $\mathbb{C}(32)$ & $\mathbb{R}(64)$ & $\mathbb{R}(64)\oplus\mathbb{R}(64)$ & $\mathbb{R}(128)$ \\

\end{tabular}}
\quad\\
\quad\\
\quad\\}

where $\mathbb{K}(n)$ denotes $n\times n$ matrices with entries in $\mathbb{K}$. The table repeats itself after dimension seven. We define
\begin{equation}
Cl^*_n\equiv Cl_{0,n}
\end{equation}
and because of the isomorphism
\begin{equation}
Cl_{n,s}\cong Cl^*_{s-n(\text{mod }8)}
\end{equation}
the Clifford chessboard is also a table for $Cl^*_{s-n(\text{mod }8)}$.

We consider real vector bundles on a manifold $M$ with a Clifford algebra structure. Hence, we have Clifford bundles $Cl^*_k$ on $M$ which means that Clifford algebras with $k$ negative generators. We can define Dirac operators (Hodge-de Rham operators) $\displaystyle{\not}D=e^a.\nabla_{X_a}$ on these bundles. Index of a Dirac operator is defined by
\begin{equation}
\text{ind}\displaystyle{\not}D_k=\text{dim}(\text{ker}\displaystyle{\not}D_k)-\text{dim}(\text{coker}\displaystyle{\not}D_k).
\end{equation}
Atiyah-Singer index theorem relates the analytic index of real Dirac operators to the topological invariants of the bundle as \cite{Atiyah Singer, Atiyah Singer2, Atiyah}
\begin{eqnarray}
\text{ind}(\displaystyle{\not}D_k)&=&\left\{
                                                                               \begin{array}{ll}
                                                                                 \text{dim}_{\mathbb{C}}\textbf{H}_k (\text{mod }2), & \hbox{ for $k\equiv 1$  $(\text{mod }8)$} \\
                                                                                 \text{dim}_{\mathbb{H}}\textbf{H}_k (\text{mod }2), & \hbox{ for $k\equiv 2$  $(\text{mod }8)$} \\
                                                                                 \frac{1}{2}\widehat{A}(M), & \hbox{ for $k\equiv 4$  $(\text{mod }8)$} \\
                                                                                 \widehat{A}(M), & \hbox{ for $k\equiv 0$  $(\text{mod }8)$}
                                                                               \end{array}
                                                                             \right.
\end{eqnarray}
where $\textbf{H}_k$ is the space of harmonic spinors and $\widehat{A}(M)$ is the $\widehat{A}$-genus of $M$ which is written in terms of Pontryagin classes. $\widehat{A}$-genus is an integer number and it is an even integer for $\text{dim }4(\text{mod }8)$. So, the index of $\displaystyle{\not}D_k$ takes values in $\mathbb{Z}_2$ for $k\equiv 1 \text{ and }2 (\text{mod }8)$ and in $\mathbb{Z}$ for $k\equiv 0 \text{ and }4 (\text{mod }8)$.

If we choose $k=s-n$, the Clifford chessboard can be turned into the index table of $\displaystyle{\not}D_{s-n(\text{mod }8)}$. Since $\text{ind}(\displaystyle{\not}D_{s-n(\text{mod }8)})$ corresponds to 0 or takes values in $\mathbb{Z}$ and $\mathbb{Z}_2$, a simple comparison shows that the Clifford chessboard turns into the peroidic table of topological phases of real classes \cite{Ertem}. These Dirac operators correspond to the Dirac Hamiltonians of topological insulators and the index of Dirac operators determine the topological invariants characterizing topological phases.

Moreover, $\mathbb{Z}$ and $\mathbb{Z}_2$ groups appearing in the index of Dirac operators are related to the K-theory groups of vector bundles. A minimal free-abelian group obtained from a monoid $A$ (group without inverse) is defines as its K-group $K(A)$.
($K(\mathbb{N})=\mathbb{Z}$ for addition operation and $K(\mathbb{Z})=\mathbb{Q}$ for multiplication operation). Isomorphism classes of vector bundles constitute a monoid structure. If we consider stable equivalence classes, that is for two vector bundles $E$ and $F$, if we have
\begin{equation}
E\oplus I^n=F\oplus I^m
\end{equation}
where $I^n$ is a $n$-dimensional trivial bundle, then we say that $E$ and $F$ are stably equivalent. Stable equivalence classes constitute a group whichis the K-group of isomorphism classes monoid. Index of Dirac operators have 1:1 correspondence with K-groups of sphere bundles and these K-groups which are denoted by $K_{\mathbb{R}}^{-(s-n)(\text{mod }8)}(pt)$ exactly gives the periodic table of real classes \cite{Lawson Michelson}. These bundles correspond to the Bloch bundles of topological insulators (in the continuum limit) and periodic table is a result of K-groups of Bloch bundles. There are also relations with Groethendieck groups and Clifford algebra extensions which give the symmetry properties with respect to $T$, $C$ and $S$ \cite{Ertem}.

Complex classes are originated from the table of complex Clifford algebras

\quad\\
{\centering{
\begin{tabular}{c c c}

$\mathbb{C}l_{s-n (\text{mod }2)}$ & $n=0$ & $1$ \\ \hline
$s=0$ & $\mathbb{C}$ & $\mathbb{C}\oplus\mathbb{C}$ \\
$1$ & $\mathbb{C}\oplus\mathbb{C}$ & $\mathbb{C}$ \\ 

\end{tabular}}
\quad\\
\quad\\
\quad\\}

and the index theorem
\begin{eqnarray}
\text{ind}(\displaystyle{\not}\mathbb{D}_k)&=&\left\{
                                                                               \begin{array}{ll}
                                                                                 \text{Td}(M), & \hbox{ for $k \text{ even}$ } \\
                                                                                 0, & \hbox{ for $k \text{  odd}$}                                                                                 
                                                                               \end{array}
                                                                             \right.
\end{eqnarray}
where $\text{Td}(M)$ denotes the Todd class which can be written in terms of Chern classes. Similar considerations in the real classes can also give the periodic table for complex classes and the whole periodic table of topological insulators and superconductors can be derived in this way \cite{Ertem}.

\end{document}